\documentclass[a4paper]{article}          
\usepackage{graphicx}
\usepackage{color}
\usepackage{amsfonts}
\usepackage{amsmath}
\usepackage{mathtools}
\usepackage[caption=false]{subfig}
\usepackage{bm}
\usepackage{booktabs}
\usepackage[toc,page]{appendix}
\usepackage{natbib}
\newcommand{\difd}{\textup{d}}
\usepackage[noblocks]{authblk}

\newcommand{\norm}[1]{\left\lVert#1\right\rVert}

\begin{document}
\title{A ``poor man's'' approach to topology optimization of natural convection problems}
\author{Janus Asmussen}
\author{Joe Alexandersen}
\author{Ole Sigmund}
\author{Casper Schousboe Andreasen}
\affil{Section for Solid Mechanics, Department of Mechanical Engineering, Technical University of Denmark, csan@mek.dtu.dk}
\date{}
\maketitle

\begin{abstract}
Topology optimization of natural convection problems is computationally expensive, due to the large number of degrees of freedom (DOFs) in the model and its two-way coupled nature. Herein, a method is presented to reduce the computational effort by use of a reduced-order model governed by simplified physics. The proposed method models the fluid flow using a potential flow model, which introduces an additional fluid property. This material property currently requires tuning of the model by comparison to numerical Navier-Stokes based solutions. Topology optimization based on the reduced-order model is shown to provide qualitatively similar designs, as those obtained using a full Navier-Stokes based model. The number of DOFs is reduced by 50\% in two dimensions and the computational complexity is evaluated to be approximately 12.5\% of the full model. We further compare to optimized designs obtained utilizing Newton's convection law.\\

\noindent\textit{topology optimization, natural convection, reduced-order model, potential flow, heat sink design}
\end{abstract}

\section{Introduction} \label{sec:intro}
Natural convection is the study of temperature-driven flow. Differences in spatial temperature cause density gradients in the fluid, which in turn cause fluid motion due to buoyancy. This then induces transport of energy in the form of convection. Natural convection is thus a strongly two-way coupled problem, in which the temperature field has great impact on the velocity field and vice versa.

In this paper, a reduced-order model is presented based on potential flow. The model can be shown to be equivalent to Darcy's law for flow in porous media, with a fictitious permeability in the fluid domain.
This flow model is coupled to the heat transport, which is modeled by a convection-diffusion equation, through the use of the Boussinesq approximation. In comparison to modeling the full Navier-Stokes equations, the reduction in system dimension is significant, as the velocities can be computed explicitly from the pressure and temperature fields, allowing for omission of all velocity degrees-of-freedom (DOFs) from the system solve and computing them as a post-processing step.  

The huge savings in CPU time come at the expense of reduced accuracy of the physical modeling. However, as will be demonstrated, optimized designs are remarkably similar to those obtained with the full Navier-Stokes model.

Topology optimization originates from solid mechanics \citep{Bendsoe1988,Bendsoe2003}, but has been extended to many other physics \citep{Deaton2014} in the recent decades. Curiously, topology optimization was applied to thermomechanical structures \citep{Rodrigues1995, Sigmund2001, Yin2002} before pure heat conduction problems \citep{Bendsoe2003,Donoso2004c,Gersborg-Hansen2006a}. 
Recently, pure heat conduction problems were revisited and it was revolutionarily shown that optimal structures for the famous volume-to-point problem are lamelar needle-like structures \citep{Yan2018}, and not the commonly observed tree-like branching structures.
Shifting perspective to fluid flow problems, topology optimization was first applied to Stokes flow by \citet{Borrvall2003} using a resistance term based on the out-of-plane channel flow thickness. This approach was later extended to Navier-Stokes flow by \citep{Gersborg-Hansen2005}. Over time, the approach has shifted to the use of a Brinkman penalization term to model an immersed solid geometry embedded in a fluid domain \citep{Brinkman1947,Angot1999,Olesen2006}. An alternative approach has been the interpolation between Stokes and Darcy flow \citep{Guest2006a,Wiker2007}. Topology optimization has been extended to passive and reactive transport problems \citep{Thellner2005,Andreasen2009,Okkels2007}. Topology optimization of conjugate heat transfer problems is an active research area, starting with the work by \citet{Dede2009} and \citet{Yoon2010}, and with many works being published the last few years \citep{Koga2013,Marck2013,Haertel2015,Yaji2016,Laniewski-Wollk2016,Haertel2017,Zeng2018,Haertel2018, 
Dugast2018, Subramaniam2018, Yaji2018, Dilgen2018}.

When considering natural convective conjugate heat transfer problems, the literature is sparser. \citet{Alexandersen2014} first presented the application of topology optimization to two-dimensional natural convective heat sinks and micropumps. The work was extended to three dimensions \citep{Alexandersen2016a} using a large scale parallel computational framework. This framework was subsequently applied to a real-life industrial problem, namely the design of passive coolers for light-emitting diode (LED) lamps \citep{Alexandersen2015,Alexandersen2018}. Their practical realization using metallic additive manufacturing (AM) \citep{Lazarov2018} and AM-assisted investment casting \citep{Lei2018} has shown topology optimized designs to be superior to standard pin-fin designs. Besides the work of Alexandersen et al, a boundary-conforming levelset approach for transient problems was presented by \cite{Coffin2016a} and recently, a density-based spectral method was presented in the work of Saglietti \citep{
Saglietti2018,Saglietti2018a}.

Due to the nature of conjugate heat transfer problems, they are computationally expensive to solve numerically. Partly due the fact that the number of unknown fields in the governing equations is large, but also due to the highly non-linear nature of the coupled equations. Furthermore, for external flow and open boundary problems, a large computational domain surrounding the structure is needed, further increasing the computational cost. In gradient-based optimization, which is an iterative process requiring hundreds, if not thousands, of field evaluations, these factors result in a significant computational burden. One way to circumvent this, has historically been to significantly simplify the problem at hand by introducing Newton's law of cooling for convective boundary conditions. By doing so, one only has to model the scalar heat transfer problem. Various approaches have been presented to introducing the design-dependency of the surface-based boundary condition into the framework of topology optimization 
\citep{Sigmund2001,Yin2002,Moon2004,Bruns2007,Alexandersen2011,Coffin2016,Zhou2016}.
A common simplification has been to use a constant average convection coefficient across all of the fluid-solid interface. This is common engineering practise for analysis, however, when introducing this to topology optimization, many problems are observed \citep{Alexandersen2011,AlexandersenThesis,Coffin2016,Zhou2016}, such as internal closed voids and over-prediction of the total heat transfer. Various ways to remedy these problems have been introduced \citep{Iga2009, Coffin2016a, Zhou2016} with varying success, but most recently, \citet{Joo2017,Joo2018} presented an approach, where the distance between features is computed based on a global search of interface elements and used to calculate a spatially-varying convection coefficient based on correlations. This approach appears to be successful to including some knowledge of the flow into the topology optimization process, but requires the choice of correlations based on assumptions of the geometry.

In this paper, an alternative approach is presented. Instead of removing the modeling of the flow field completely, a simplified reduced-order flow model is used in place of the full Navier-Stokes equations. The inspiration comes from the paper by \citet{Zhao2018}, which considers turbulent forced convective channel cooling modeled by a Darcy flow approximation. This work uses a similar concept, to simplify the flow modeling. However now, natural convection (fully-coupled conjugate heat transfer) is considered and includes a stronger physical foundation for the reduced-order model.

The paper is organized as follows: the reduced-order model is introduced in Section 2; the discretized system is presented in Section 3; topology optimization is discussed in Section 4; calibration of the reduced-order model and comparison to the full model is presented in Section 5; numerical results for two examples are presented in Section 6; Section 7 finally covers a discussion and conclusions of the paper.

\section{Governing equations}
\subsection{Heat transfer}
In order to model conjugate heat transfer driven by natural convection, a domain, $\Omega$, is considered consisting of two non-overlapping subdomains: a fluid subdomain, $\Omega_f \subset \Omega$; and a solid subdomain, $\Omega_s \subset \Omega$. In the fluid domain, both the flow and heat transfer is to be modeled. In the solid domain, only the heat transfer is to be modeled. 

The temperature field is modeled by the convection-diffusion equation:
\begin{equation}
\rho c_p u_i \frac{\partial T}{\partial x_i} - \frac{\partial}{\partial x_i} \left( k(\boldsymbol{x}) \frac{\partial T}{\partial x_i} \right) = Q(\boldsymbol{x})
 \label{eq:govEq_advDiff}
\end{equation}
where $\rho$ is the density, $c_p$ is the specific heat capacity, $u_i$ are the velocity components, $T$ is temperature, $x_i$ are the spatial coordinates, $k$ is the spatially-varying conductivity and $Q$ is a spatially-varying volumetric heat source term. The first term accounts for convective heat transfer, while the second term accounts for diffusive heat transfer. The above unified equation models the heat transfer in both the fluid and solid domains, by assuming $u_i = 0 \text{ for } \boldsymbol{x}\in\Omega_s$ and by varying the thermal conductivity in the two subdomains:
\begin{equation}
k(\boldsymbol{x}) = \left\{\begin{array}{lr}
        k_f & \text{ for } \boldsymbol{x}\in\Omega_f \\
        k_s & \text{for } \boldsymbol{x}\in\Omega_s 
        \end{array} \right.
\end{equation}
Furthermore, the volumetric heat source is only active in a subset of the solid domain, $\Omega_Q \subset \Omega_s$:
\begin{equation}
Q(\boldsymbol{x}) = \left\{\begin{array}{lr}
        0 & \text{ for } \boldsymbol{x}\in\Omega_s \\
        Q_0 & \text{for } \boldsymbol{x}\in\Omega_Q 
        \end{array} \right.
\end{equation}

The following boundary conditions are appended to the governing equation to achieve a well-posed system:
\begin{align}
&& T &= T^* &\textrm{ on } S_T \\
&&  k \frac{\partial T}{\partial x_i}n_i &= q_h &\textrm{ on } S_h \label{eq:govEq_heatBC}
\end{align}
where $T^*$ is a prescribed temperature on the boundary $S_T$ and $q_h$ is a prescribed heat flux on the boundary $S_h$ with $n_i$ being the unit normal vector to the boundary.

\subsection{Fluid flow}
The reduced-order, or simplified, flow model will be derived starting from the Navier-Stokes equations and reducing it based on various assumptions.
The incompressible Navier-Stokes equations are given as:
\begin{equation}
\rho u_{j}\frac{\partial u_i}{\partial x_j} - \mu \frac{\partial}{\partial x_j}\left( \frac{\partial u_i}{\partial x_j} + \frac{\partial u_j}{\partial x_i} \right) + \frac{\partial p}{\partial x_i} = \rho g_{i} \label{eq:app_NS}
\end{equation}
where $g_i$ is the gravitational acceleration.
In order to model natural convection due to density variations, the Boussinesq approximation is used:
\begin{equation}
\rho g_{i} \approx \rho_{0}(1-\beta(T-T_0))g_{i}
\end{equation}
where $\rho_0$ is the density at the reference temperature, $T_0$, and $\beta$ is the coefficient of volumetric expansion. This approximation is introduced into the volumetric gravity force of Eq. \ref{eq:app_NS}, while using the reference density for the inertial term, giving:
\begin{multline}
\rho_{0} u_{j}\frac{\partial u_i}{\partial x_j} - \mu \frac{\partial}{\partial x_j}\left( \frac{\partial u_i}{\partial x_j} + \frac{\partial u_j}{\partial x_i} \right) + \frac{\partial p}{\partial x_i} \\ = \rho_{0}(1-\beta(T-T_0)) g_{i} \label{eq:app_boussNS}
\end{multline}
Next, buoyancy is assumed to be the dominant forcing of the system and, thus, inertia is negligible:
\begin{equation}
\left| \rho_{0} u_{j}\frac{\partial u_i}{\partial x_j} \right| \ll \left| \rho_{0}(1-\beta(T-T_0)) g_{i} \right| \label{eq:app_negligble}
\end{equation}
Introducing this assumption into Eq. \eqref{eq:app_boussNS}, yields:
\begin{equation}
- \mu \frac{\partial}{\partial x_j}\left( \frac{\partial u_i}{\partial x_j} + \frac{\partial u_j}{\partial x_i} \right) + \frac{\partial p}{\partial x_i} = \rho_{0}(1-\beta(T-T_0)) g_{i} \label{eq:app_midpoint}
\end{equation}
To further simplify the flow equation, it is assumed that the viscous resistance force is linearly dependent on the velocity\footnote{While this simplification may seem unusual, comparing the two expressions for various test problems has shown remarkable similarity away from the viscous boundary layer.} and can be described as:
\begin{equation}
\mu \frac{\partial}{\partial x_j}\left( \frac{\partial u_i}{\partial x_j} + \frac{\partial u_j}{\partial x_i} \right) \approx -\bar{\mu}u_i
\end{equation}
where $\bar{\mu}$ is a new material parameter (with the unit of $\frac{\textup{Pa}}{\textup{m}^2 \textup{s}}$) for the reduced-order model.
Inserting this into Eq. \eqref{eq:app_midpoint} and rearranging gives:
\begin{equation}
u_i  = -\frac{1}{\bar{\mu}(\boldsymbol{x})} \left( \frac{\partial P}{\partial x_i} + \rho_{0}\beta(T-T_0) g_{i} \right) \label{eq:gov_velExp}
\end{equation}
where the constant term has been absorbed into the pressure gradient term with $P = p - \rho_{0}g_{i}x_{i}$, where $P$ is the modified pressure including the gravitational head at constant density $\rho_0$.
Theoretically, $\bar{\mu}$ must be infinity in the solid domain, to ensure non-existent velocities\footnote{Numerically, this requirement must be relaxed as discussed in Section \ref{sec:matinterp}.}. In the fluid domain, $\bar{\mu}$ is a new governing material property for the simplified model, that must be tuned in order to use the simplified model. Thus, the reduced-order material parameter varies in the domain, $\Omega$, as:
\begin{equation}
\bar{\mu}(\boldsymbol{x}) = \left\{\begin{array}{lr}
        \bar{\mu}_{f} & \text{ for } \boldsymbol{x}\in\Omega_f \\
        \infty & \text{for } \boldsymbol{x}\in\Omega_s 
        \end{array} \right.
\end{equation}

The above can be reposed as a velocity potential. However, due to the non-homogeneous nature of the temperature-dependent buoyancy, an additional term must be included:
\begin{equation}
u_i  = \frac{\partial \phi}{\partial x_i} + f_i
\end{equation}
where the velocity potential, $\phi$, is given by the expression:
\begin{equation}
\phi  = -\frac{1}{\bar{\mu}(\boldsymbol{x})}{P} \label{eq:gov_velpotential}
\end{equation}
and the forcing, $f_i$, is given by:
\begin{equation}
f_i = -\frac{1}{\bar{\mu}(\boldsymbol{x})} \rho_0 \beta (T-T_0) g_i
\end{equation}

In order to get a governing equation for the pressure, $P$, the velocity expression in Eq. \eqref{eq:gov_velExp} is inserted into the incompressibility condition, $\frac{\partial u_i}{\partial x_i} = 0$, giving:
\begin{equation}
\frac{\partial}{\partial x_i}\left( \frac{1}{\bar{\mu}(\boldsymbol{x})} \frac{\partial P}{\partial x_i} \right) = - \frac{\partial}{\partial x_i} \left( \frac{\rho_{0}\beta}{\bar{\mu}(\boldsymbol{x})}(T-T_0) g_{i} \right) \label{eq:presEqu}
\end{equation}
This is a Poisson equation for the pressure, $P$, with a spatially-varying coefficient and a spatially-varying forcing, essentially based on the gradient of the temperature field.
The following boundary conditions are appended to the pressure equation, Eq. \eqref{eq:presEqu}, to achieve a well-posed system:
\begin{align}
&& P &= P^* &\textrm{ on } S_p \\
&& u_i n_i &= q_f &\textrm{ on } S_u  \label{eq:govEq_BCs}
\end{align}
where $P^*$ is a prescribed pressure on the boundary $S_p$ and $q_f$ is a prescribed normal velocity (or potential flux) on the boundary $S_u$.
\\ \indent
By setting $\bar{\mu} = \frac{\mu}{\kappa}$, Darcy's law for porous media flow under buoyancy is recovered:
\begin{equation}
u_i  = -\frac{\kappa}{\mu} \left( \frac{\partial P}{\partial x_i} + \rho_{0}\beta(T-T_0) g_{i} \right)
\end{equation}
where $\kappa$ is the permeability of a porous media. The permeability can be seen as an artificial material parameter used to tune the reduced-order model. This interpretation leads to the comparability with the approach presented by \citet{Zhao2018} that serves as inspiration for this work.

\subsection{Navier-Stokes-Brinkman model for comparison}
To calibrate the reduced-order model and as a performance reference for optimised designs, the full-order model is based on the work by \citet{Alexandersen2014}. A Navier-Stokes-Brinkman (NSB) formulation is used, where the governing equations for the fluid flow are given as:
\begin{align}
\rho_0 u_j&\frac{\partial u_i}{\partial x_j}-\mu\frac{\partial}{\partial x_j}\left( \frac{\partial u_i}{\partial x_j}+\frac{\partial u_j}{\partial x_i}
\right)
+\frac{\partial p}{\partial x_i}
+\alpha(\boldsymbol{x}) u_i\nonumber\\
\qquad&=-\rho_0\beta g_i (T-T_0) \label{eq:NSB}\\
\frac{\partial u_i}{\partial x_i}&=0
\end{align}
where $\alpha(\boldsymbol{x})$ is the Brinkman penalisation coefficient:
\begin{equation}
\alpha(\boldsymbol{x}) = \left\{\begin{array}{lr}
        0 & \text{ for } \boldsymbol{x}\in\Omega_f \\
        \infty & \text{for } \boldsymbol{x}\in\Omega_s 
        \end{array} \right.
\end{equation}
As for the reduced-order model, the heat transfer is governed by the convection-diffusion equation given in Eq. \eqref{eq:govEq_advDiff}.

\subsection{Dimensionless numbers}
In the above developments, dimensional quantities were used. However, to aid direct comparison with the previous work by \citet{Alexandersen2014}, the dimensionless Grashof number will be used. Herein, the Grashof number is defined as:
\begin{equation}
\textrm{Gr} = \frac{\textrm{Ra}}{\textrm{Pr}}
\end{equation}
where $\textrm{Ra}$ is the Rayleigh number and $\textrm{Pr}$ is the Prandtl number.
The Rayleigh number describes the relation between natural convection and diffusion and is defined as:
\begin{equation}
\textrm{Ra} = \frac{g \beta \Delta T H^3 {\rho_0}^{2} c_p}{\mu k_{f}}
\end{equation}
where $\Delta T$ is a reference temperature difference. In the contrary, the Prandtl number is based solely on the fluid material properties:
\begin{equation}
\textrm{Pr} = \frac{c_{p} \mu}{k_{f}}
\end{equation}
This gives the following final expression for the Grashof number:
\begin{equation}
\textrm{Gr} = \frac{g \beta \Delta T H^3 {\rho_{0}}^{2}}{\mu^2},
\label{eq:dimless_grashof}
\end{equation}

\section{Numerical solution}
The presented methodology is implemented in MATLAB with the following implementation details.

\subsection{Discretization}
The governing pressure equation, Eq. \eqref{eq:presEqu}, is discretized using  the Galerkin method. The strong form is multiplied by a weight function, $w$, integratation is performed over the domain and integration-by-parts is used to introduce the natural boundary condition:
\begin{align}
\int_\Omega \frac{1}{\bar{\mu}} \left( \frac{\partial w}{\partial x_i}  \left( \frac{\partial P}{\partial x_i}  +  \beta \rho_0 (T-T_{0}) g_i \right) \right) d \Omega  =  \int_{S_u} w q_f dS ,
\end{align}
\\ \indent
Due to well-known stability issues for highly convective problems, the convection-diffusion equation is discretized using a Streamline-Upwind Petrov-Galerkin (SUPG) method  \citep{Brooks1982} with a modified weight function:
\begin{equation}
w^* = w + \tau u_j \frac{\partial v}{\partial x_j}
\end{equation}
where $\tau$ is a stabilisation parameter.
The perturbation can be interpreted as an addition of artificial diffusion to the problem \citep{Brooks1982,Fries2004}. The strong form is multiplied by the modified weight function, integration is performed over the domain and integration-by-parts is used to introduce the natural boundary condition:
\begin{align}
\begin{split}
- \int_\Omega w^* \frac{\rho_0 c_p}{\bar{\mu}} \left(  \frac{\partial P }{\partial x_i} + \beta\rho_0(T-T_0)g_i \right) \frac{\partial T}{ \partial x_i} d \Omega  \\
 +\int_\Omega k \frac{\partial w}{\partial x_i} \frac{\partial T}{\partial x_i} d \Omega =
\int_{S_h} w q_h dS + \int_\Omega w^* Q d \Omega,
\end{split}
\end{align}
where second order terms, arising from the multiplication of the perturbation with the diffusion term, are neglected due to the use of bilinear shape functions in the following.
The stabilization parameter, $\tau$, has been chosen as (\citet{Donea2003,Shakib1991}):
\begin{align}
\tau = \left(\left(\frac{2\| \mathbf{u}\|_2}{h}\right)^2+9\left(\frac{4\nu}{h^2}\right)^2\right)^{-1/2}
\end{align}
where the velocity expression in Eq. \eqref{eq:gov_velExp} is used for calculating the local velocity.

The field variables, $p$ and $T$, are discretized using bilinear shape functions. The velocity given by Eq. \eqref{eq:gov_velExp} is evaluated in the element centroid and is thus elementwise constant. The design field is discretized using elementwise constant variables, in turn rendering the material parameters to be elementwise constant. The monolithic finite element discretization of the problem ensures continuity of the fields, as well as fluxes across fluid-solid interfaces.

\subsection{Non-linear solver}
Newton's method is used to solve the non-linear system of equations, where the residual of the discretized system is of the form:
\begin{equation}
\mathbf{R}(\mathbf{s}) = \mathbf{K}(\mathbf{s}) \mathbf{s}-\mathbf{f} = \mathbf{0},
\end{equation}
where $\mathbf{K}$ is the non-linear system matrix, $\mathbf{f}$ is the load vector, and:
\begin{equation}
\mathbf{s} =
\begin{Bmatrix}
\mathbf{p} \\
\mathbf{t}
\end{Bmatrix}
\end{equation}
is the global solution vector composed of pressure and temperature variables.

Based on a previous vector $\mathbf{s}_r$, the solution is iteratively updated based on a linearization of the residual. The solution at step $r + 1$ is then:
\begin{equation}
\mathbf{s}_{r+1} = \mathbf{s}_r - \lambda \left(\frac{\partial \mathbf{R}(\mathbf{s}_r)}{\partial \mathbf{s}}\right)^{-1}\mathbf{R}(\mathbf{s}_r),
\end{equation}
where $\lambda$ is a damping coefficient. The solution is accepted when $||\mathbf{R}||_2 / ||\mathbf{R}_0||_2 \leq 10^{-4}$ where $\mathbf{R}_0$ is the initial residual. 
For tightly coupled highly non-linear systems, this rather relaxed requirement is used because convergence to tighter tolerances can be difficult.
The damping coefficient is updated in each Newton iteration by a second-order polynomial fit as described by \citet{Alexandersen2016a}.

The described algorithm performs well when the solver is initiated with a good initial point. This is generally the case, as each non-linear solve is started with the solution from the previous design iteration. Thus, if the problem converges for the initial design, it tends to converge throughout all iterations. If this is not the case, it is possible to recover by performing a ramping of either the heat flux or the Grashof number.

\subsection{Computational cost}
Under the assumption that the performance of the Newton solver is independent of the problem size, $n$, then the computational complexity can be bounded by:
\begin{equation}
c_\textrm{total} \leq c_\textrm{Newton}m
\end{equation}
where $m$ is the number of Newton iterations and $c_\textrm{Newton}$ is the complexity of a single iteration.
The computational complexity of a single iteration can be decomposed as follows:
\begin{equation}
c_\textrm{Newton} = c(\mathbf{R}) + c\left(\frac{\partial \mathbf{R}(\mathbf{s})}{\partial \mathbf{s}}\right) + c_\textrm{direct}
\end{equation}
where $c(\mathbf{R})$ is the complexity of the residual assembly, $c\left(\frac{\partial \mathbf{R}(\mathbf{s})}{\partial \mathbf{s}}\right)$ is the complexity of assemblying the Jacobian, and $c_\textrm{direct}$ is the complexity of the linear solve (herein using a direct solver):
\begin{equation}
c_\textrm{direct} = \mathcal{O}\left(\frac{2}{3}n^3\right)
\end{equation}

The computational complexity of the assembly routines are difficult to ascertain. However, it seems reasonable to assume that the complexity of the assembly procedures for the reduced-order and full-order models to be of similar order, although it must of course be lower for the reduced-order model due to the fewer DOFs and fewer element-level matrices. The number of Newton iterations, $m$, is related to the wanted precision and assumed to be constant and the same for both methods. The limiting complexity of the two models is assumed to be the cost of the direct solution of the linear systems. For two-dimensional problems, the number of DOFs per node is 2 and 4 for the reduced-order and full-order models, respectively. Therefore, the estimated reduction in computational cost is:
\begin{equation}
\frac{c^\textrm{RO}_\textrm{total}}{c^\textrm{FO}_\textrm{total}} =  \frac{1}{8}
\end{equation}
That is, the cost of the reduced-order model is estimated to be $12.5\%$ of the full-order model.

\section{Topology optimization}

\subsection{Material interpolation} \label{sec:matinterp}
A density-based topology optimization approach is applied, where a continuous design field $\gamma(\textbf{x}) \in [0; 1]$ is introduced. In the discretized setting, each finite element is attributed a piecewise constant design variable, $\gamma_i$. The design variables take the value $\gamma = 0$ in the fluid domain, $\Omega_f$, and $\gamma = 1$ in the solid domain, $\Omega_s$. In order to allow for the continuous transition between the two phases, the reduced-order material parameter and the conductivity are interpolated using the following functions:
\begin{align}
\frac{1}{\bar{\mu}(\gamma)} &= \frac{1}{\bar{\mu}_s} + \left(1-\gamma\right)^{p_{\bar{\mu}}} \left( \frac{1}{\bar{\mu}_f}-\frac{1}{\bar{\mu}_s} \right) \\
k(\gamma) &= k_f + \gamma^{p_k} (k_s-k_f),
\end{align}
where the subscript $s$ denotes the pure solid material property, the subscript $f$ denotes pure fluid and $p_{\bar{\mu}}$ and $p_k$ are penalization factors.

The problems considered in this paper generally require incrementing the penalization factors gradually to achieve binary designs. The utilized continuation sequence of penalization factors are chosen as \\ $p_{k} = \{2, 8, 16, 16\}$ and $p_{\bar{\mu}} = \{8, 8, 8, 20\}$ with an increment conducted every 50th iteration\footnote{{This continuation strategy has proven to be beneficial for the problem at hand and was chosen to allow for direct comparison with the work of \citet{Alexandersen2014}.}} or if the maximum design change is less than 1\%, which is also the final stopping criterion for the optimizer. These parameters corresponds to the settings used for the full problem \citep{Alexandersen2014} and may possibly be adapted to match the reduced-order model better.

\subsection{Optimization problem}
The goal of the considered examples is to optimize a heat sink structure with regard to the thermal compliance of the system. The thermal compliance is given as:
\begin{equation}
\psi = \int_{S_{h}} q_h T dS,
\end{equation}
Minimising the thermal compliance is equivalent to minimising the average temperature at the applied heat flux and has been successfully used as objective functional in the past \citep{Yoon2010a,Alexandersen2014,Alexandersen2016a,Alexandersen2018}.
The optimization problem is stated as:
\begin{align*}
\begin{cases}
\underset{ \bm{\gamma} \in  \mathbb{R}^n }{\text{minimize}} &: \psi \\
\textrm{subject to} &: \mathbf{R} = \mathbf{0}  \\
& \,\,\,\, \bm{\gamma}^{\text{T}} \mathbf{v} \leq V^ *\\ 
& \,\,\,\, 0 \leq \gamma_i \leq 1 \textrm{ for } i = 1 \ldots n,   
\end{cases}
\end{align*}
where $\gamma_i$ are the design variables, $\mathbf{v}$ contains the element volumes, $V^*$ is the maximum allowable solid volume and $n$ is the number of design elements.
The optimization problem is solved using the Method of Moving Asymptotes by \citet{Svanberg1987} in a nested approach, as stated above. An outer move limit on the maximum design variable change per iteration is set to 20\% in order to stabilise the design progression.

\subsection{Filtering}
Topology optimization is known to yield mesh-dependent solutions for heat conduction problems, due to numerical artifacts such as checkerboard patterns and mesh dependence. To alleviate these problems, a filter is commonly applied. Herein density filtering \citep{Bourdin2001,BruTor01} is used for regularization and ensuring a minimal lengthscale of the design. A minimal lengthscale is important for several reason, one of which is to ensure solid features that are thick enough to effectively block the flow and fluid channels that are thick enough to resolve the flow \citep{Evgrafov2006,Alexandersen2013}.

\section{Comparison and calibration} \label{sec:comparison}
\begin{figure}
\centering
\includegraphics[width=1\textwidth]{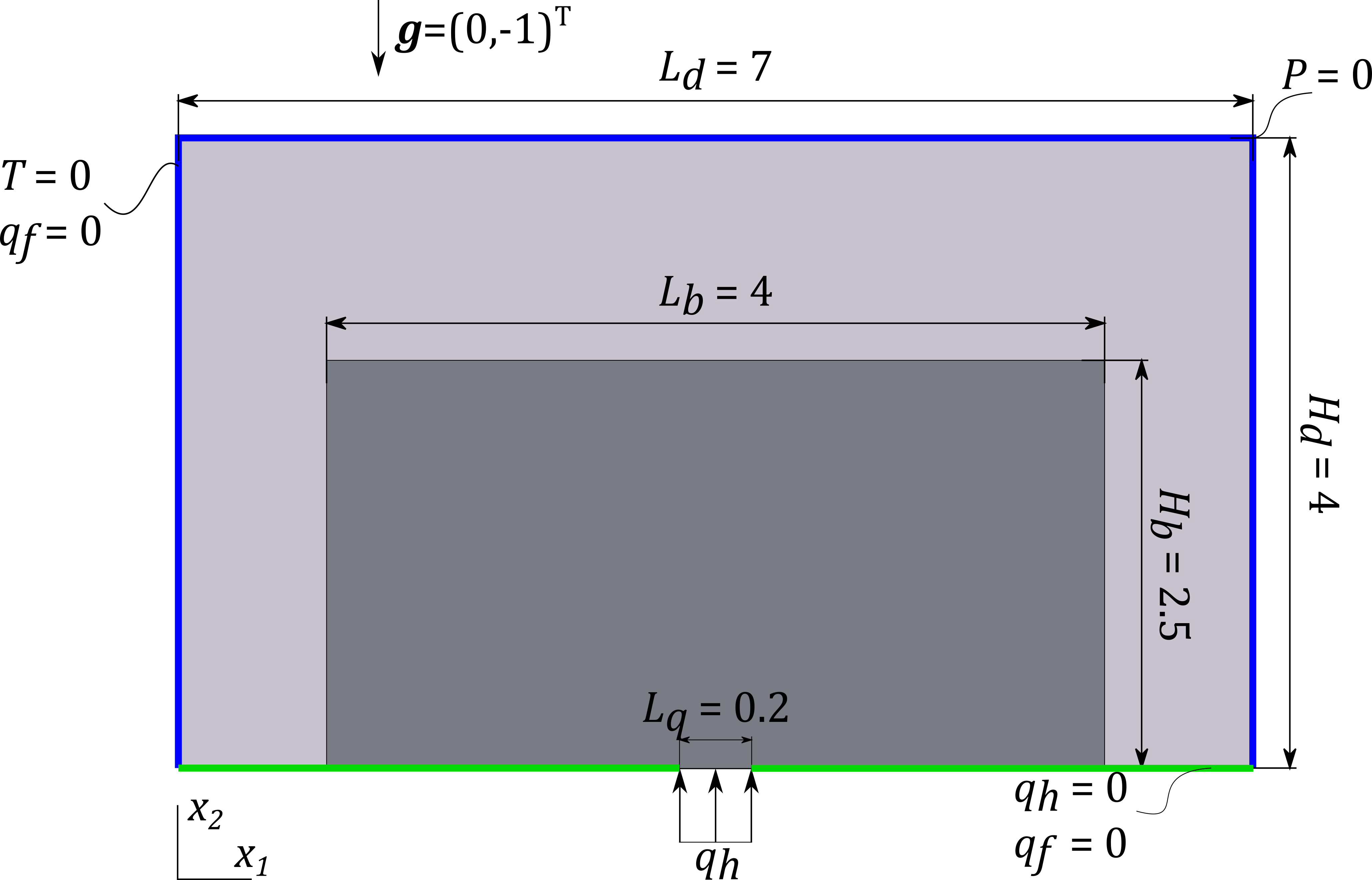}
\caption{Domain used for comparison of reduced- and full-order models. Light gray denotes fluid domain while darker gray indicates solid domain.
No fluid flow is allowed across the outer boundary $(q_f=0)$.}
\label{fig:comparison_domain}
\end{figure}
The presented potential flow model is intended to approximate flows governed by the full Navier-Stokes equations. The introduction of the reduced-order material property, $\bar{\mu}$, requires tuning in order to use the reduced-order model. Thus, it is of interest to investigate the difference between the results achieved using the proposed reduced-order model and the full-order model. For the full-order model, the laminar flow module of COMSOL Multiphysics 5.3 is used. The Brinkman term is added by appending a volume force to the Navier-Stokes equations. The problem used for tuning is shown in Fig. \ref{fig:comparison_domain}.

\begin{figure}
	\includegraphics[width=1\columnwidth]{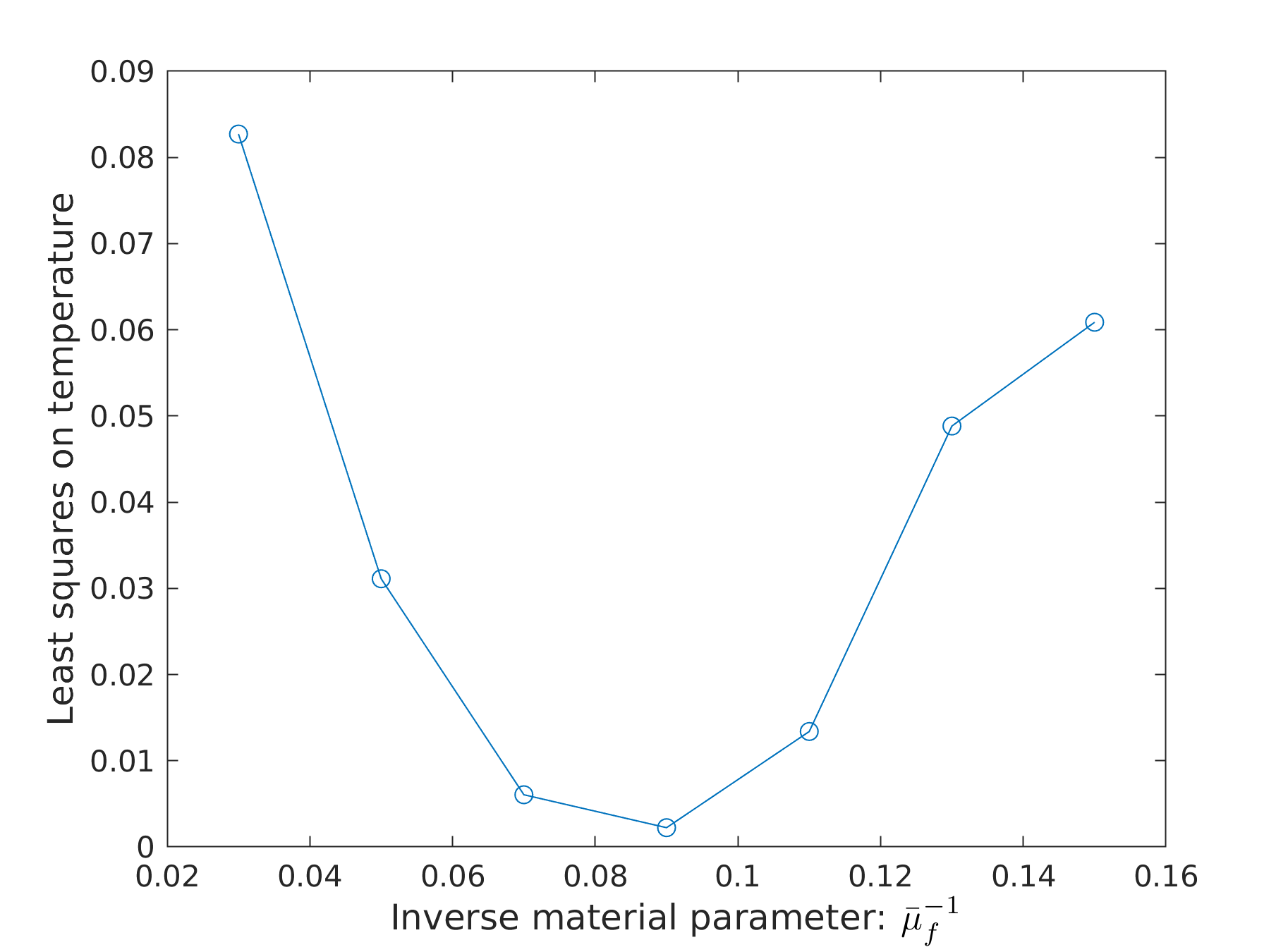}
	\caption{Least-squares error of temperature DOFs in the potential flow and NSB models for varying $\bar{\mu}_{f}$, while simulating the problem given in Fig. \ref{fig:comparison_domain}.} 
	\label{fig:LSQ}
\end{figure}

\begin{figure*}
\centering
     \subfloat[Reduced-order, temperature]{%
             \includegraphics[width=0.48\textwidth,trim=50 80 30 60 ,clip]{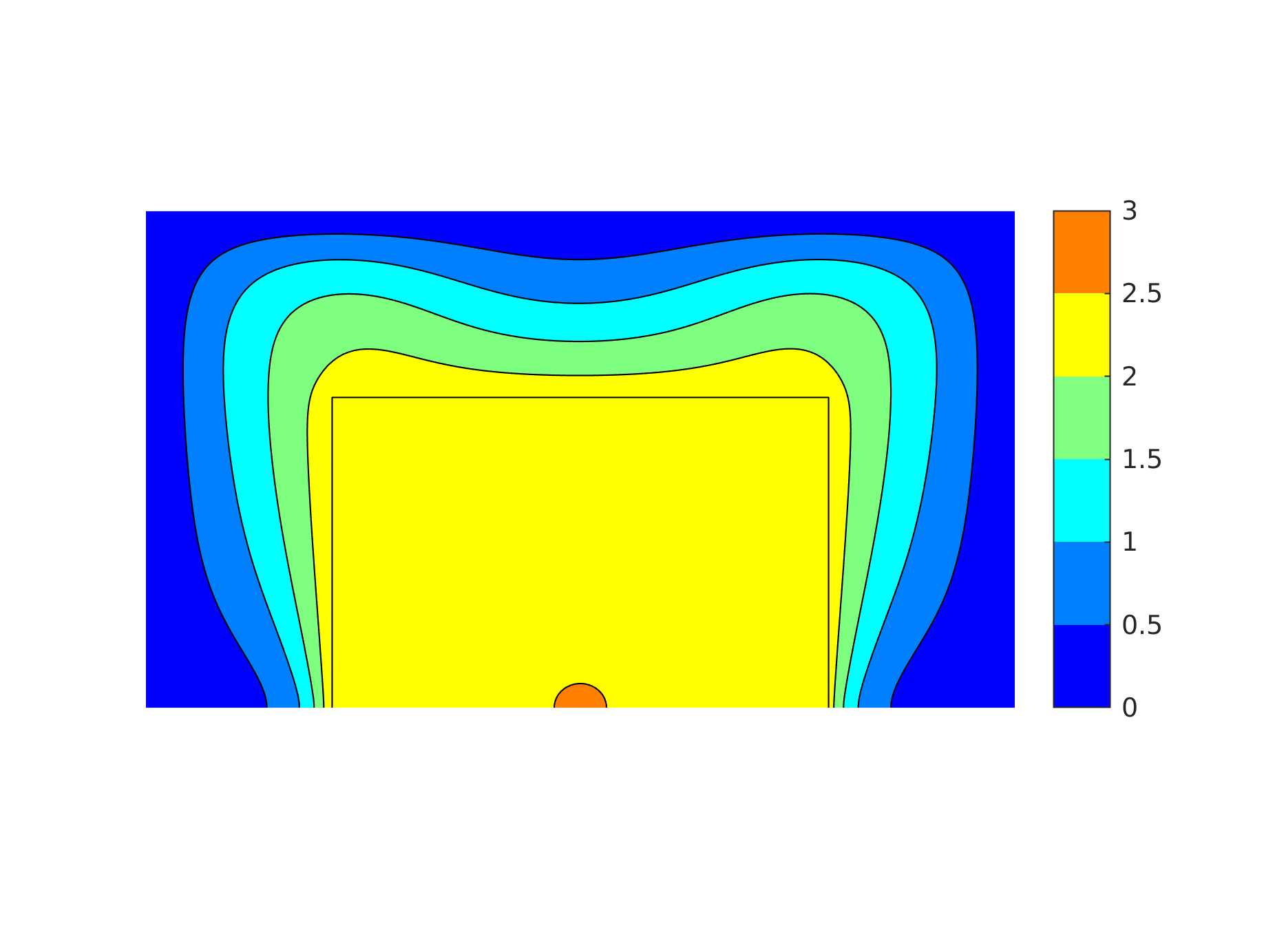}
       \label{fig:comparison_darcyTemp}
     }\hfill
     \subfloat[Navier-Stokes-Brinkman, temperature]{%
                     \includegraphics[width=0.48\textwidth,trim=50 80 30 60 ,clip]{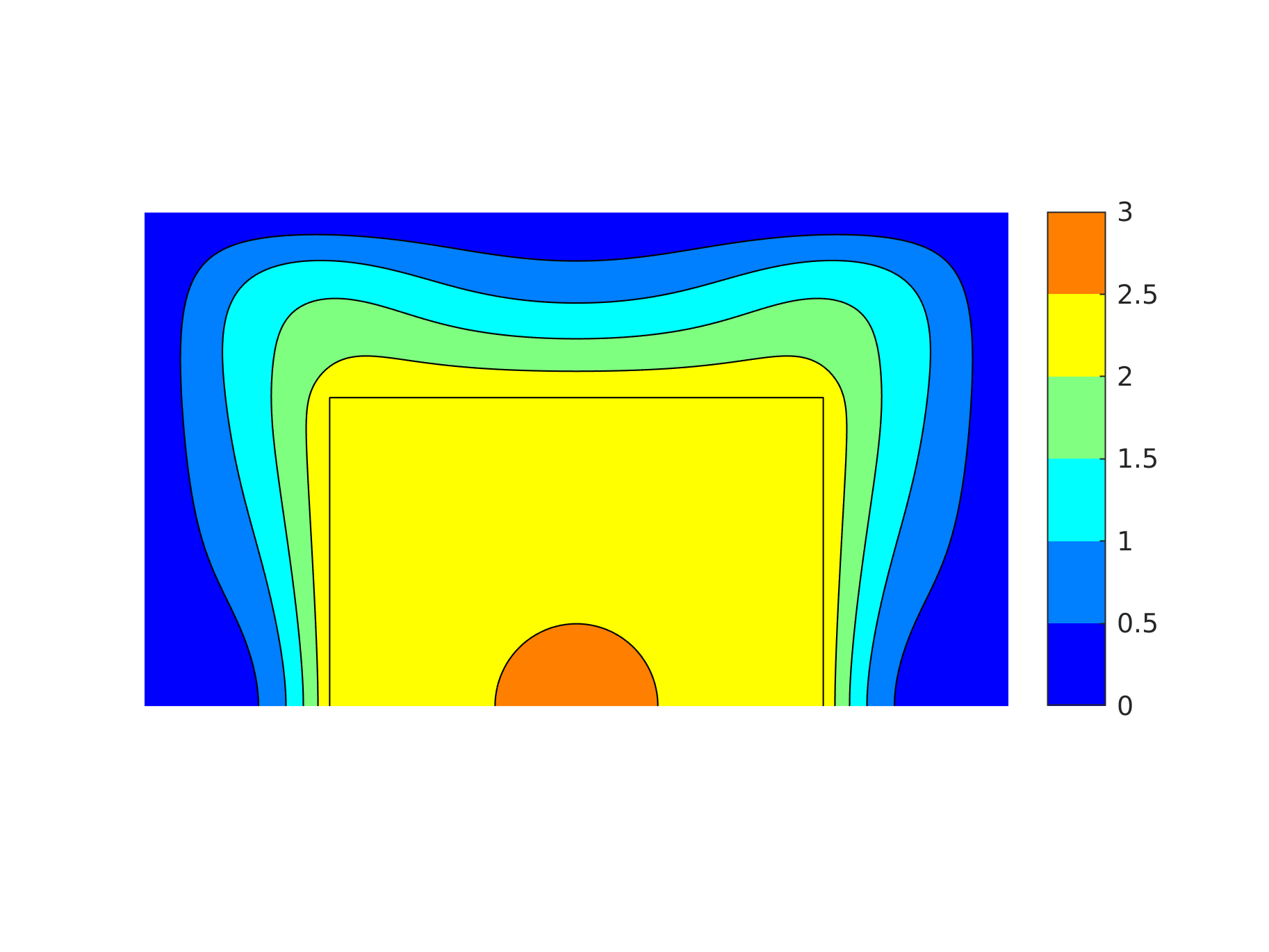}
       \label{fig:comparison_NSBTemp}
     }
     
     \subfloat[Reduced-order, velocity]{%
      \includegraphics[width=0.48\textwidth,trim=50 80 30 60 ,clip]{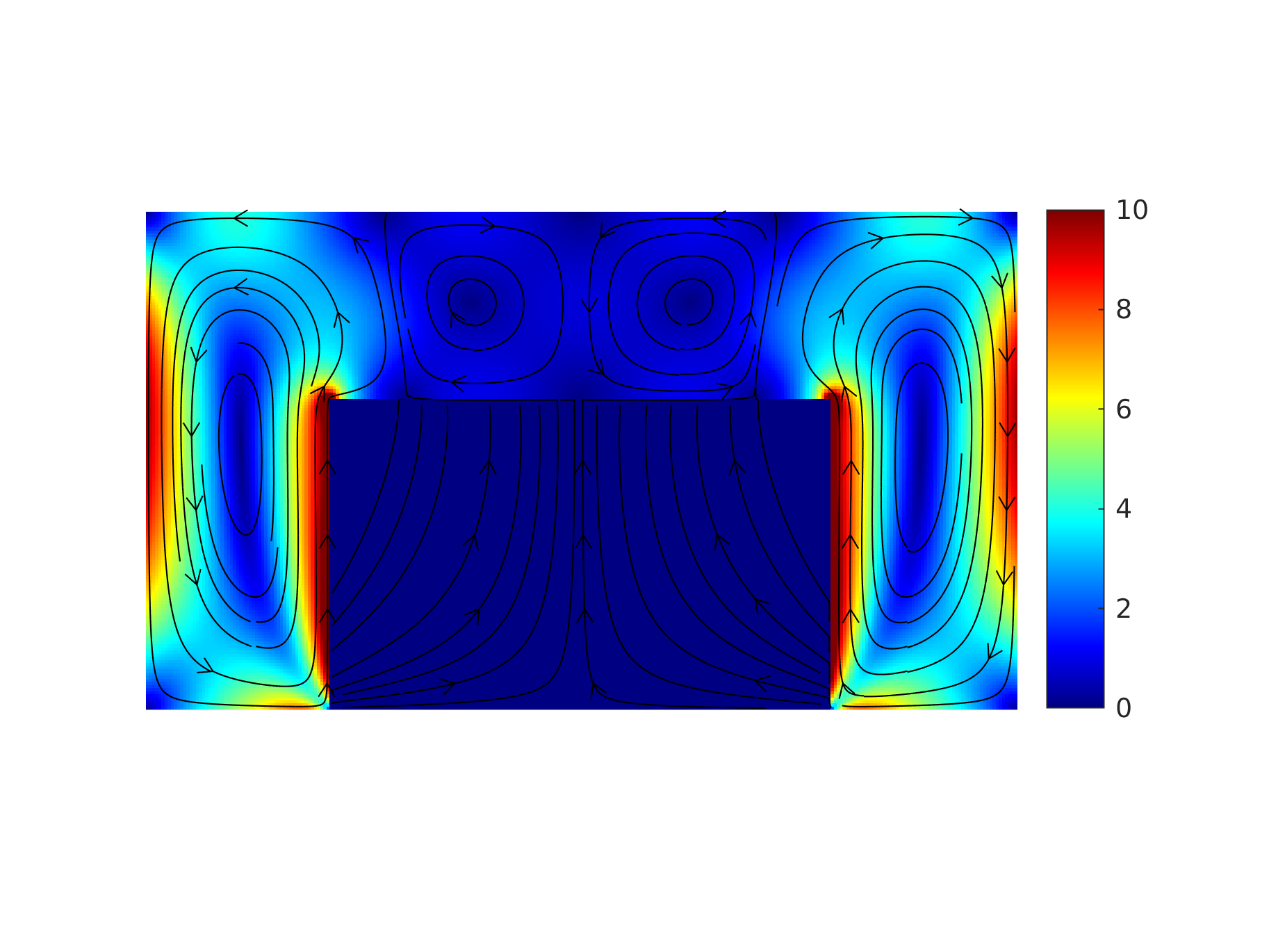}
       \label{fig:comparison_darcyVelMag}
     }\hfill
     \subfloat[Navier-Stokes-Brinkman, velocity]{%
       \includegraphics[width=0.48\textwidth,trim=50 80 30 60 ,clip]{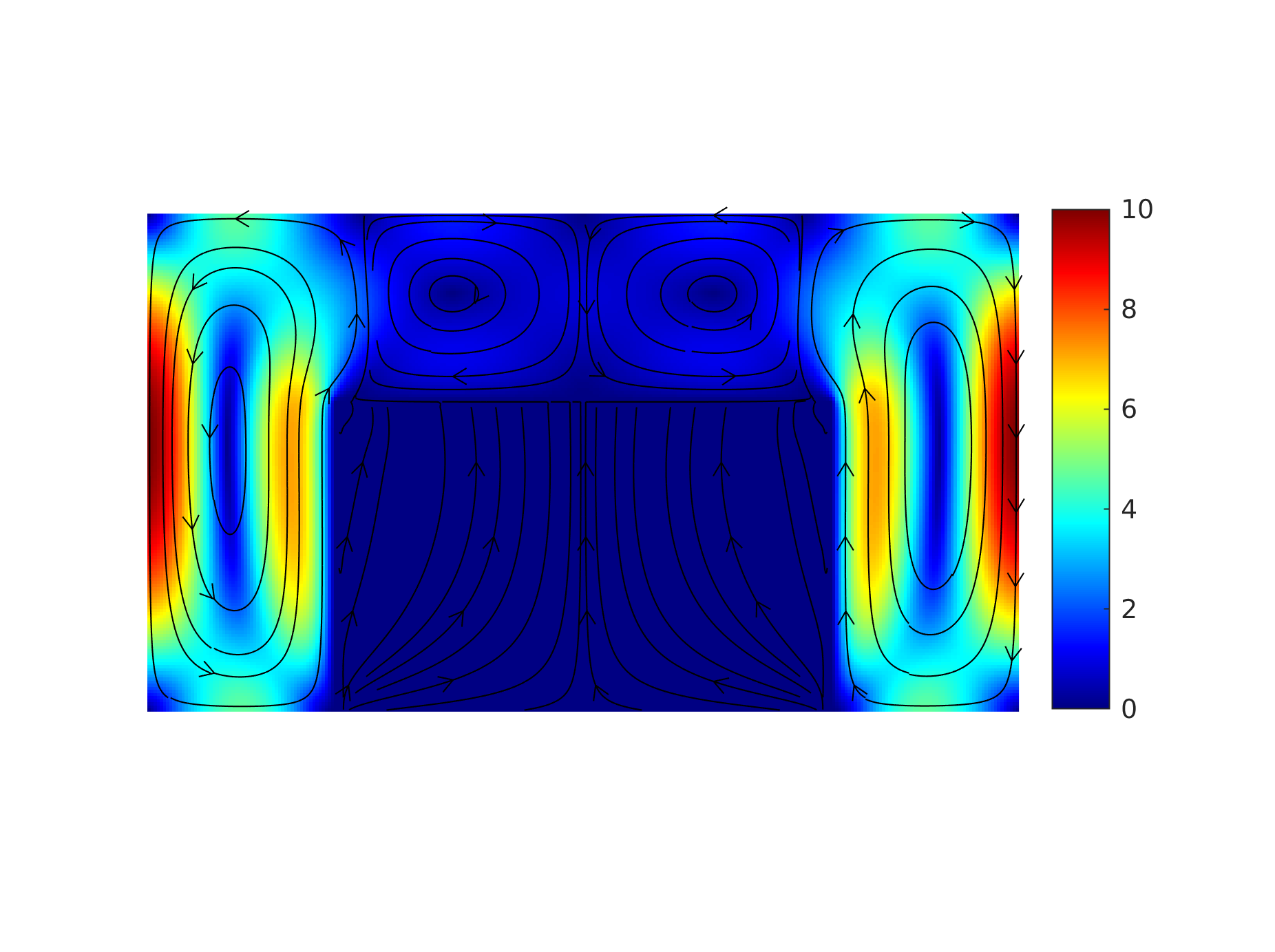}
       \label{fig:comparison_NSBVelMag}
     }
     \caption{Comparison of state fields for the test problem of \ref{fig:comparison_domain} between the potential flow and NSB models. Upper row shows temperatures and streamlines, while bottom row shows the velocity magnitude.}     
     \label{fig:comparison_designAndStreamlines}
\end{figure*}

A box of solid material (dark gray) is located in a closed cavity surrounded by fluid (light gray). The domain has unit out-of-plane thickness, i.e. $t_{\textrm{D}} = 1$. A distributed heat flux of size $q_{\textrm{h}} = 110$ is applied to the center part of the bottom boundary. The vertical and top walls are subject to temperature conditions $T = 0$, while the lower wall is insulated, $-k\frac{\partial T}{\partial x_{i}}n_{i} = 0$. A reference pressure $P = 0$ is applied at the top right corner. The gravitational acceleration is vertical and of size $\mathbf{g} = \{0, -1\}^{\textrm{T}}$ and the thermal expansion coefficient is $\beta=100$, yielding an effective $\textrm{Gr} = 6400$. The domain is discretized using 280x160 square elements. The natural boundary condition for the potential flow model is no flux, i.e. a no-penetration slip condition on the velocities. Hence, in order to ensure maximum comparability, the outer walls in the full-order model are modeled as no-penetration slip-boundaries, i.e. $u_i 
n_i=0$. 
The 
material 
properties used are shown in Tab. \ref{tab:comparison_boxMprop}. 
Ideally, the velocities inside the solid should be zero, requiring $\bar{\mu}_{s} = \infty$. However, numerically this is not practical and thus a high, but finite, value is chosen, $\bar{\mu}_{s}=10^{7}$. Correspondingly, in the NSB model, the Brinkman coefficient is set to 0 in the fluid, $\alpha_{f}=0$, and to a high, but finite, number in the solid, $\alpha_{s} = 10^7$. Furthermore, the fluid viscosity in the reference NSB model is set to 1, $\mu = 1$.
\begin{table}[]
\centering
\caption{Material properties for the calibration example.}
\label{tab:comparison_boxMprop}
\begin{tabular}{lcccccc}
      & $k$ & $\bar{\mu}$  & $\beta$ & $c_p$ & $\rho_0$ \\ \hline
Solid & 100 & $10^{7}$ & 1      & 1     & 1        \\
Fluid & 1   & \textit{variable} & 1        & 1     & 1        \\ \hline
\end{tabular}
\end{table}
A calibration process is conducted to find the value of $\bar{\mu}_f$, such that the temperature results in the best fit between the reduced-order and full-order models. The choice of using the temperature field as a quality measure is because this is the field of interest in the optimization, rather than the flow field directly. The value has been identified by minimizing the least-squares error of the temperature DOFs using the two models, i.e. $1/N_{dof}\sum(T_{NSB}-T_{ROM})^2$. A plot is shown in Fig. \ref{fig:LSQ}, from which a value of $\frac{1}{\bar{\mu}_f} = 0.09$ is chosen based on the figure of merit. Fig. \ref{fig:comparison_designAndStreamlines} also shows that a qualitatively good comparison is obtained. Fig. \ref{fig:comparison_darcyTemp} and \ref{fig:comparison_NSBTemp} show the temperature fields and streamlines for the two models. From this it is seen that the reduced-order model produces qualitatively similar results when tuned appropriately.

Velocity magnitudes are illustrated in Figs. \ref{fig:comparison_darcyVelMag} and \ref{fig:comparison_NSBVelMag} and are seen to be generally in the same range. However, major discrepancies are observed at the fluid-solid interface, which is due to the lack of a viscous boundary layer in the potential flow model. Thus, the local flow velocity, as well as the width of the recirculation cells, at the sides are over-predicted by the potential flow model.  

The chosen value of $\bar{\mu}_f$ is assumed to be constant with respect to convection/diffusion dominance and will be used for other Grashof numbers as well. It is the authors experience that the best $\bar{\mu}_f$ is mostly dependent on problem geometry for the problems investigated herein.

\section{Results} \label{sec:results}
\subsection{Heat Sink}
Topology optimization is now applied to the problem depicted in Fig. \ref{fig:comparison_domain}. The dark gray area is considered as the design domain. The problem was originally treated by \citet{Alexandersen2014}, although with no-slip boundary conditions at the outer walls. Only half the domain is modeled due to symmetry. The model domain is discretized with 140x160 square elements. The filter radius is set to $r_{\textrm{min}} = 0.06$, corresponding to 2.4 elements. The allowed solid volume fraction is $50\%$ of the design domain.

In order to compare to the results of \citet{Alexandersen2014}, the problem is investigated with $\textrm{Pr} = 1$ and at varying Grashof numbers. Note that the choice of $\textrm{Pr} = 1$ and all material parameters set to 1, except $k_{s}=100$ and a varying $\beta$, corresponds to $\textrm{Gr} = \textrm{Ra} = \beta H^3$ cf. \eqref{eq:dimless_grashof}. The treated problem only has a single prescribed temperature, which makes it difficult to determine the Grashof-number \textit{a priori}. Thus, an \textit{a priori} Grashof-number is defined assuming a temperature difference of $\Delta T = 1$. This \textit{a priori} Grashof-number is solely used to scale the convective to diffusive heat transfer. The physical Grashof-number can \textit{a posteriori} be determined based on the resulting temperature field (\cite{Alexandersen2016a}). 

\begin{figure*}\centering
\includegraphics[width=0.32\linewidth,trim=80 30 50 20,clip]{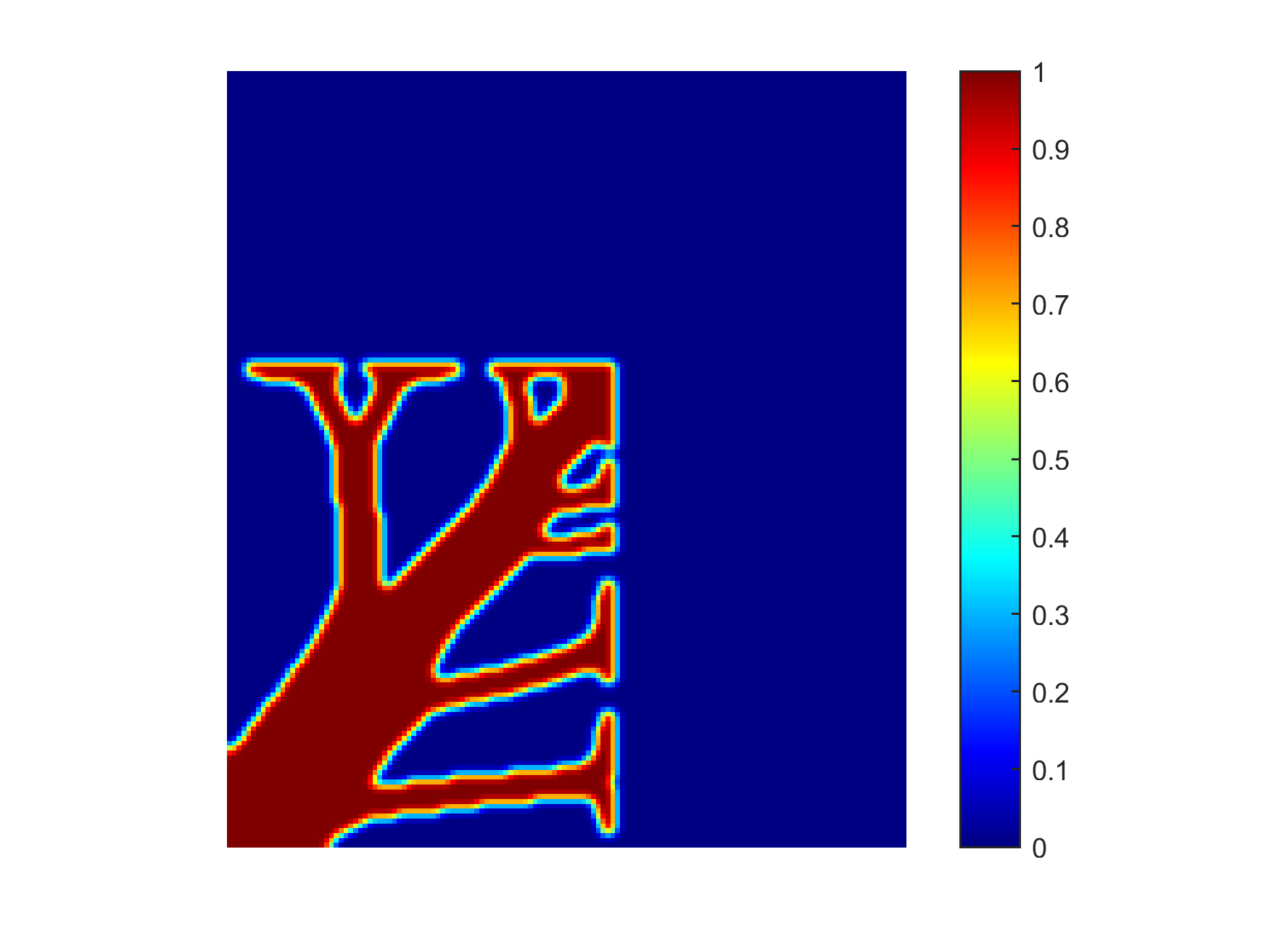}\hfill
    \includegraphics[width=0.32\linewidth,trim=80 30 50 20,clip]{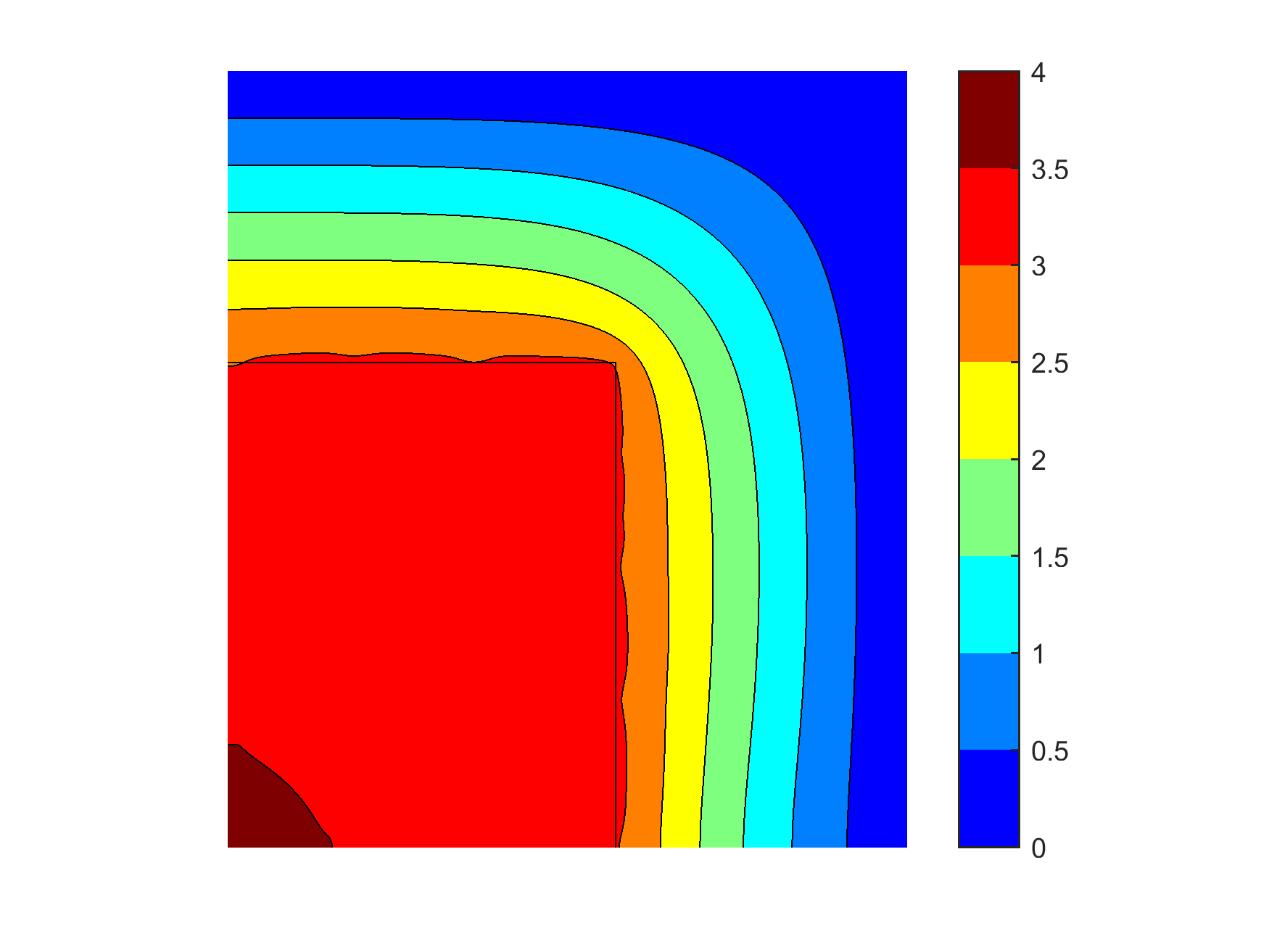}\hfill
      \includegraphics[width=0.32\linewidth,trim=80 30 50 20,clip]{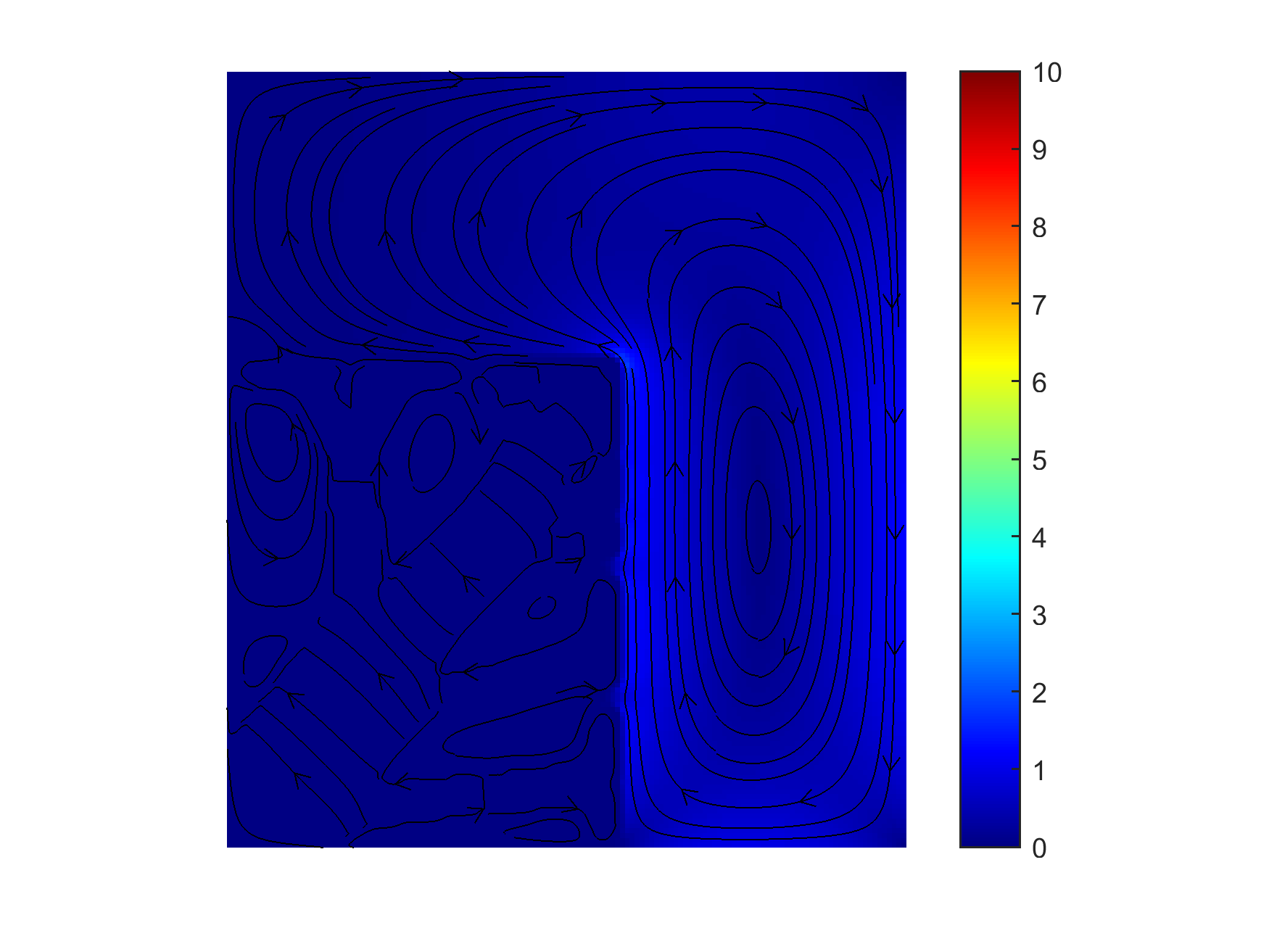}\\
      {$\textrm{Gr}=640$}\\ 
      \includegraphics[width=0.32\linewidth,trim=80 30 50 20,clip]{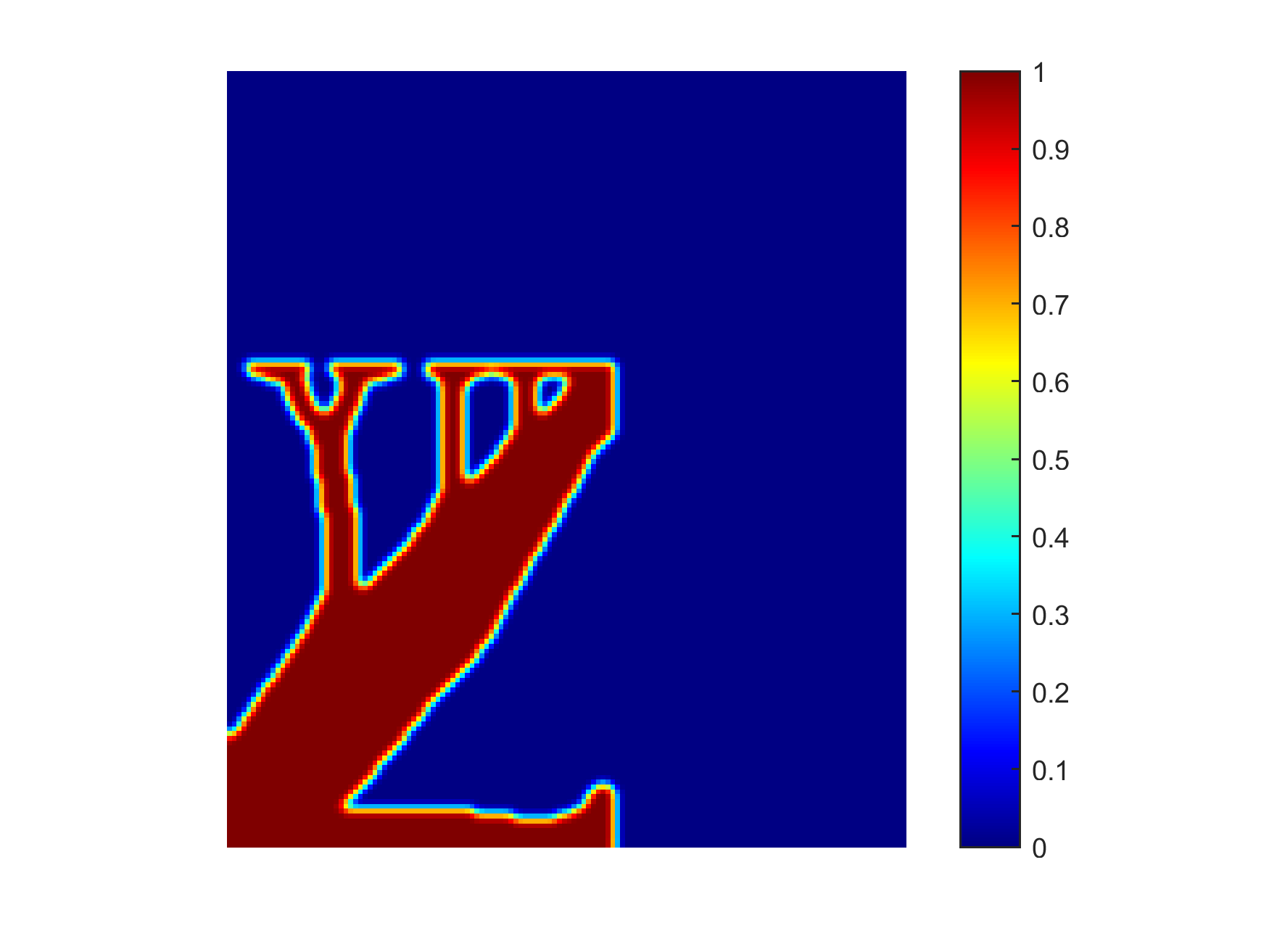}\hfill
      \includegraphics[width=0.32\linewidth,trim=80 30 50 20,clip]{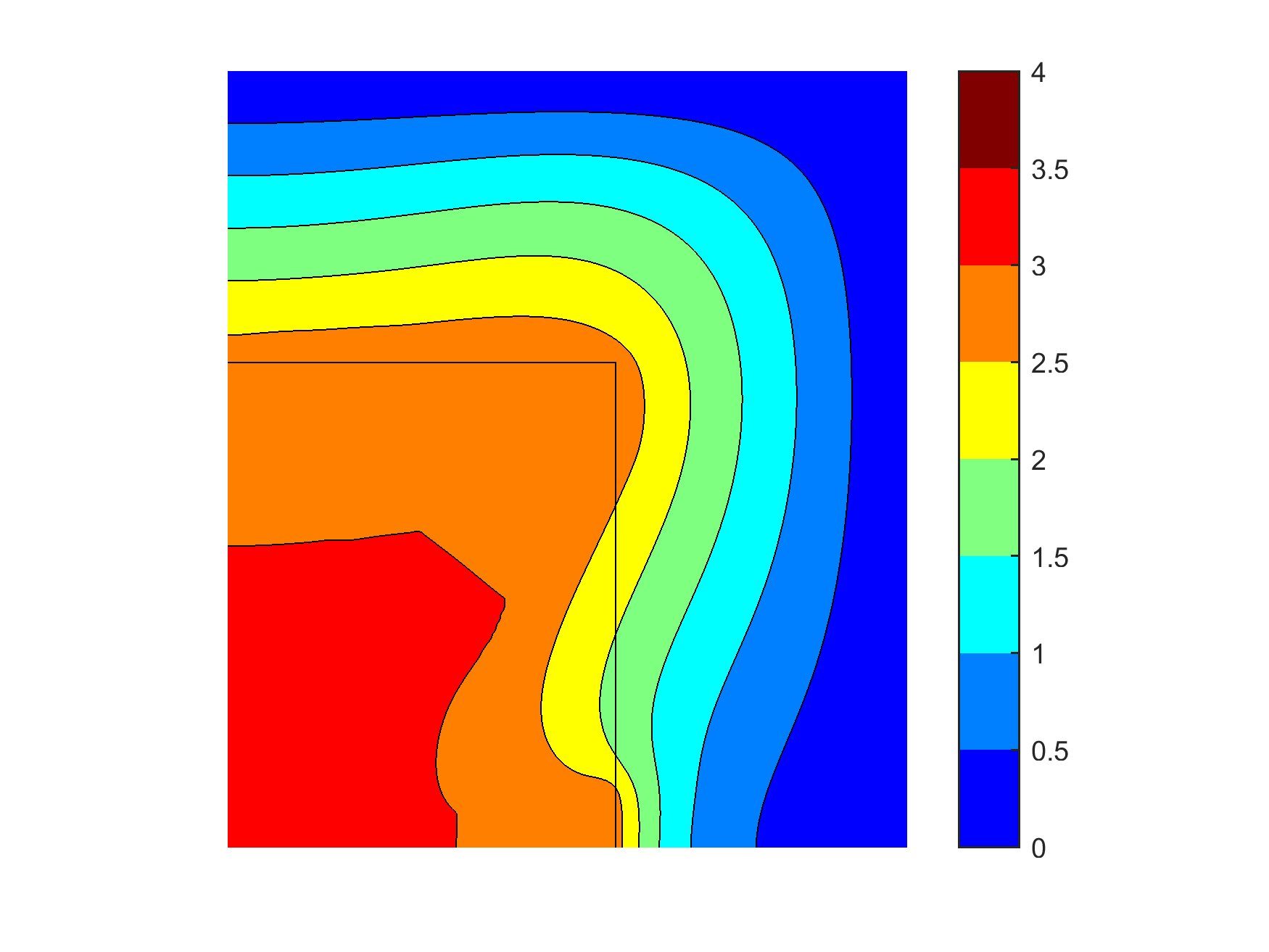}\hfill
      \includegraphics[width=0.32\linewidth,trim=80 30 50 20,clip]{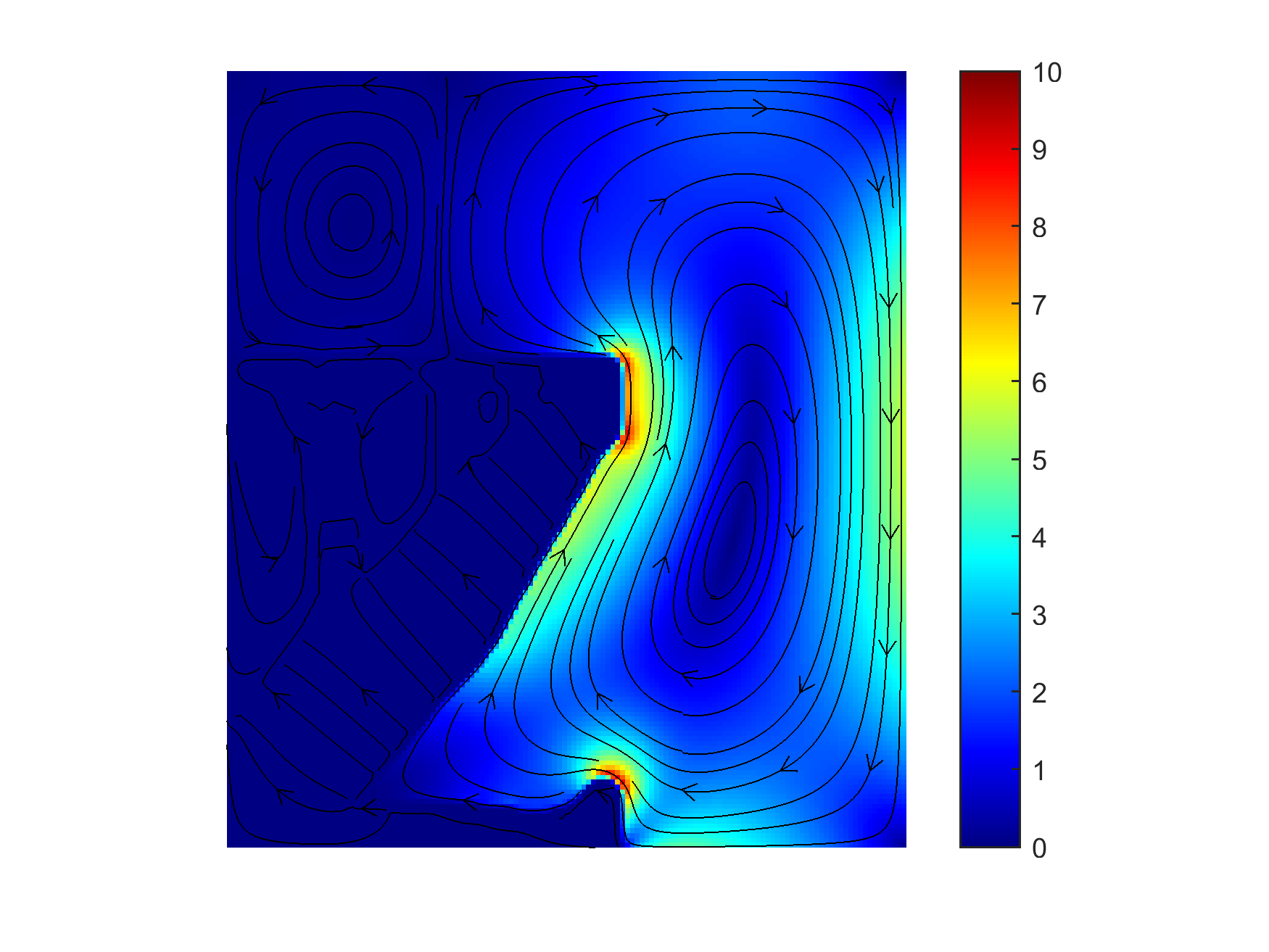}\\
      {$\textrm{Gr}=3200$}\\
            \includegraphics[width=0.32\linewidth,trim=80 30 50 20,clip]{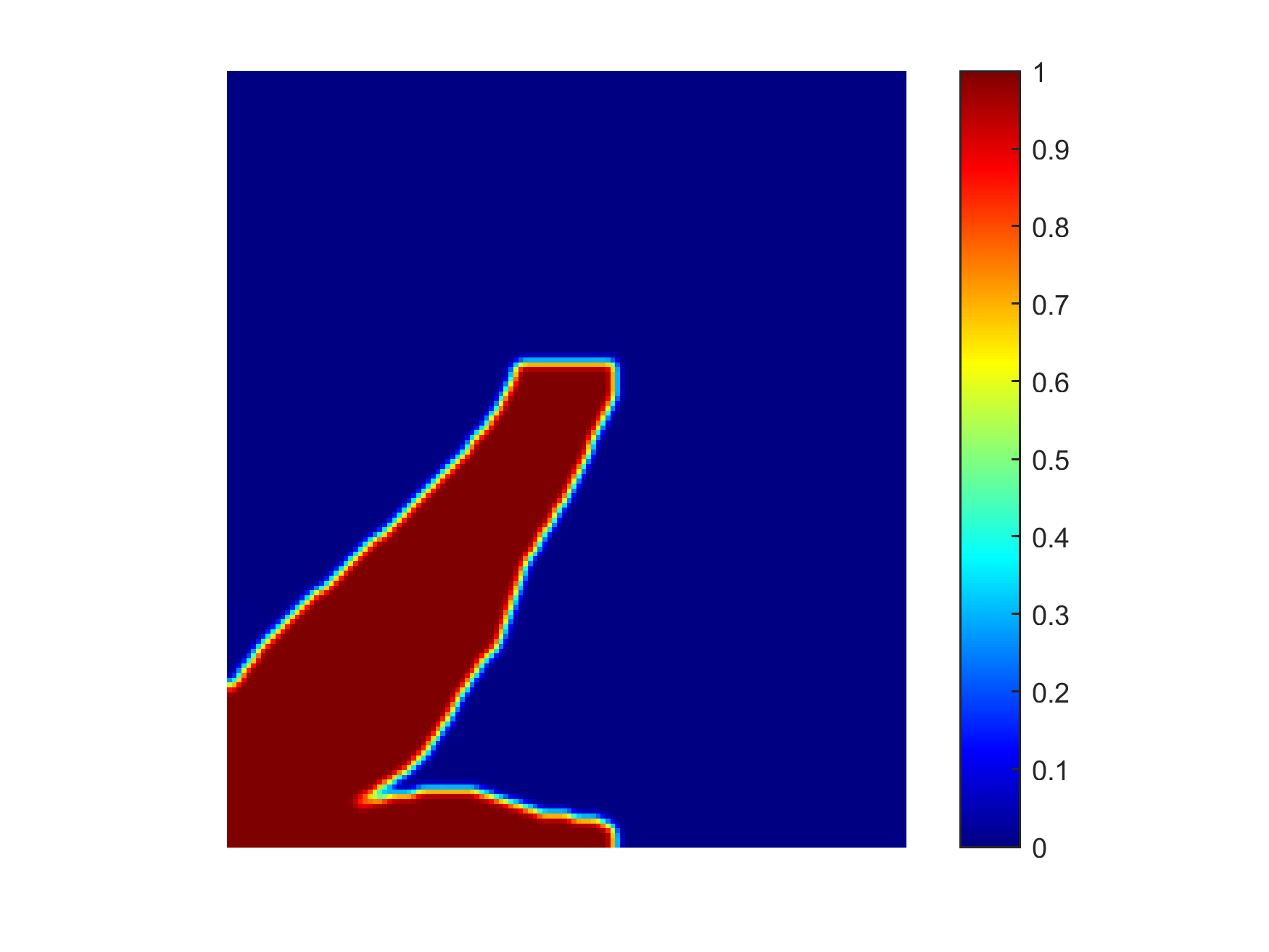}\hfill
      \includegraphics[width=0.32\linewidth,trim=80 30 50 20,clip]{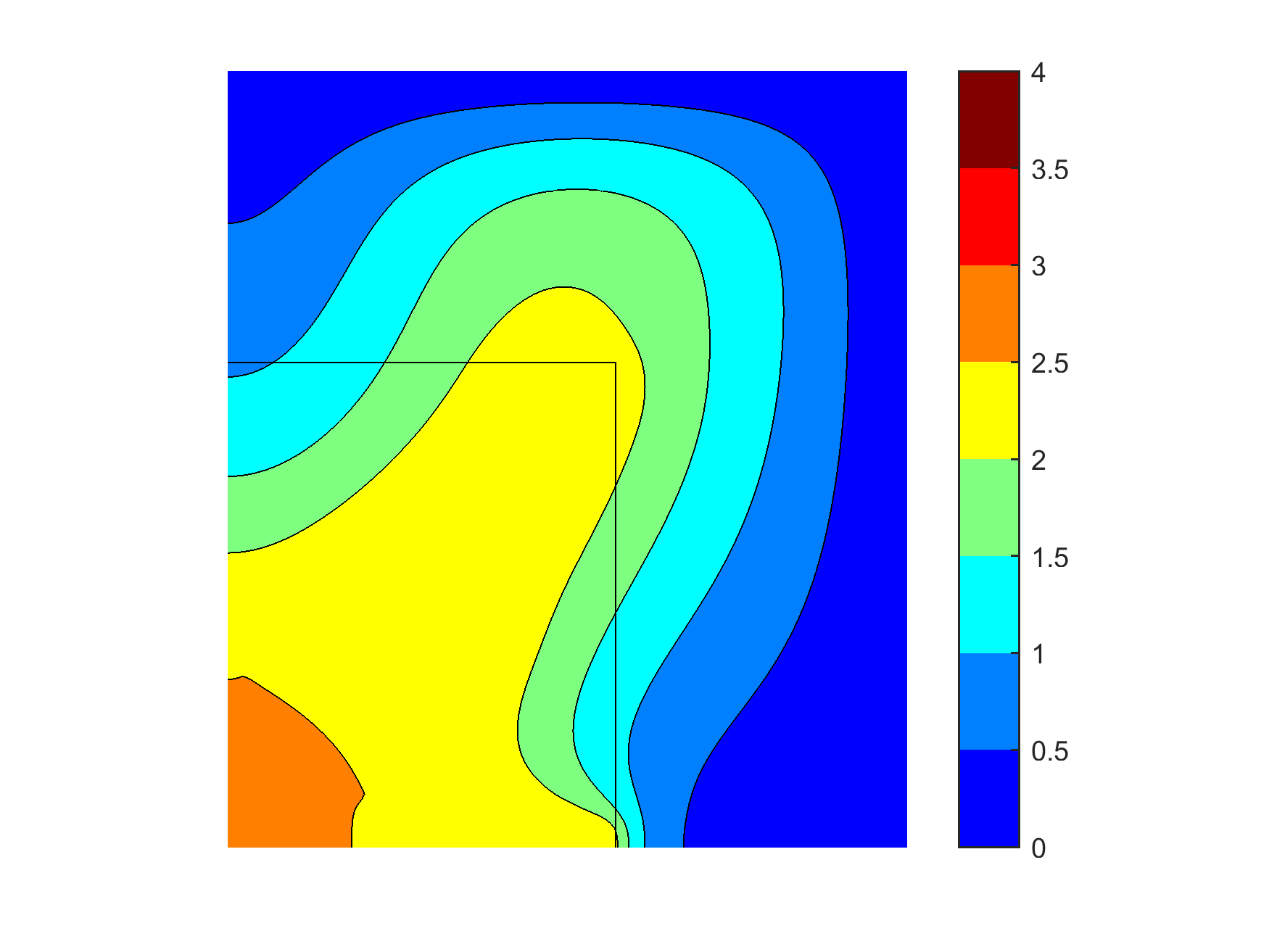}\hfill
      \includegraphics[width=0.32\linewidth,trim=80 30 50 20,clip]{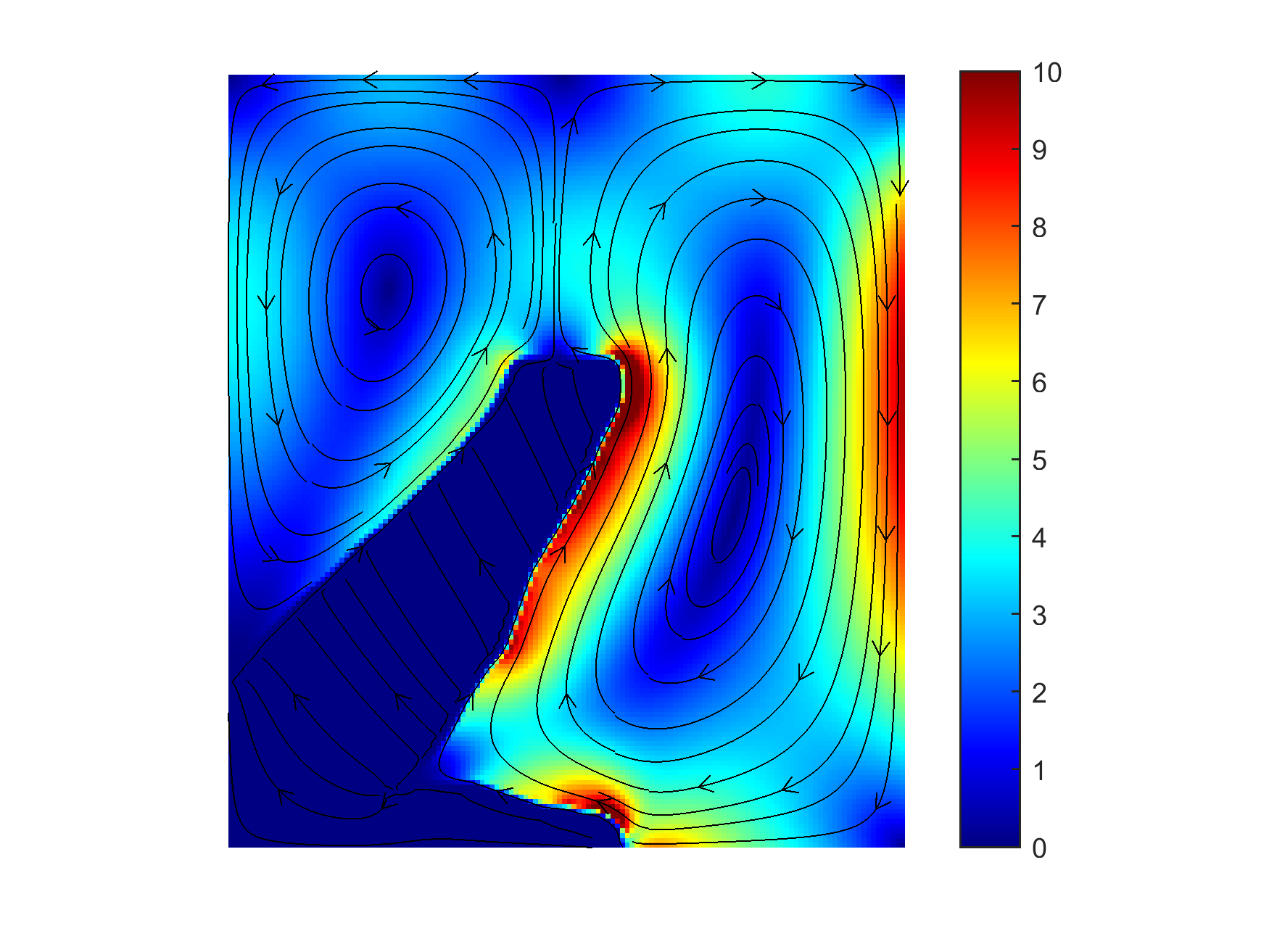}\\
      {$\textrm{Gr}=6400$}
     \caption{Design, temperature and velocity fields for three cases obtained using the reduced-order model.}
     \label{fig:results_DarcyDesign}
 \end{figure*}

Three cases are considered with \textit{a priori} Grashof numbers $\textrm{Gr} = \{640, 3200, 6400 \}$. The lowest case yields a solution which is heavily diffusion-dominated, while the latter yields a case in which convective heat transfer plays a significant role. The resulting designs are presented in Fig. \ref{fig:results_DarcyDesign}. For the diffusion-dominated case of $\textrm{Gr} = 640$, shown in Fig. \ref{fig:results_DarcyDesign}(a), the optimized design shows significant branching extending towards the boundary of the design domain. The design shows little tendency to accomodate the flow. Optimizing for a problem with no flow yielded qualitativly similar designs. Increasing the Grashof-number to $\textrm{Gr = 3200}$ removes most of the branching at the vertical design domain boundaries, see Fig. \ref{fig:results_DarcyDesign}(b), allowing for more heat to be convected away. For the case of $\textrm{Gr} = 6400$, shown in Fig. \ref{fig:results_DarcyDesign}(c), all branching has disappeared and the 
obtained designs has smoothly-varying walls.

The observed design trend is exactly the same as seen when using the full Navier-Stokes model \citep{Alexandersen2014}. The design starts with significant branching for the diffusion-dominated problem and then branching becomes less dominant as convection increases. In fact, comparing one-to-one with the designs obtained by \citet{Alexandersen2014}, the resemblence is surprisingly good. Although extremely similar, there are subtle differences. For example, the design for $\textrm{Gr} = 6400$ (Fig. \ref{fig:results_DarcyDesign}(c)) features a less smooth surface with a deeper dimple along the rightmost interface. These differences are most likely due to the lack of viscous friction close to the design interfaces, in the potential flow model.

To evaluate the performance of each design, a cross-check of the final objective function is conducted. The thermal compliance is evaluated for each design at each flow state. The thermal compliance is evaluated at the final penalization factors, $p_{k} = 16$ and $p_{\bar{\mu}} = 20$, and the results are listed in Tab. \ref{tab:results_darcyObjectiveFunction}\footnote{\label{footnote1} In order to make the comparison to the values presented by \cite{Alexandersen2014}, the obtained objective values have been multiplied by 2 (due to half-domain model) and divided by 100 (due to difference in domain thickness). Furthermore, a scaling error has been found in the original values presented by \cite{Alexandersen2014}, which must be multiplied by $10^3$.}.  The cross-check confirms that the designs, relative to each other, are best at the Grashof number at which they were optimized.

\begin{table}
\centering
\caption{Cross-check of design performance . All designs are evaluated at $\textrm{Gr}={640,3200,6400}$ and the related thermal compliance ($[\times 10^{-1}]$) is stated in the table. Bold text indicates best performance for a given setting (column). Designs evaluated using $p_k = 16$ and $p_{\kappa} = 20$.}
\begin{tabular}{lccc}
	&    \multicolumn{3}{c}{Evaluated at Gr}  \\
	\hline 
	Designed for Gr & 640 & 3200 & 6400\\
	\noalign{\smallskip}\hline\noalign{\smallskip}
	$640$	 & \textbf{8.06} & 7.36 & 6.42\\
	$3200$ & 8.28 & \textbf{7.30} & 6.22 \\
	$6400$ & 8.80 & 7.36 & \textbf{5.94} \\
	\noalign{\smallskip}\hline
\end{tabular}
\label{tab:results_darcyObjectiveFunction} 
\end{table}

\subsection{Comparison of performance under the full-order model}
\begin{figure*}
\centering
\subfloat[Temperature, $\textrm{Gr}=640$]{\includegraphics[height=5cm,trim=30 5 90 5,clip]{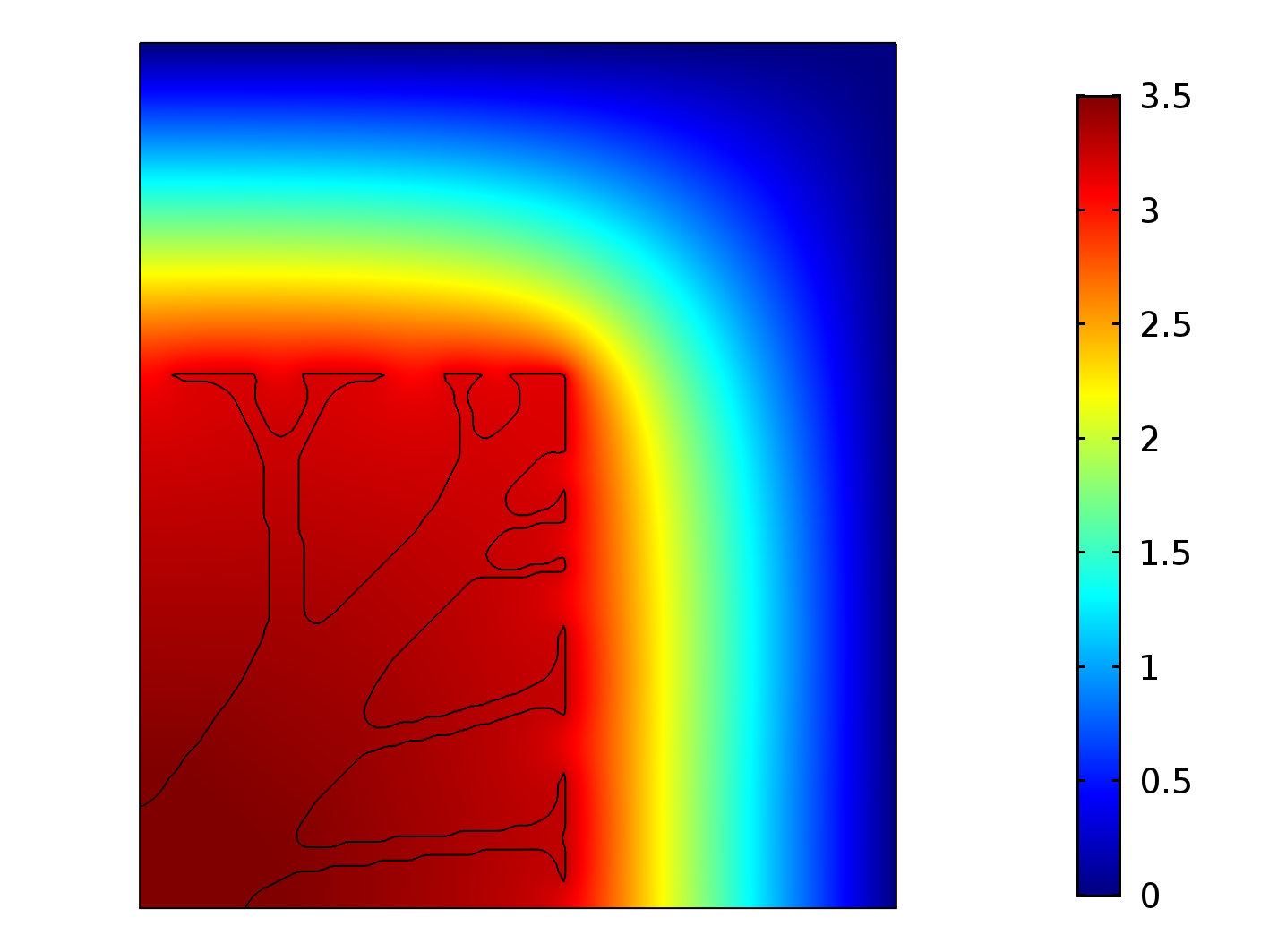}
\includegraphics[height=5cm,trim=280 5 10 15,clip]{fig/csan/NS_Gr640_temp_upd_01}
}\hfill
\subfloat[Velocity, $\textrm{Gr}=640$]{\includegraphics[height=5cm,trim=30 5 90 5,clip]{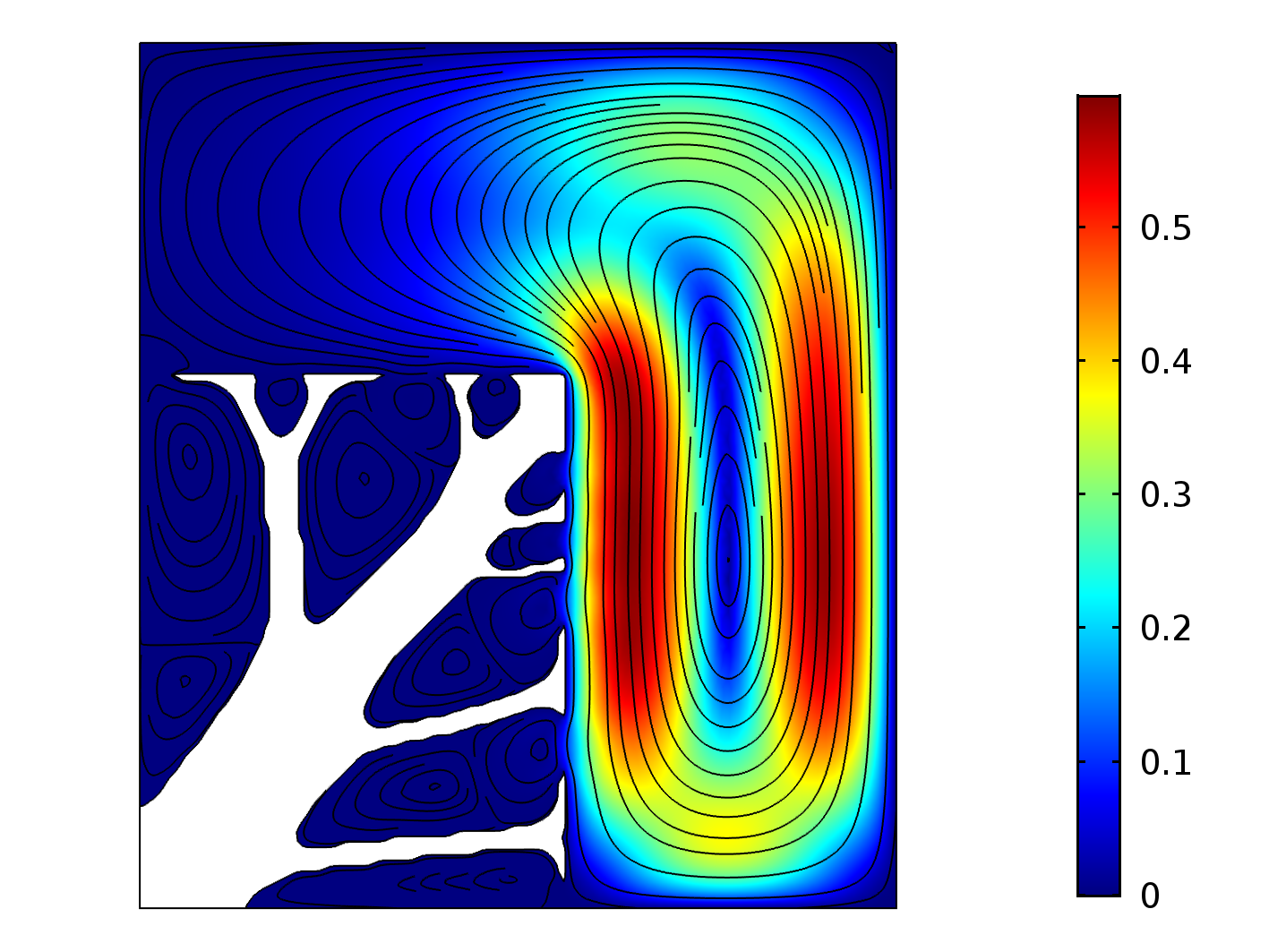}
\includegraphics[height=6cm,trim=280 5 10 15,clip]{fig/csan/NS_Gr640_vel_upd_01}}\\

\subfloat[Temperature, $\textrm{Gr}=3200$]{\includegraphics[height=5cm,trim=30 5 90 5,clip]{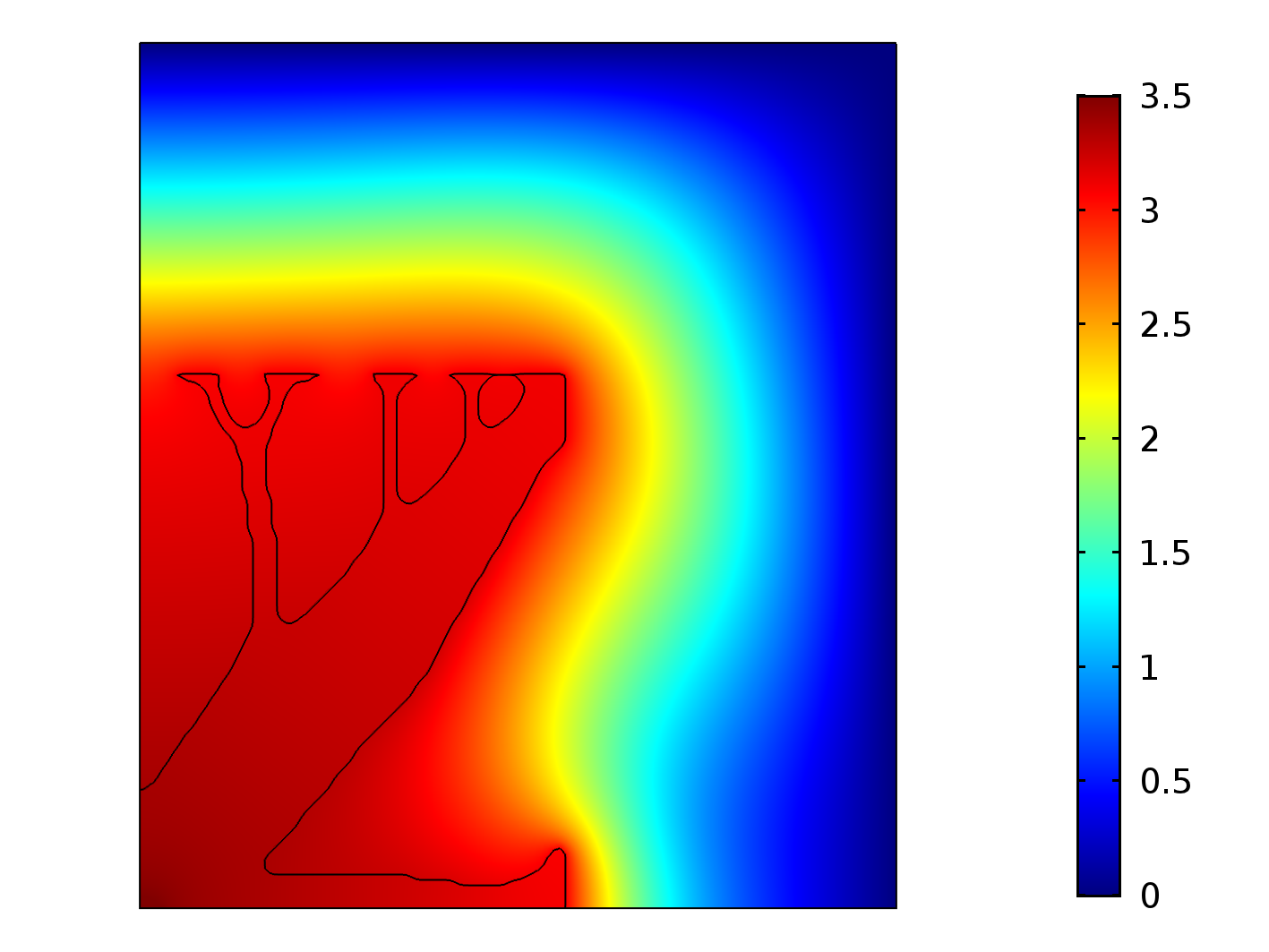}
\includegraphics[height=5cm,trim=280 5 10 15,clip]{fig/csan/NS_Gr3200_temp_upd_01}
}\hfill
\subfloat[Velocity, $\textrm{Gr}=3200$]{\includegraphics[height=5cm,trim=30 5 90 5,clip]{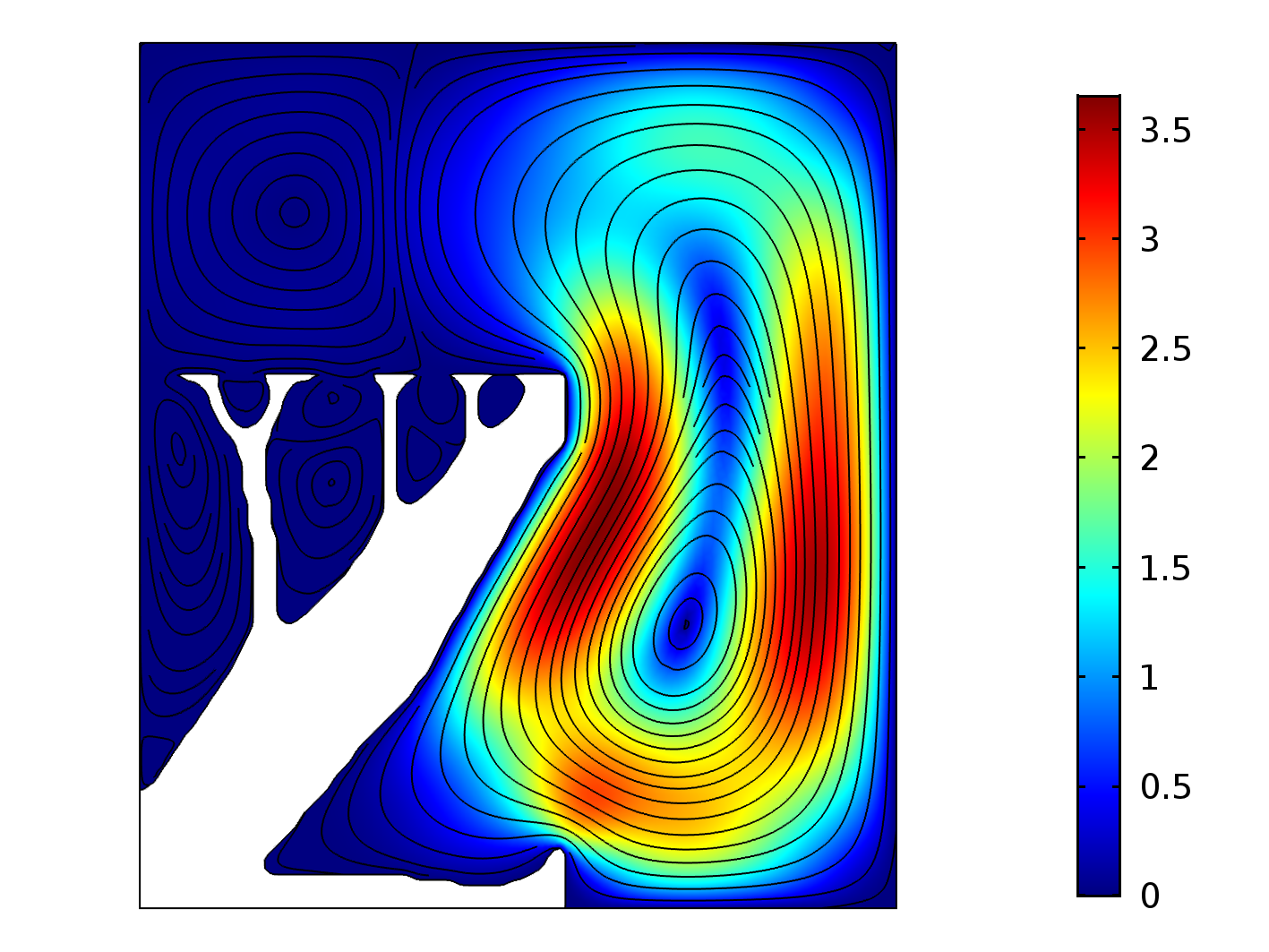}
\includegraphics[height=5cm,trim=280 5 10 15,clip]{fig/csan/NS_Gr3200_vel_upd_01}}\\

\subfloat[Temperature, $\textrm{Gr}=6400$]{\includegraphics[height=5cm,trim=30 5 90 5,clip]{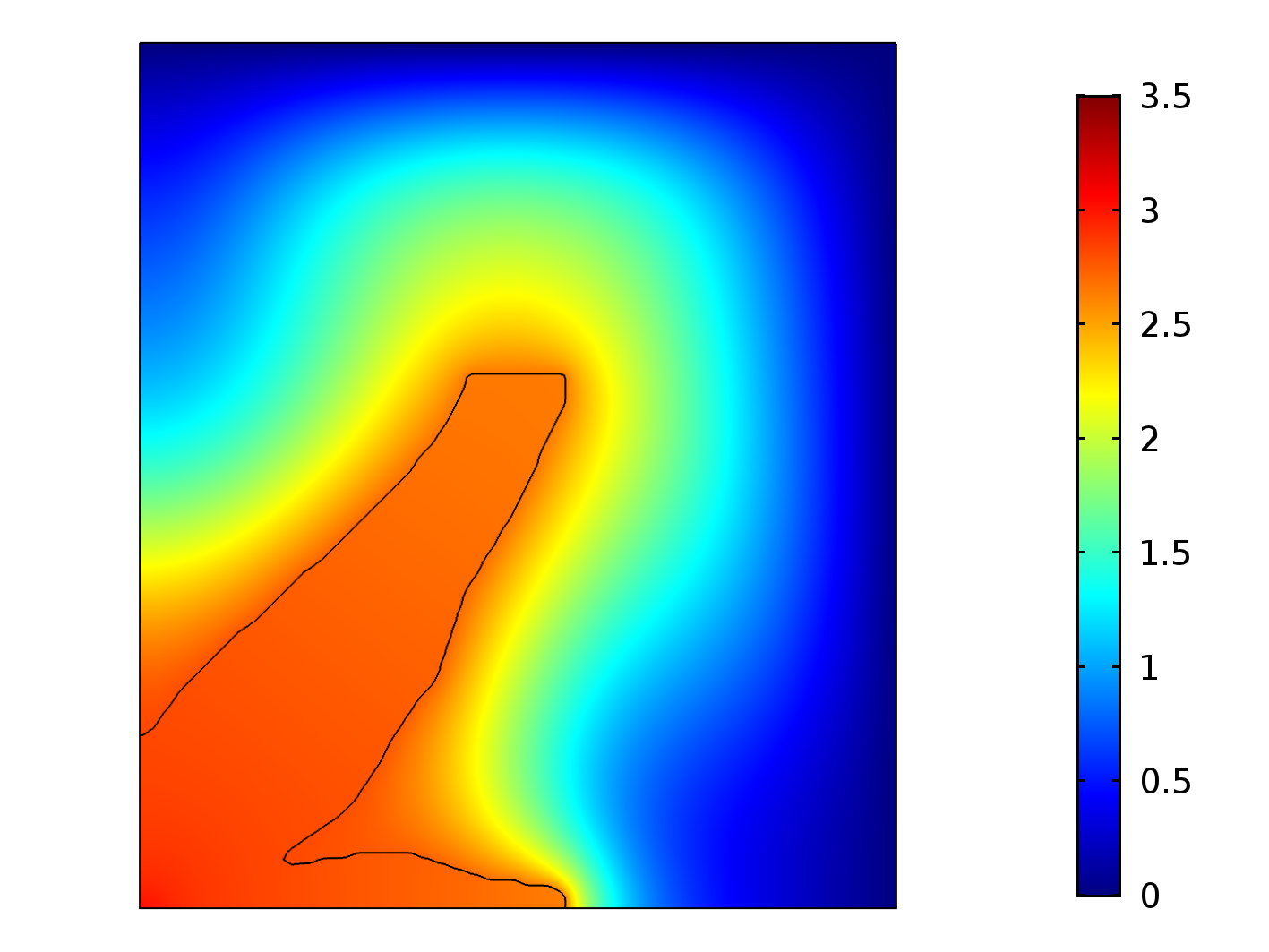}
\includegraphics[height=5cm,trim=280 5 10 15,clip]{fig/csan/NS_Gr6400_temp_upd_01}
}\hfill
\subfloat[Velocity, $\textrm{Gr}=6400$]{\includegraphics[height=5cm,trim=30 5 90 5,clip]{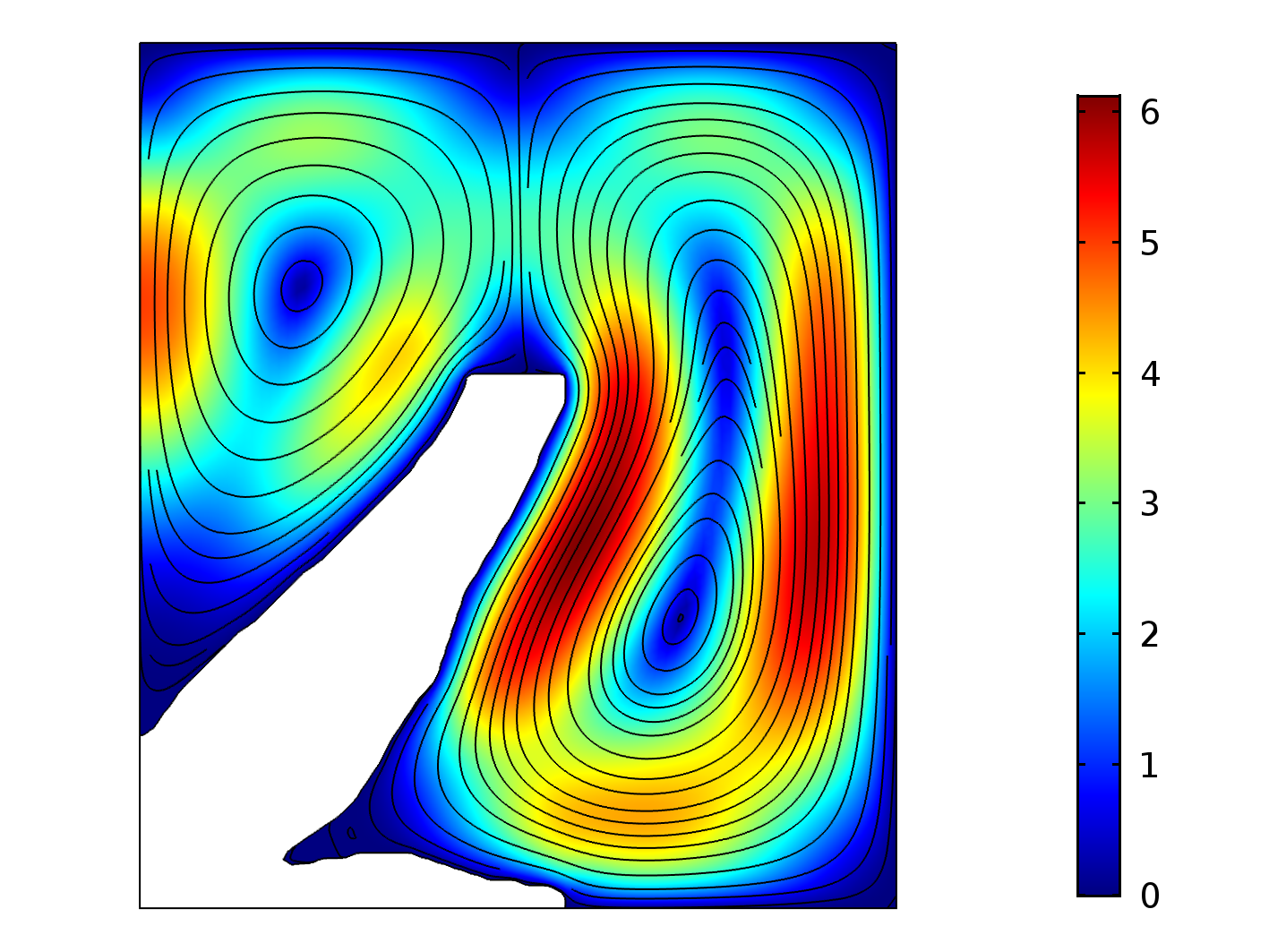}
\includegraphics[height=5cm,trim=280 5 10 15,clip]{fig/csan/NS_Gr6400_vel_upd_01}}\\
     \caption{Heat sink results when modeling the fluid flow using a NS flow model. Note that no-slip walls were used at the domain boundaries.}
     \label{fig:results_comsol}
   \end{figure*}

The obtained designs are tested using the full-order flow model, where designs are thresholded, to obtain truly black/white designs, using a smooth isocontour of the design field. The threshold value is set such that design values satisfying $\gamma \geq 0.1$ are solid, while design values satisfying $\gamma < 0.1$ are fluid\footnote{This threshold value is chosen to allow for direct comparison with the results presented by \cite{Alexandersen2014}, due to the reasoning presented therein.}. The thresholded designs along with temperature and velocity profiles are shown in Fig. \ref{fig:results_comsol}.

For the reevaluation, no-slip conditions are used on the outer walls, as this is considered the true reference case \citep{Alexandersen2014}. The stream profiles are qualitatively similar to those produced by the reduced-order model, with an equal number of vortices produced. 

A cross-check of the objective functions is shown in Tab. \ref{tab:results_NSobjectiveFunction}\footnotemark[4].
It can be seen that the cross-check is still passed for the thresholded designs. The values using the Navier-Stokes flow model are slightly higher than when using the potential flow model, most likely due to a combined effect of the design thresholding and the no-slip conditions along the interface. Furthermore, compared to the reference results presented by \citet{Alexandersen2014}, the reduced-order results give a slightly lower thermal compliance. This is likely due to convergence to local minima and small differences in the implementations and the thresholding procedures. The maximum temperature shows that the true Grashof numbers are a factor of 3 to 4 higher than the \textit{a priori} computed.

\begin{table}
\centering
\caption{Thermal compliance values ({$[\times 10^{-1}]$}) from the NS model with no-slip conditions at all solid-fluid interfaces. $\Delta T_{max}$ is the maximum observed temperature difference.}
\label{tab:results_NSobjectiveFunction}       
\begin{tabular}{lcccccc}
 &    \multicolumn{3}{c}{Evaluated at Gr} &\multicolumn{3}{c}{$\Delta T_{max}$ at Gr}  \\
\hline 
Designed for Gr & 640 & 3200 & 6400 & 640 & 3200 & 6400\\
\noalign{\smallskip}\hline\noalign{\smallskip}
$640$  & \textbf{8.12} & 7.82 & 7.24 & 3.7 & 3.6 & 3.3 \\
$3200$ & 8.38 & \textbf{7.78} & 7.00 & 3.8 & 3.6 & 3.2\\
$6400$ & 8.94 & 7.88 & \textbf{6.62} & 4.1 & 3.6 & 3.0\\
\noalign{\smallskip}\hline
\end{tabular}
\end{table}

\subsection{Comparison to naive simplified convection model}
\begin{figure*}
\centering
\subfloat[$\textrm{Gr}=640$]{\includegraphics[height=0.25\textheight]{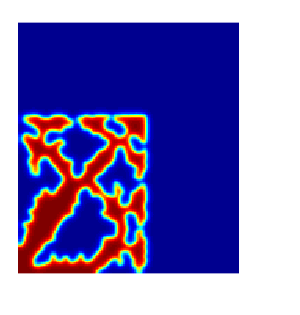}}
\subfloat[$\textrm{Gr}=3200$]{\includegraphics[height=0.25\textheight]{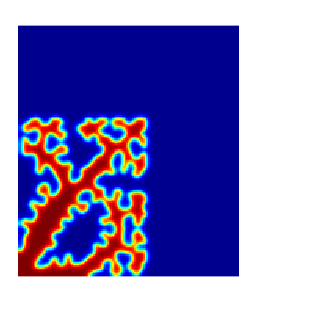}}
\subfloat[$\textrm{Gr}=6400$]{\includegraphics[height=0.25\textheight]{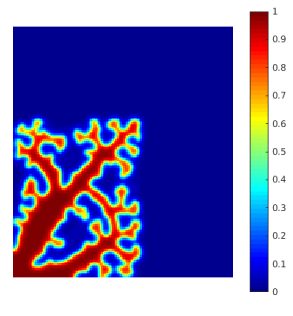}}
\caption{Heat sink designs obtained using the simplified convection boundary model.}
\label{fig:results_simplconv}
\end{figure*}

\begin{figure*}
\centering
\subfloat[$\textrm{Gr}=640$]{
\includegraphics[width=0.4\textwidth]{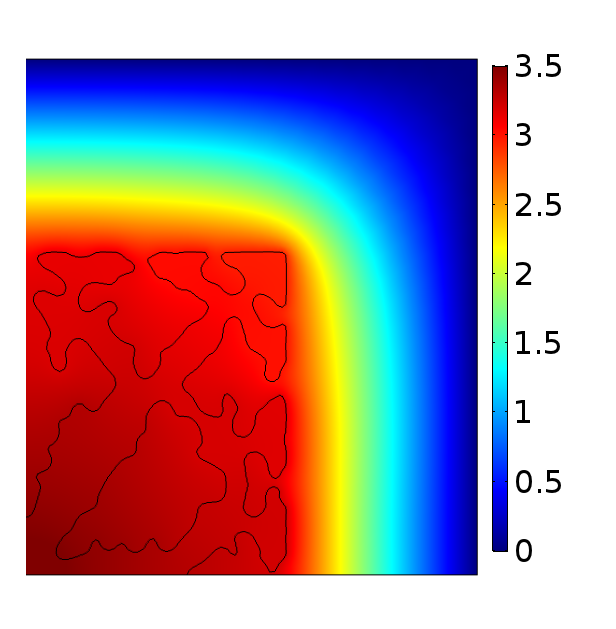}
\hspace{0.05\textwidth}
\includegraphics[width=0.4\textwidth]{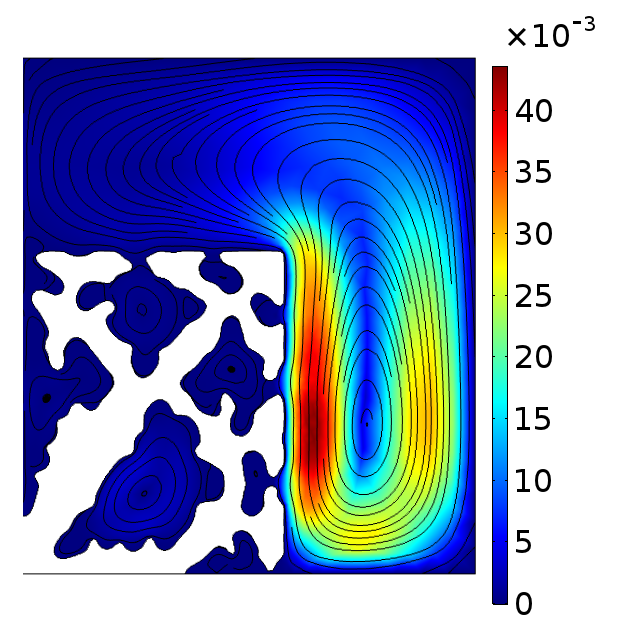}}\\
\subfloat[$\textrm{Gr}=640$]{
\includegraphics[width=0.4\textwidth]{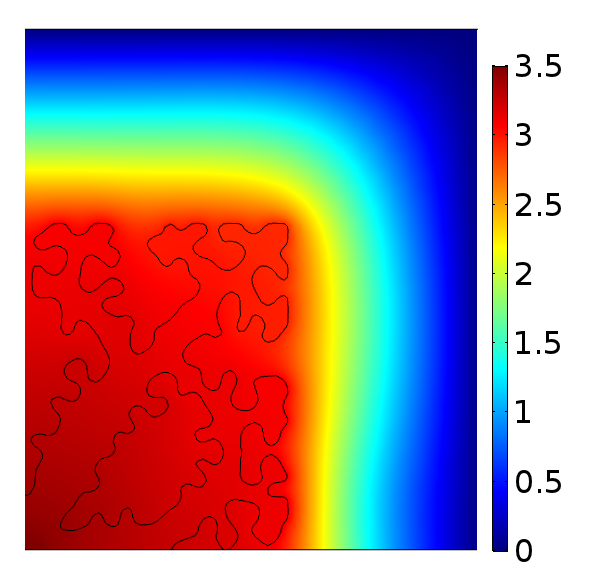}
\hspace{0.05\textwidth}
\includegraphics[width=0.4\textwidth]{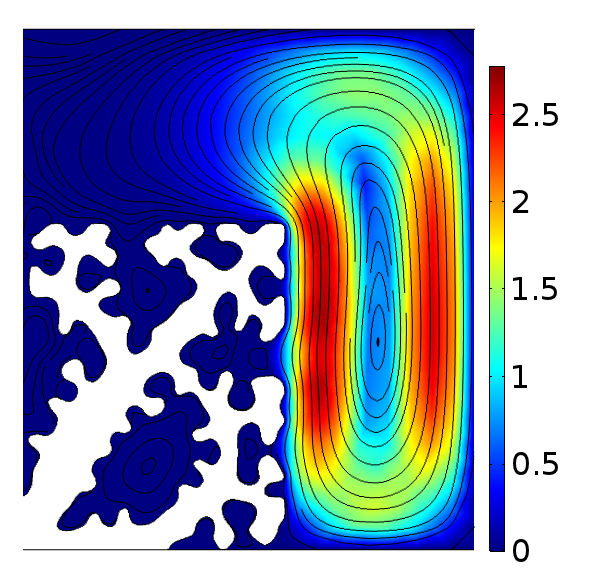}}\\
\subfloat[$\textrm{Gr}=6400$]{
\includegraphics[width=0.4\textwidth]{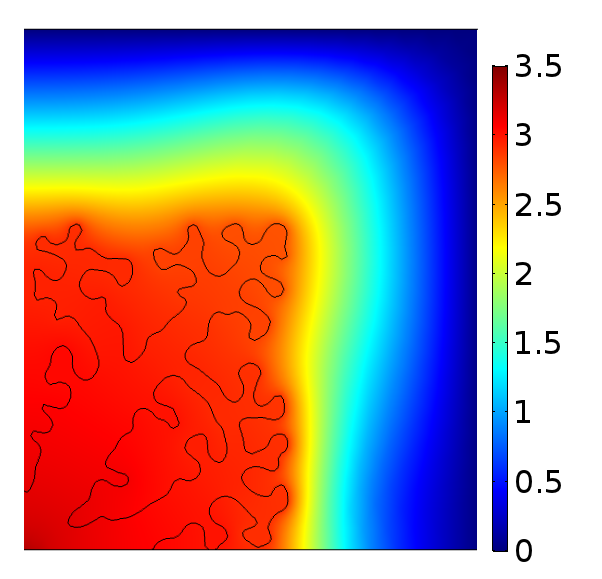}
\hspace{0.05\textwidth}
\includegraphics[width=0.4\textwidth]{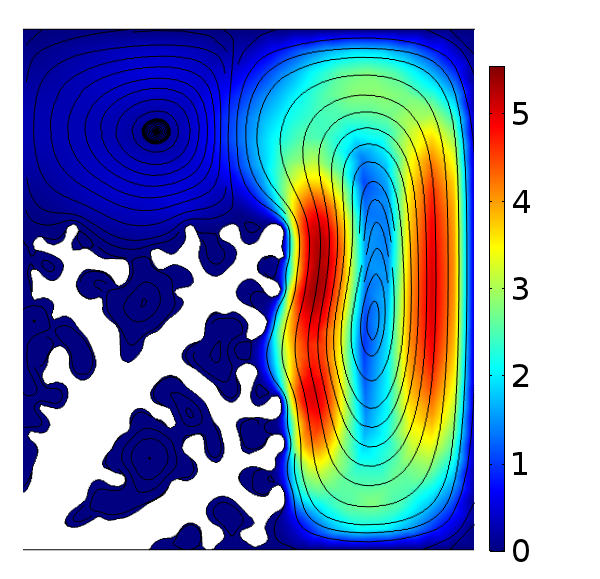}}
\caption{Heat sinks designed using the simplified convection boundary model evaluated using the full NS model. Left column: Temperature distribution. Right column: Velocity magnitude and streamlines.}
\label{fig:results_simplconvCOMSOL}
\end{figure*}

\begin{figure}
\centering
\includegraphics[width=0.5\textwidth]{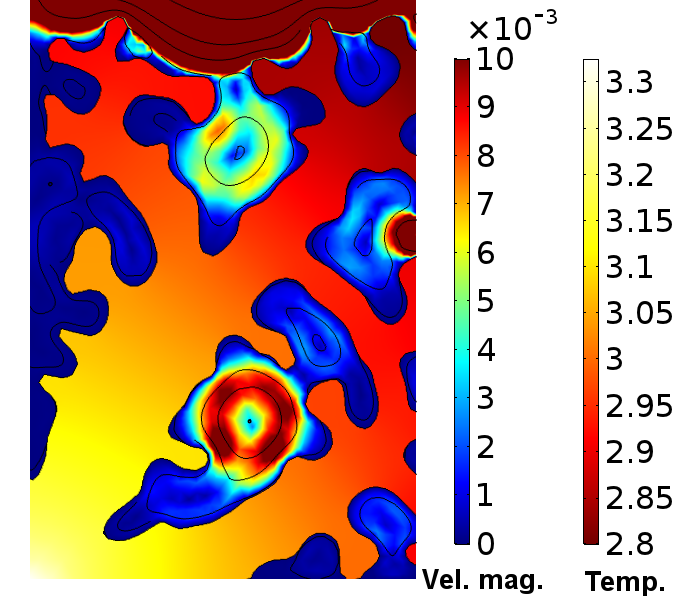}
\caption{Close-up of the very low-velocity convection cells inside close cavities for the $\textrm{Gr}=6400$ simplified model design. The velocity field and the solid temperature distribution are shown using two seperate color schemes and scales.}
\label{fig:closeup_simplconvCOMSOL}
\end{figure}

The proposed reduced-order model is further compared to using a naive simplified convection model based on Newton's law of cooling, with a constant convection coefficient on the solid-fluid interface (\cite{Alexandersen2011,Coffin2016,Zhou2016,Lazarov2014}). This model does not model the flow at all and the specific approach is described in Appendix \ref{app:simplified}. The convection coefficient for the three Grashof numbers are calculated from COMSOL models of the reference designs from \citep{Alexandersen2014} using the full Navier-Stokes flow model, by computing the surface average of the local convection coefficient:
\begin{equation}
\bar{h} = \frac{1}{A_{fs}} \int_{\Gamma_{fs}} \frac{q_{n}}{T} ds
\end{equation}
where $\bar{h}$ is the average convection coefficient, $A_{fs}$ is the area of the fluid-solid interface ($\Gamma_{fs}$) and $q_{n}$ is the normal flux at the interface.
\begin{table}
\centering
\caption{Non-dimensional effective convection coefficients for the three Grashof numbers, calculated from the full-order model for the reference designs from \citep{Alexandersen2014}.}
\label{tab:simplified_convectionCoefficients}       
\begin{tabular}{lccc}
Grashof & 640 & 3200 & 6400 \\
\noalign{\smallskip}\hline\noalign{\smallskip}
$\bar{h}$ & 0.17883 & 0.27820 & 0.76345\\
\noalign{\smallskip}\hline\noalign{\smallskip}
\end{tabular}
\end{table}
\begin{table}
\centering
\caption{Final thermal compliance values ({$[\times 10^{-1}]$}) for the optimised design using naive simplified convection model, evaluated using the simplified and full-order models for the three Grashof numbers. The average convection coefficient calculated using the full-order model is also shown.}
\label{tab:simplified_postValues}       
\begin{tabular}{lccc}
Gr & 640 & 3200 & 6400 \\
\noalign{\smallskip}\hline\noalign{\smallskip}
Simplified conv. model & 6.9825 & 4.2579 & 2.4703\\
Full model & 7.9778 & 7.8008 & 7.2695\\
\noalign{\smallskip}\hline\noalign{\smallskip}
$\bar{h}$ & 0.16089 & 0.14301 & 0.15703\\ 
\noalign{\smallskip}\hline\noalign{\smallskip}
\end{tabular}
\end{table}
The non-dimensional convection coefficient for the three cases is shown in Tab. \ref{tab:simplified_convectionCoefficients}. These values are then used in the optimisation procedure using the simplified convection model. The obtained designs are shown in Fig. \ref{fig:results_simplconv}. It can be seen that the obtained designs are significantly different from those using the two flow models. The designs contain closed cavities and small-scale features. The internal cavities appear due to the global nature of the simplified convection model, where the convective boundary condition is applied on all fluid-solid interfaces and does not discriminate between internal and external surfaces. Similarly, the small-scale features are attractive due to a combination of the problem being purely conductive (although with a convection boundary condition) and the fact that more surface area yields 
better heat transfer.
The obtained designs are imported into COMSOL and evaluated using the full Navier-Stokes flow model and the results of this analysis is shown in Fig. \ref{fig:results_simplconvCOMSOL}. As expected, it is clear that the fluid in the internal parts of the heat sinks (partially- and fully-closed cavities) is moving at near zero velocities and, thus, almost no cenvective heat transfer is taking place. This is further exemplified by Fig. \ref{fig:closeup_simplconvCOMSOL}, which shows a close-up of the velocities in the internal (partially- and fully-closed) cavities for the $\textrm{Gr}=6400$ design, when analysed using the full-order model. Here it can be seen, that although convection cells do form due to temperature variations in the solid, the resulting velocities are significantly lower than those of the outer flow due to the temperature variations being very small. 
The thermal compliance values using both the simplified convection model and the full Navier-Stokes models are presented in Tab. \ref{tab:simplified_postValues}. The effective convection coefficient calculated using the full-order model is also shown. By comparing the thermal compliance values using the simplified and full-order models in Tab. \ref{tab:simplified_postValues}, it is clear that the heat transfer is vastly over-predicted using the simplified convection model. Furthermore, it can be seen that all three designs perform more or less similarly when evaluated using the full Navier-Stokes model. However, using the simplified convection model, the higher the Grashof number / convection coefficient, the more over-predicted the heat transfer is due to internal cavities. The fact that the convective heat transfer is vastly over-estimated, is further illustrated by comparing the actual effective convection coefficient values in the bottom row of Tab. \ref{tab:simplified_postValues} with the original 
values 
in Tab. \ref{tab:simplified_convectionCoefficients}. 

\subsection{Centrally heated domain}
\begin{figure}
\centering
\includegraphics[width=0.5\linewidth]{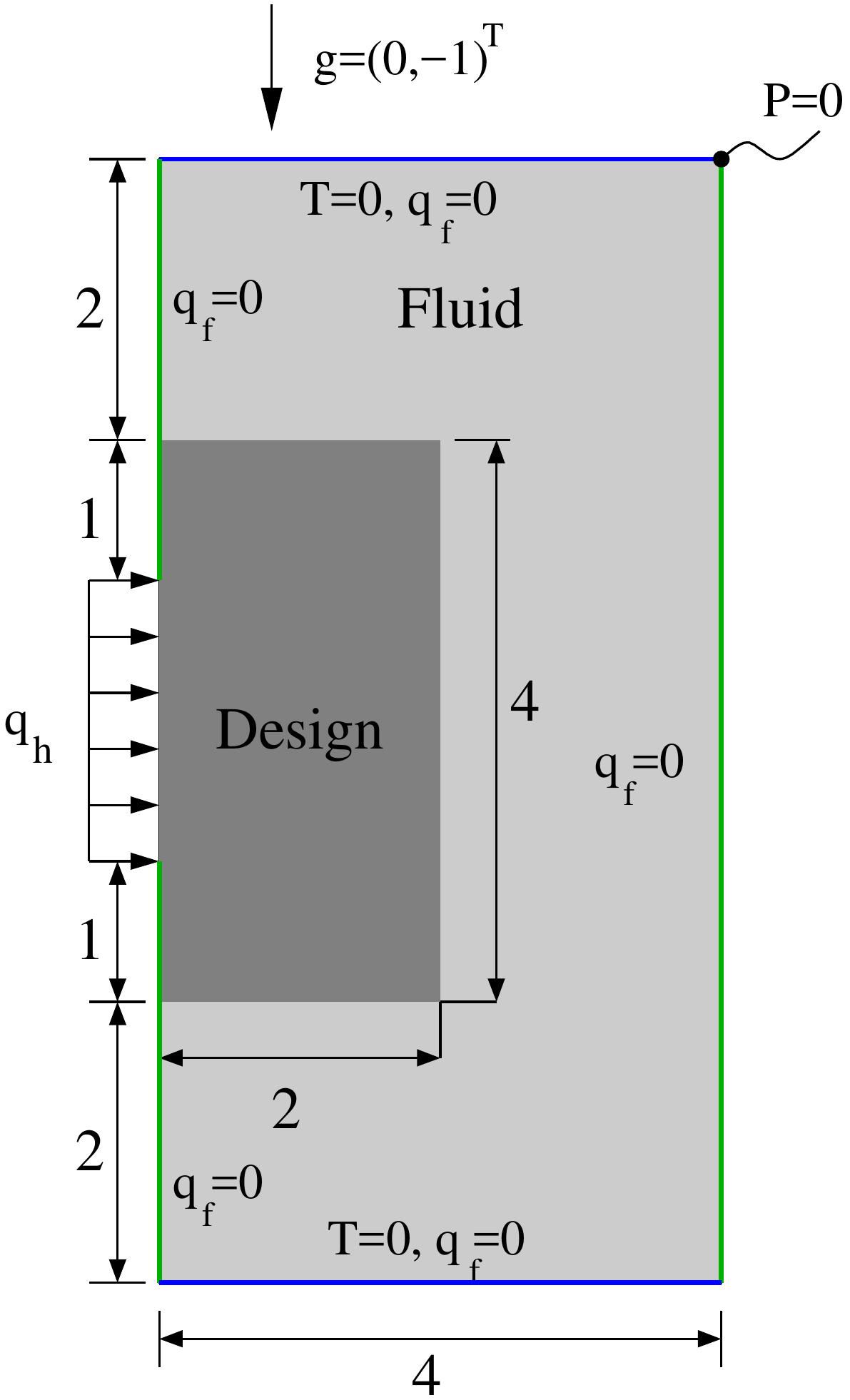}
 \caption{Rectangular closed cavity heated through the centre part of the left otherwise insulated boundary. Right boundary is insulated while top and bottom are isothermal, $T=0$. All boundaries are modeled as slip $u_in_i=0$ and the pressure is constrained $p=0$ in the lower left corner.}
 \label{fig:example2}
\end{figure}

The second example considers a rectangular cavity as shown in Fig. \ref{fig:example2}. The vertical cavity walls are assumed adiabatic except, for the centre part of the left wall which is heated by a constant heat flux $q_{h}=3$. The top and bottom are assumed isothermal $T=0$. All boundaries are assumed closed with slip condition $u_in_i=0$ and the pressure in the top right corner is constrained to zero. The gravity is pointing downwards $g=(0,-1)^T$. The domain is discretised using $120\times 240$ elements using a filter with radius $r_{\textrm{min}}= 0.08$ ($2.4$ elements).

The objective is to minimize the thermal compliance in the domain subject to a volume constraint for solid material of $30\%$ of the design domain. In the first example, a large part of the heat was transported from the structure to the right vertical wall by convection. In this example, the vertical wall is insulated forcing the optimizer to design optimized cooling through two horizontal boundaries which is more challenging. The low temperature of the top boundary can be utilized with relative ease while the low temperature of the bottom boundary is more difficult to utilize as it requires that hot fluid is pushed downwards while relatively colder fluid must be pushed upwards. This results in separate diffusion and convection dominated areas as will be discussed below.

The results in this section are obtained using only the first penalization step in the previously mentioned continuation scheme and an initial uniform material distribution of $10\%$ solid in the design domain. The continuation has been omitted as the level of discreteness in the solution after the first penalization step was satisfactory.

The parameter $\bar{\mu}$ has been obtained using the same method as for the previous example. The design domain is considered solid and the temperature distribution is obtained both by the NSB and the reduced-order model. The value of $\bar{\mu}$ that minimizes the least squares temperature difference was found to be $\bar{\mu}^{-1}=0.15$ under the condition $\textrm{Gr}=51200$ using a sweep with intervals of $\Delta (\bar{\mu}^{-1})=0.02$.

\begin{figure*}
\subfloat[Design, $\textrm{Gr}=5120$]{\includegraphics[width=0.3\textwidth]{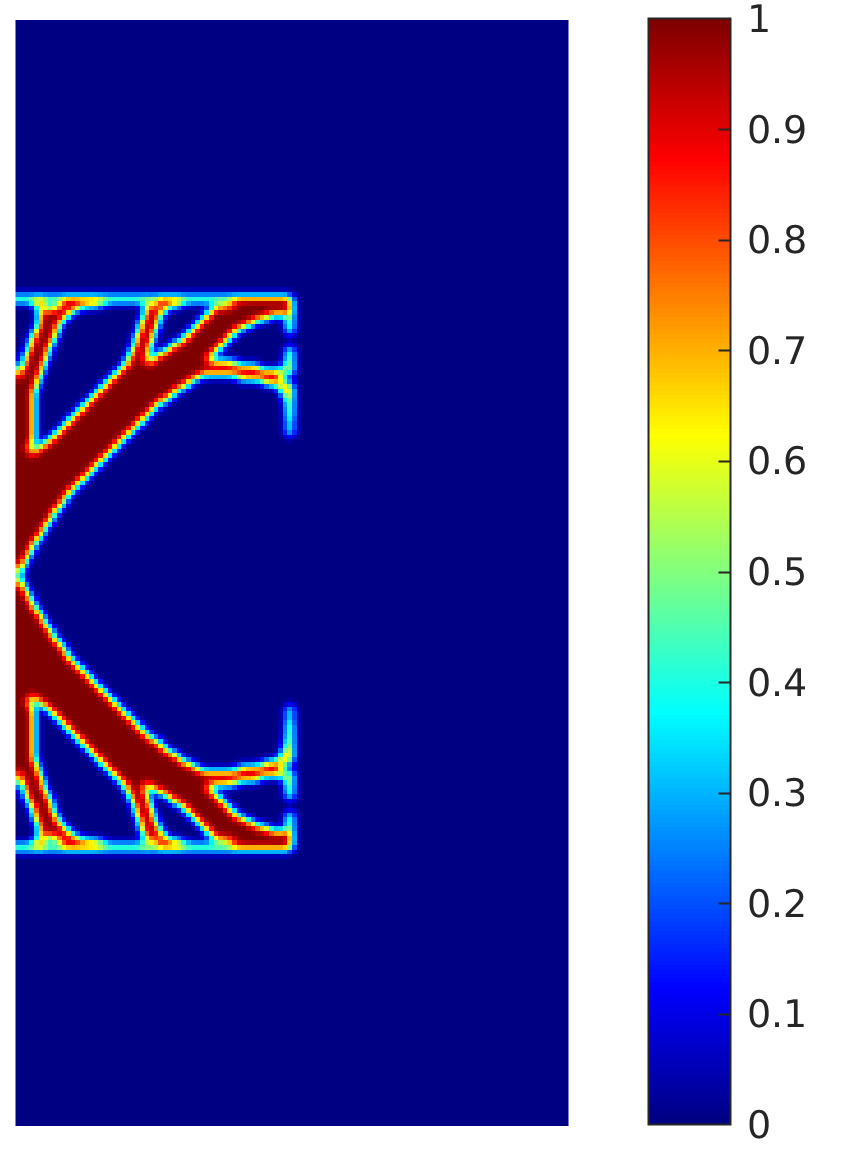}}\hfill
\subfloat[Temperature, $\textrm{Gr}=5120$]{\includegraphics[width=0.3\textwidth]{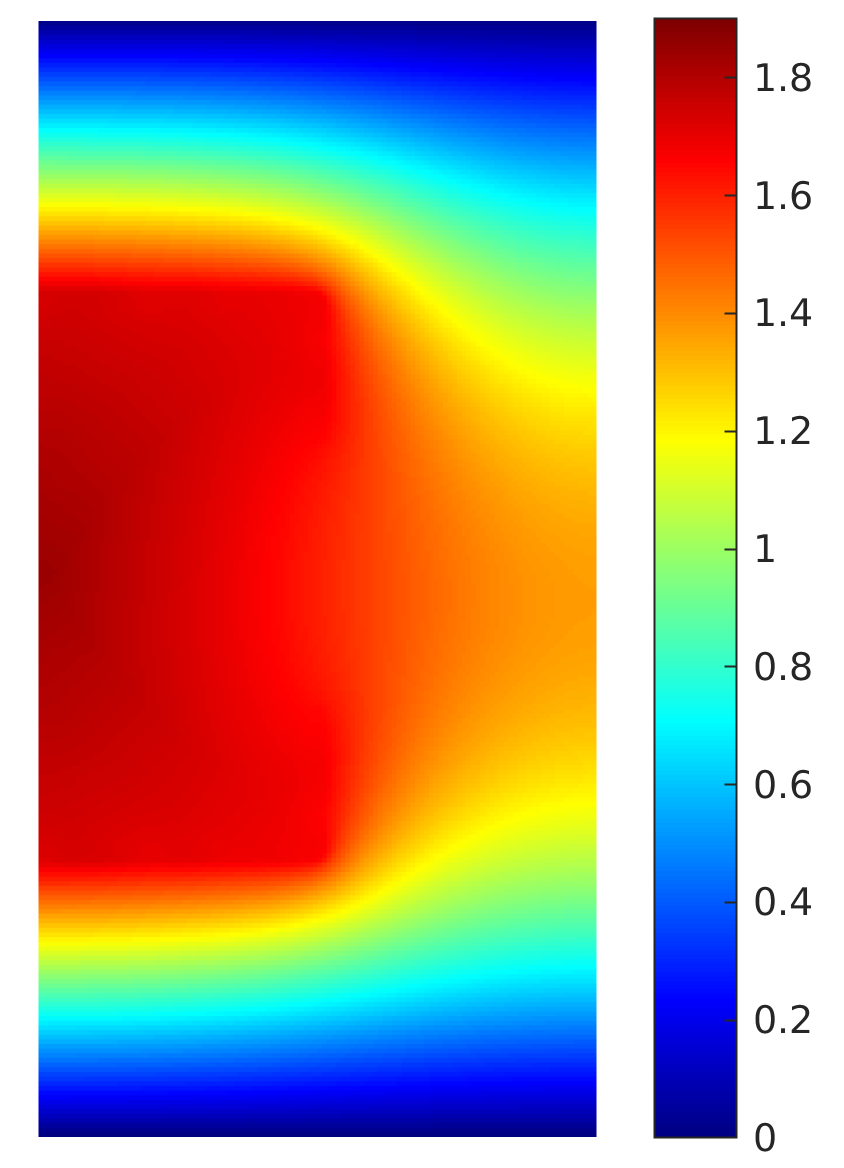}}\hfill
\subfloat[Velocity, $\textrm{Gr}=5120$]{\includegraphics[width=0.3\textwidth]{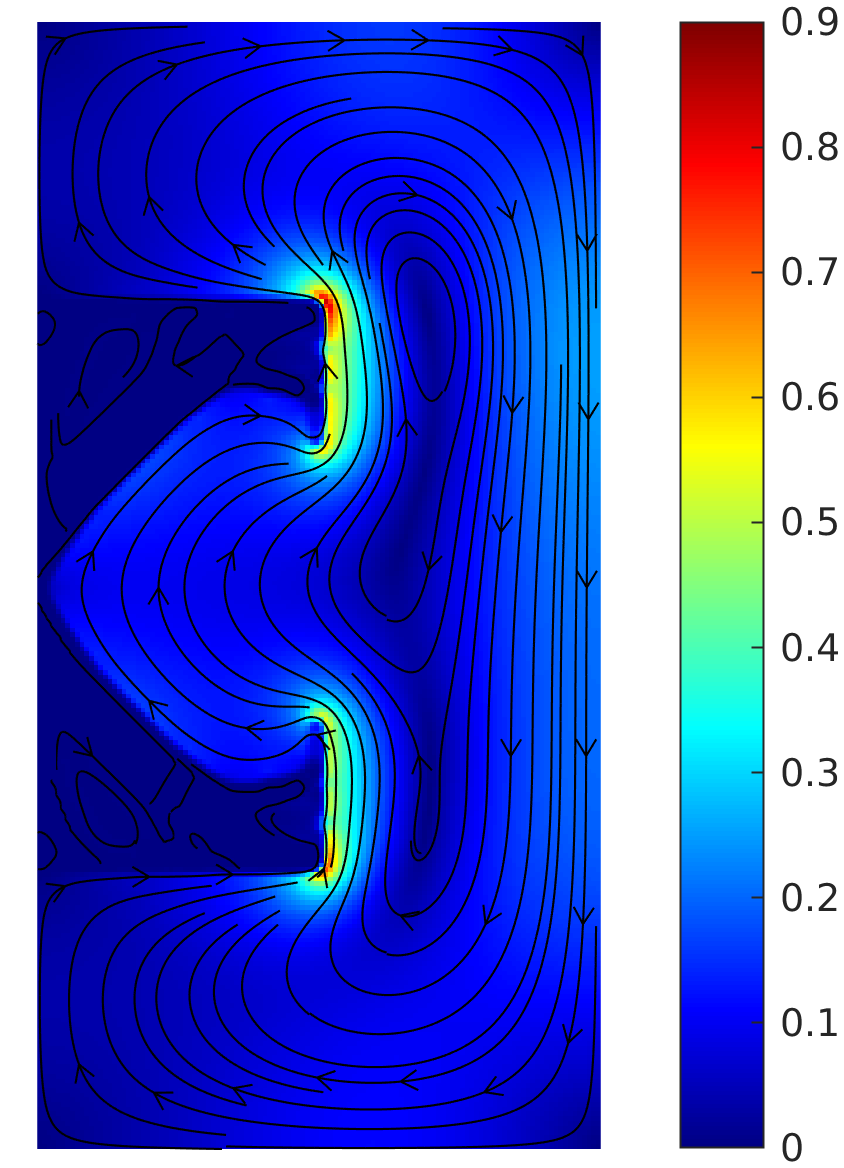}}\\
\subfloat[Design, $\textrm{Gr}=10240$]{\includegraphics[width=0.3\textwidth]{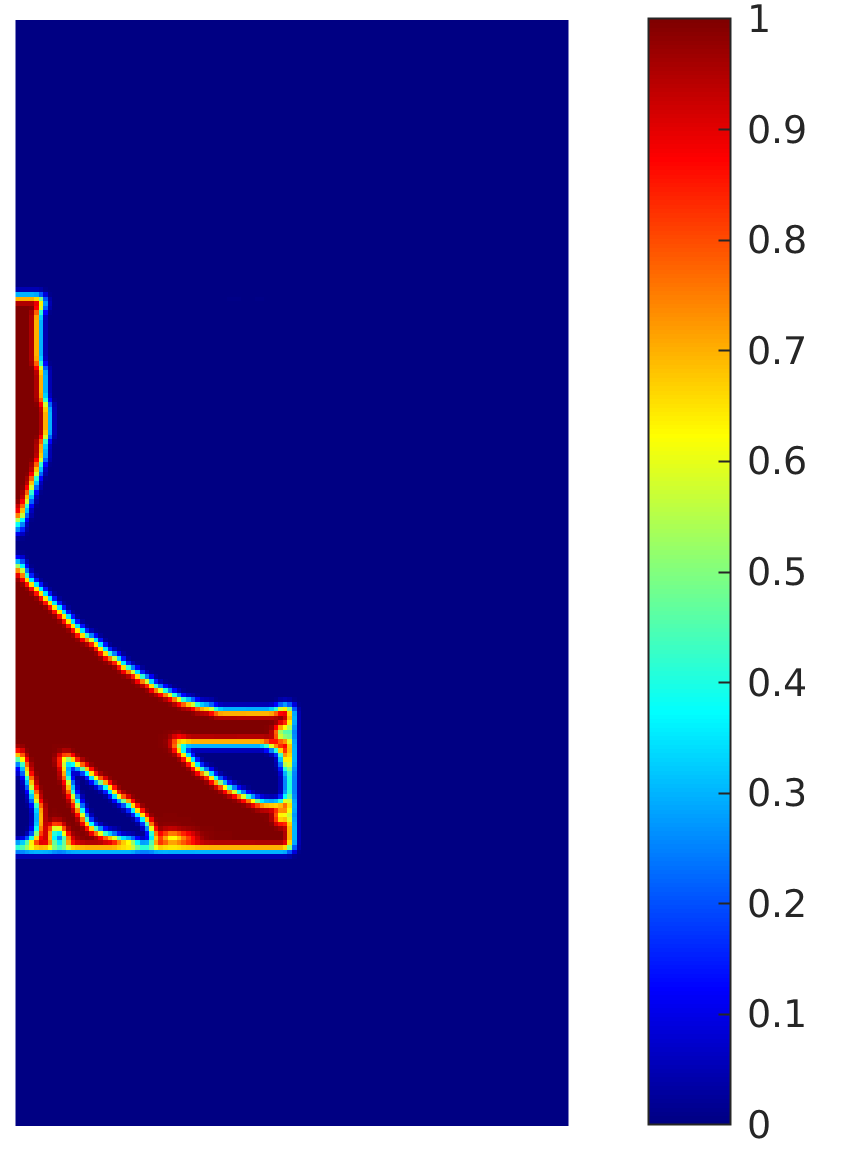}}\hfill
\subfloat[Temperature, $\textrm{Gr}=10240$]{\includegraphics[width=0.3\textwidth]{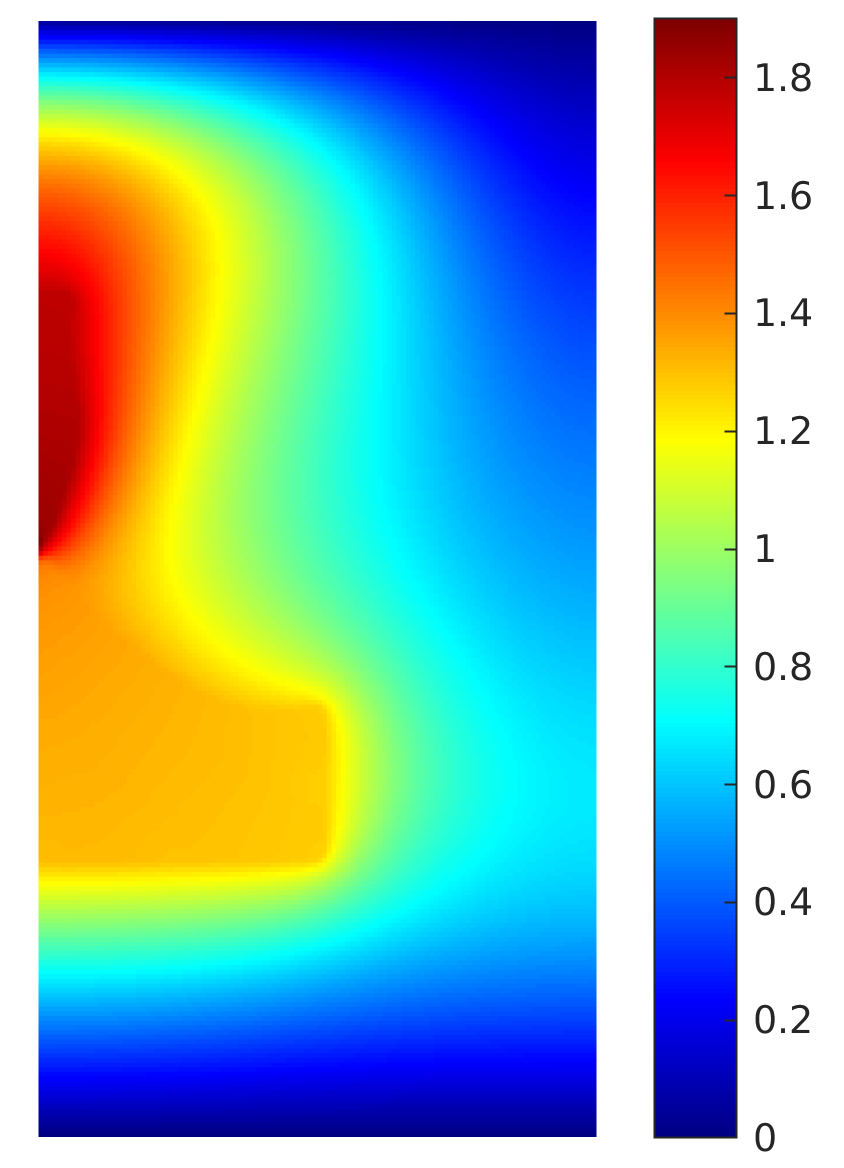}}\hfill
\subfloat[Velocity, $\textrm{Gr}=10240$]{\includegraphics[width=0.3\textwidth]{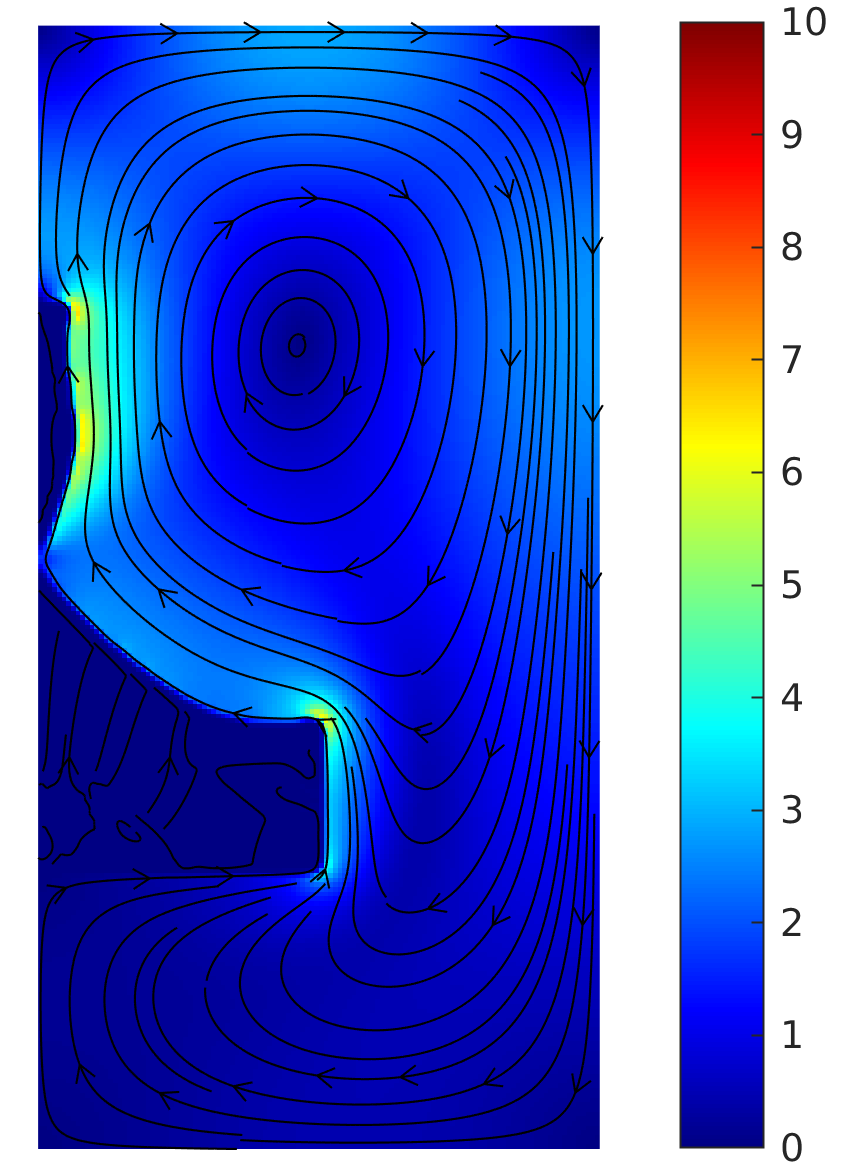}}\\
\subfloat[Design, $\textrm{Gr}=51200$]{\includegraphics[width=0.3\textwidth]{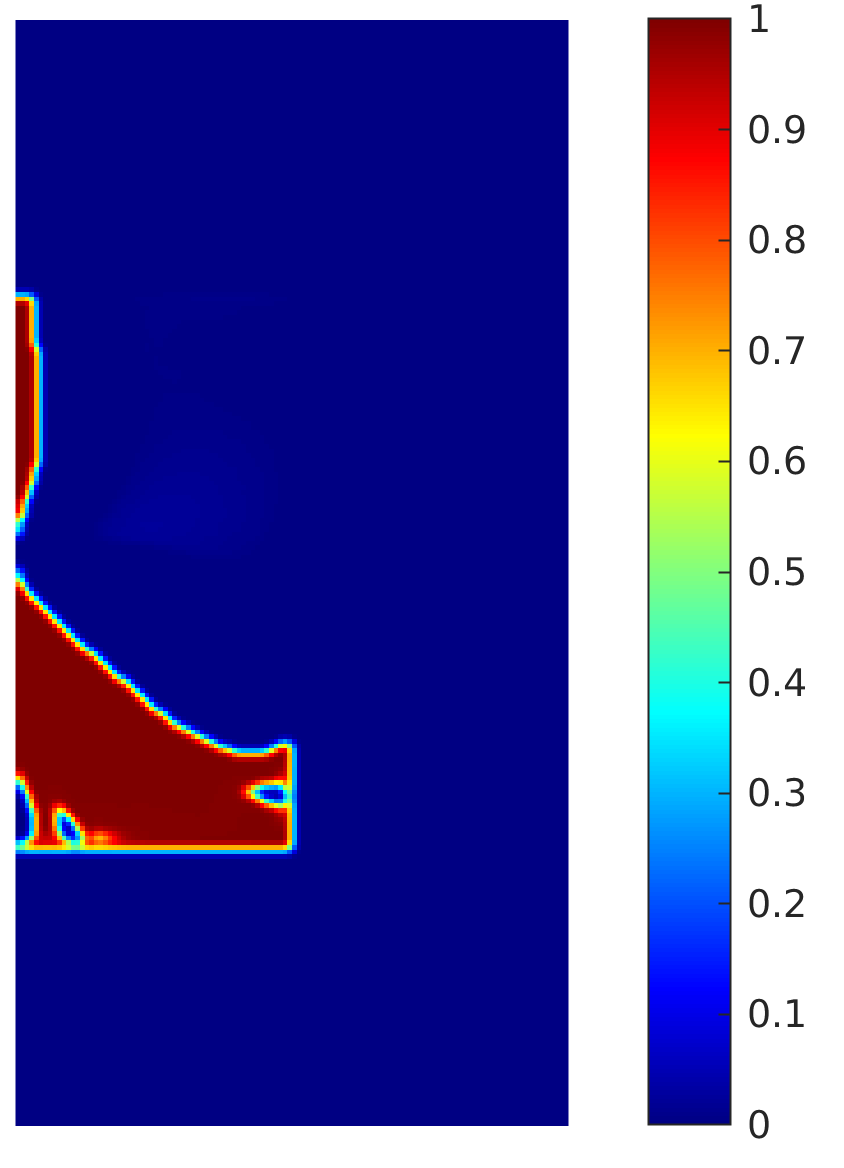}}\hfill
\subfloat[Temperature, $\textrm{Gr}=51200$]{\includegraphics[width=0.3\textwidth]{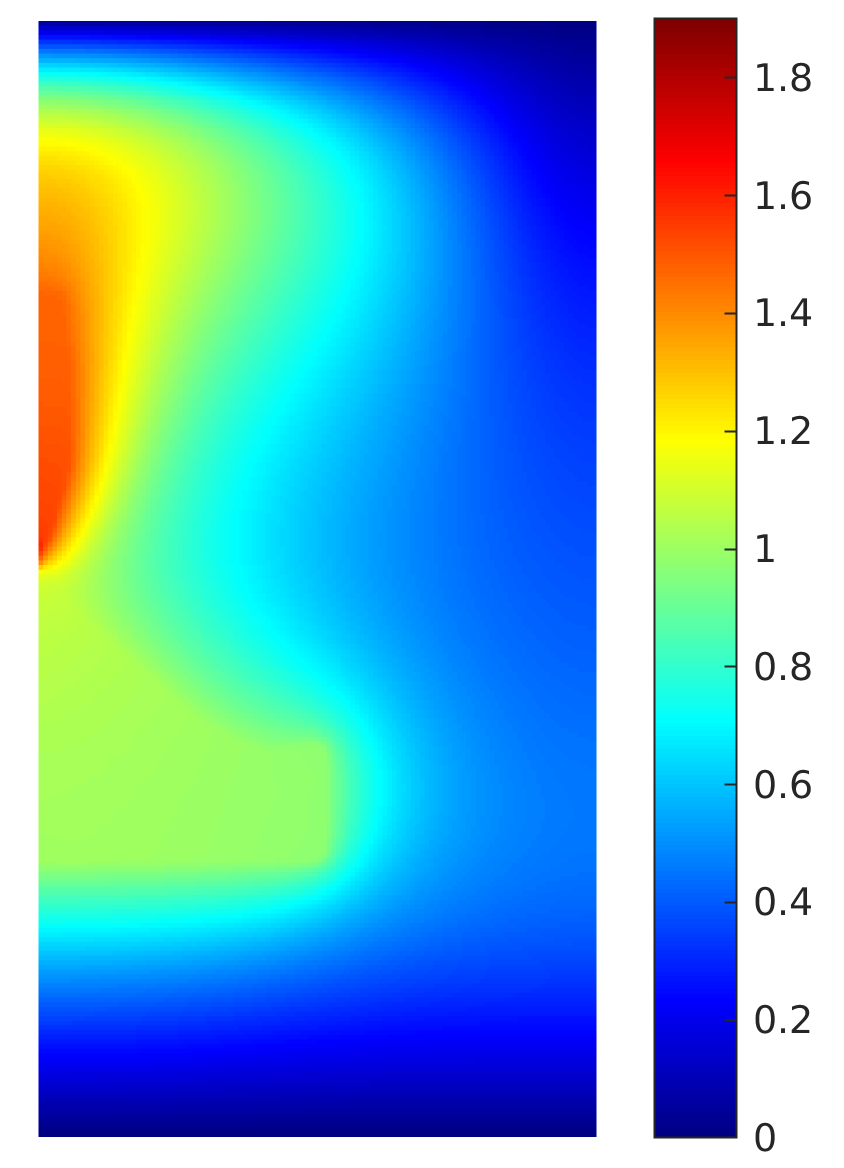}}\hfill
\subfloat[Velocity, $\textrm{Gr}=51200$]{\includegraphics[width=0.3\textwidth]{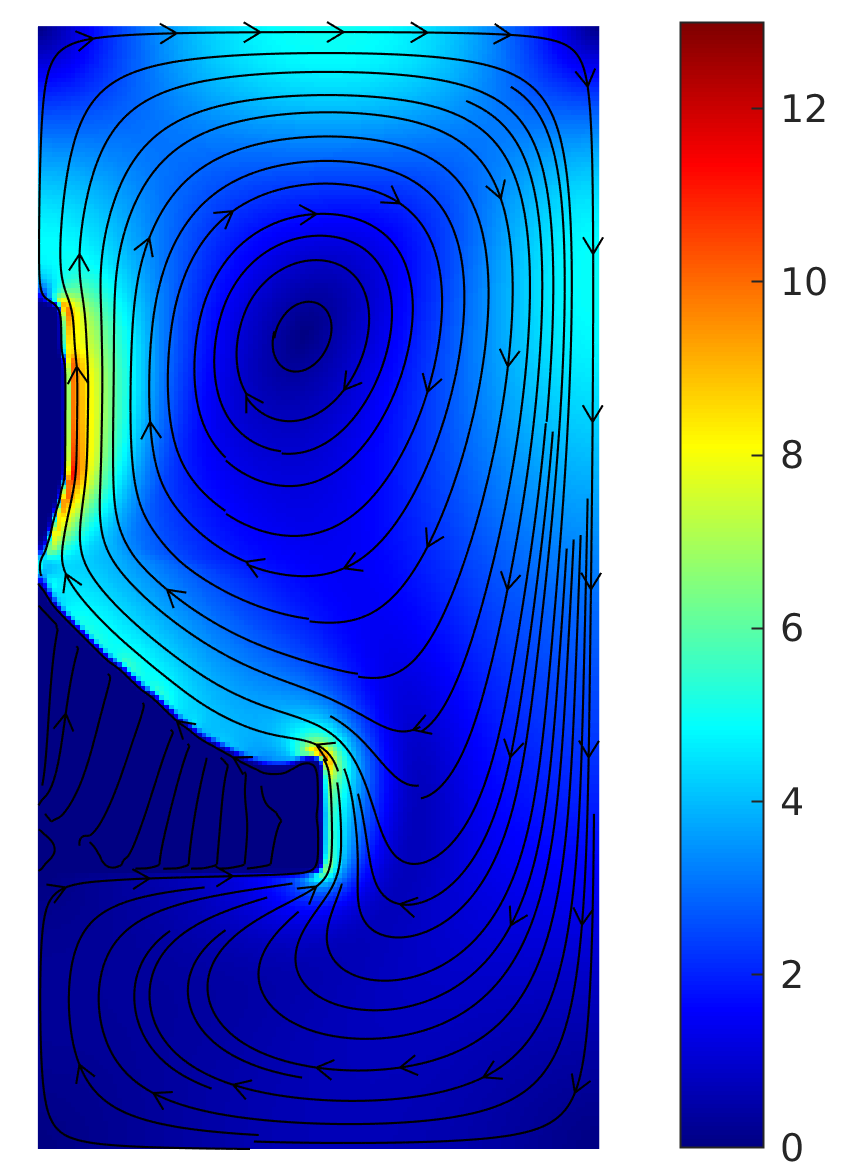}}
 \caption{Optimized designs and state plots for Grashof numbers $\textrm{Gr}=\{5120,10240,51200\}$. Note that  velocity color scale varies for increased visual reading.}
 \label{fig:ex2_designs}
\end{figure*}

The optimized designs for the thermal expansion coefficients $\beta=\{10, 50, 100\}$ are shown in Fig. \ref{fig:ex2_designs}. As in the previous example, material parameters are chosen such that $\textrm{Gr}=\textrm{Ra}=\beta H^3$ which yields $\textrm{Gr}=\{5120, 10240, 51200\}$. It is clear that for the lowest Grashof number the conductive transport is dominating, which is clear from the symmetric design with branching and the corresponding temperature plot. The velocities are moderate and stay below 1, having the highest magnitude at the vertical interface between heat sink and fluid. 
Increasing the Grashof number to  $\textrm{Gr}=10240$ and even further to $\textrm{Gr}=51200$ results in a very convection influenced design, at least for the upper part of the design domain. A heat sink that allows a large convection roll to form at the top part appears. The lower part of the heat sink has more in common with the previous diffusion dominated design, where branching is formed to support the heat diffusion away from the source towards the cold bottom boundary. The heat sink is divided centrally by a small fluid spacing. This insulates the bottom part from the top and restricts the temperature in this part of the sink. The bottom part is cooled primarily by diffusion at the bottom face and by convection at the upper face. At this face the heat is exchanged to the cold fluid which makes the fluid gain velocity. Further up, in the top part of the heat sink, the fluid momentum is boosted by the higher temperature in the top part of the heat sink. This difference in heat transport is also visible 
from the velocity magnitude plot, 
which clearly shows 
that the velocities are in general low in the bottom part of the domain where diffusion dominates. The velocity is high at the sink's vertical interface with the fluid domain, where the hot sink is heating the fluid and making it flow in the direction of gravity. The largest velocities are obtained near the upper hot part of the heat sink. 

\begin{table*}[h]
 \begin{tabular}{l c c c c c c}
 &\multicolumn{3}{c}{Compliance at Gr} & \multicolumn{3}{c}{$\Delta T_{max}$ at Gr}\\
\noalign{\smallskip} \hline\noalign{\smallskip}
  Designed at Gr & 5120 & 10240 & 51200  &  5120 & 10240 & 51200 \\
  \noalign{\smallskip}\hline\noalign{\smallskip}
   5120&  \textbf{10.75} & 10.07 &  8.50 & 1.85& 1.74 & 1.48\\
  10240 & 12.54 & \textbf{9.24} & 7.52 & 2.42& 1.88& 1.55\\
  51200 & 12.83 & 9.27 &\textbf{7.48} & 2.54 & 1.92 & 1.56\\
  \noalign{\smallskip}\hline
 \end{tabular}
\caption{Performance for optimized designs evaluated under all conditions. The Grashof numbers refer to an assumed temperature difference $\Delta T =1$. The maximum temperature is for all cases identified slightly above the centre of the heat source boundary.}
\label{tab:ex2_performance}
\end{table*}

The performance of the optimized designs can be compared in-between the different designs and values are listed in Tab. \ref{tab:ex2_performance}. Each of the designs perform best at their respective design condition. The maximum temperature reveals that the true Grashof number is about the double value of the a priori computed.

\begin{figure*}
\subfloat[$\textrm{Gr}=5120$]{\includegraphics[width=0.3\textwidth]{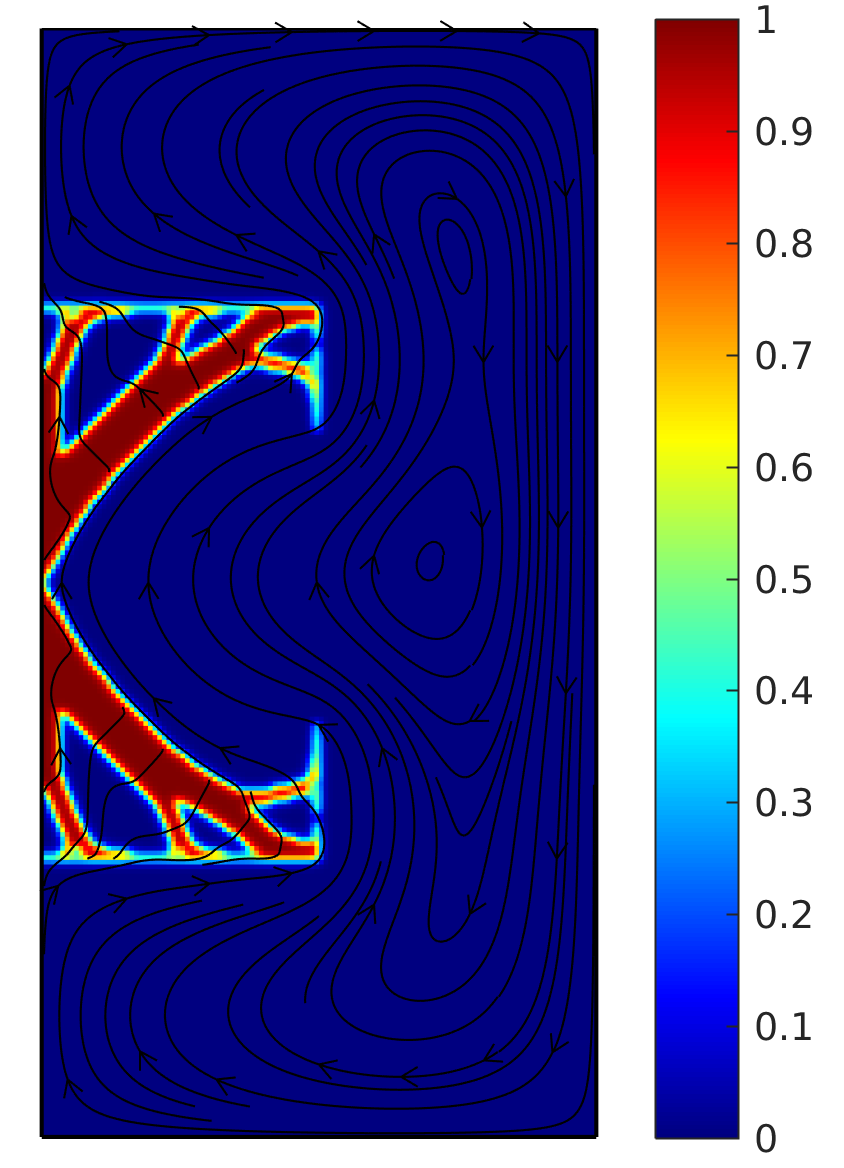}}\hfill
\subfloat[$\textrm{Gr}=10240$]{\includegraphics[width=0.3\textwidth]{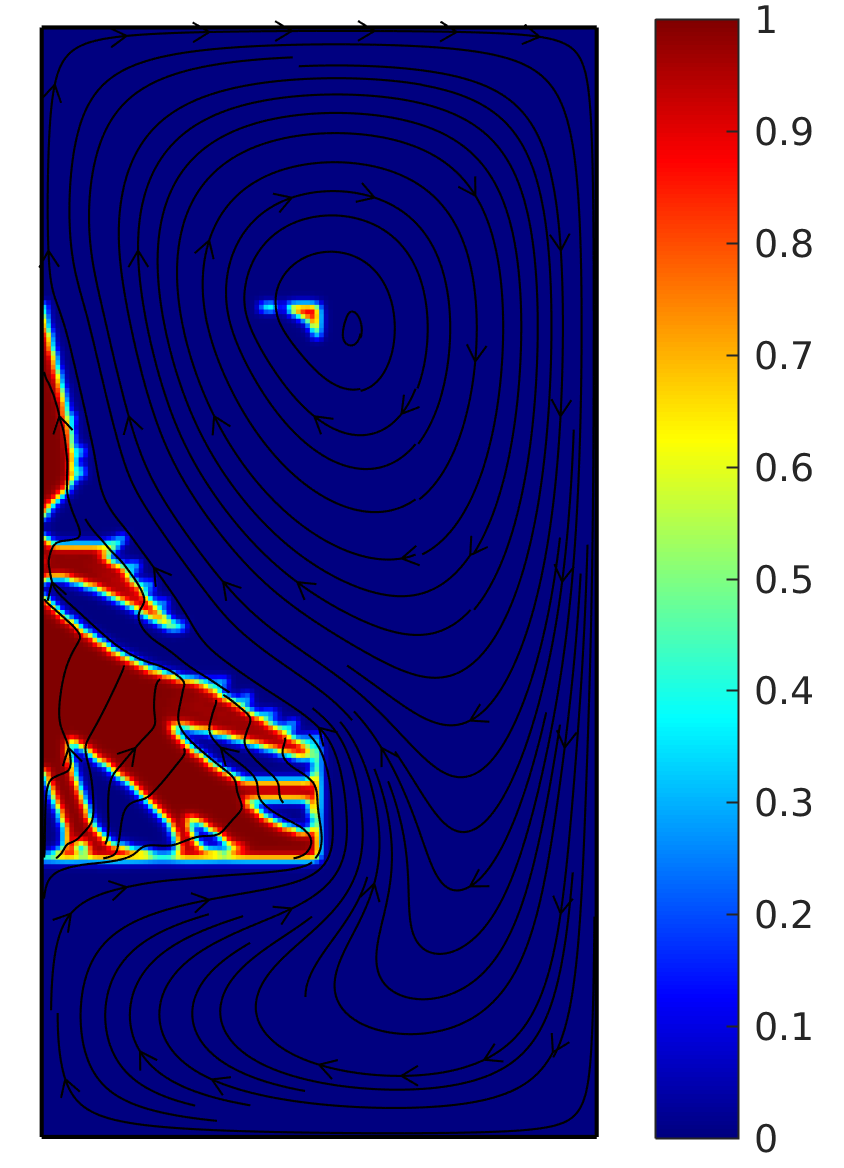}}\hfill
\subfloat[$\textrm{Gr}=51200$]{\includegraphics[width=0.3\textwidth]{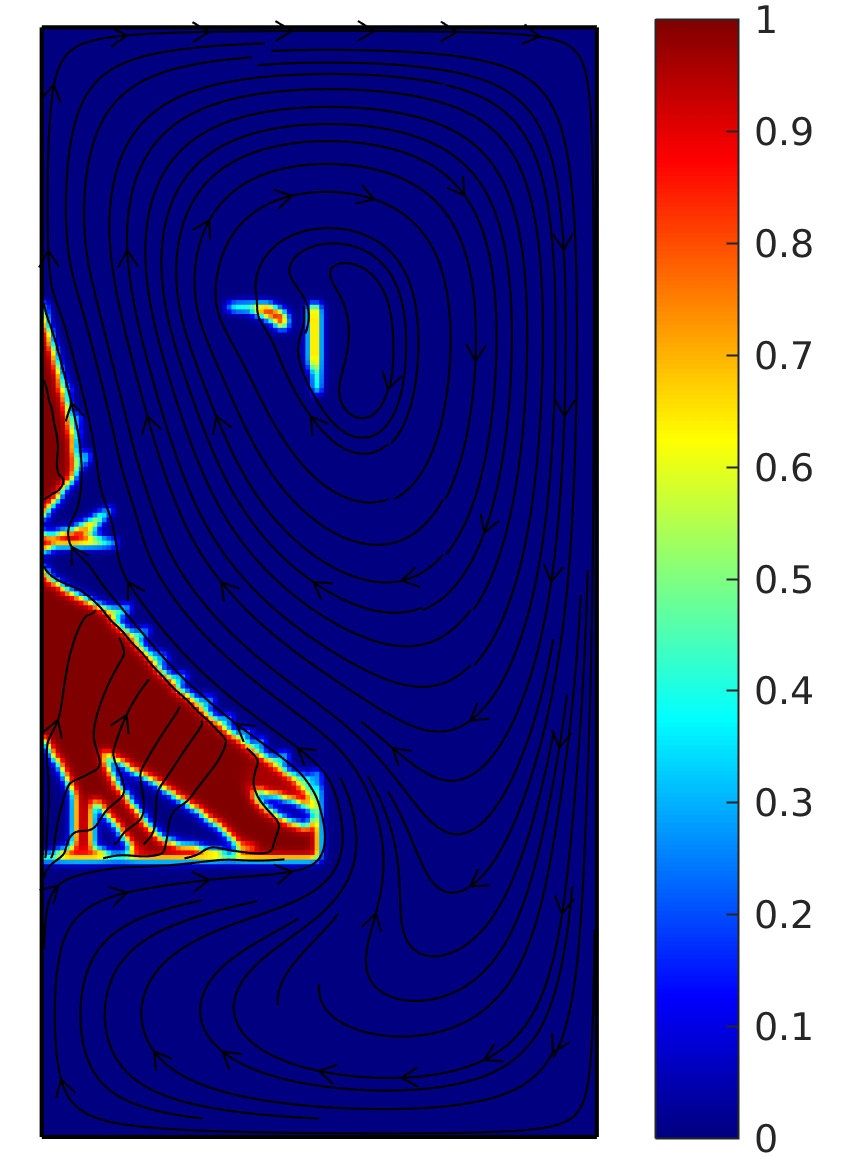}}

 \caption{Comparative designs obtained using the fully coupled Navier-Stokes-Brinkman equations for the same Grashof numbers $\textrm{Gr}=\{5120,10240,51200\}$. Designs obtained using a MATLAB-COMSOL implementation of the formulation used in \cite{Alexandersen2014}. The objective values read $\phi_{5120}=11.10$, $\phi_{10240}=7.94$, $\phi_{51200}=6.93$}
 \label{fig:ex2_NSB}
\end{figure*}

The designs obtained by the reduced-order model can be compared to those obtained by a full NSB model in order to verify that the obtained designs are somewhat similar and the design trend stepping from a diffusion dominated to a convection dominated setting results in comparable designs. Such designs are shown in Fig. \ref{fig:ex2_NSB} and it is clear that the results match very well for the diffusion-dominated design. 

The convection-dominated design differs slightly as the NSB design seems to facilitate a smooth low-curvature flow path at the fluid-solid interface of the heat sink. This is different from the reduced order design, where the curvature is steeper, as was also observed for the first example. This may be attributed to the neglected viscous dissipation, which favours smooth directional changes. In the designs obtained for $\textrm{Gr}=10240$ and $\textrm{Gr}=51200$, a small amount of material is left near the eye of the convection cell and the reason for placing it here is attributed purely to minor flow guidance effects as the conductive properties are poorly 
exploited at this position. It 
is likewise possible to achieve such small amounts of material at this location using the reduced-order model. However, it is the authors experience that the occurrence depends on initial material distribution and penalization scheme and has limited influence on the end performance.

\begin{figure*}

\includegraphics[width=0.3\textwidth]{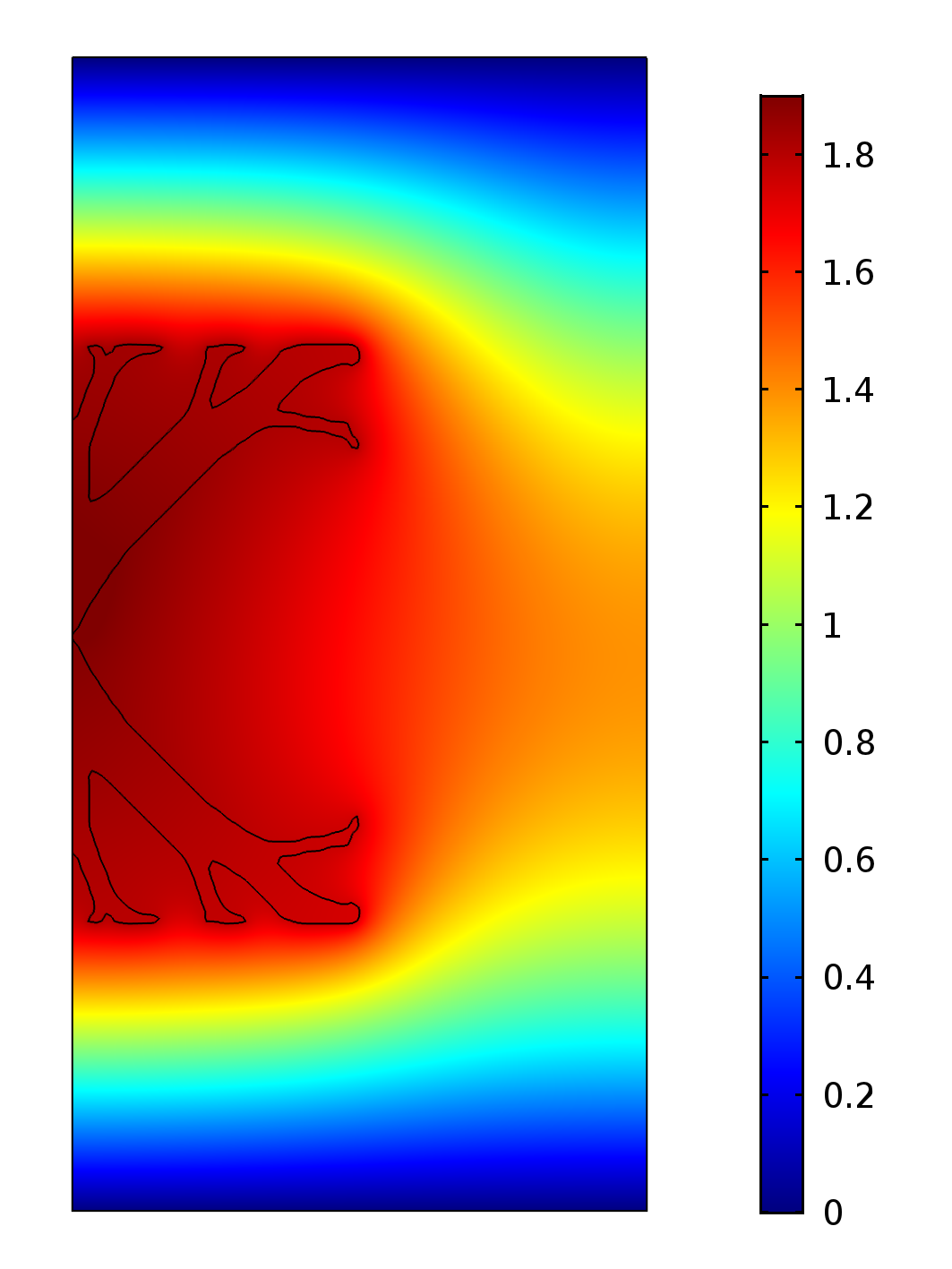}\hfill
\includegraphics[width=0.3\textwidth]{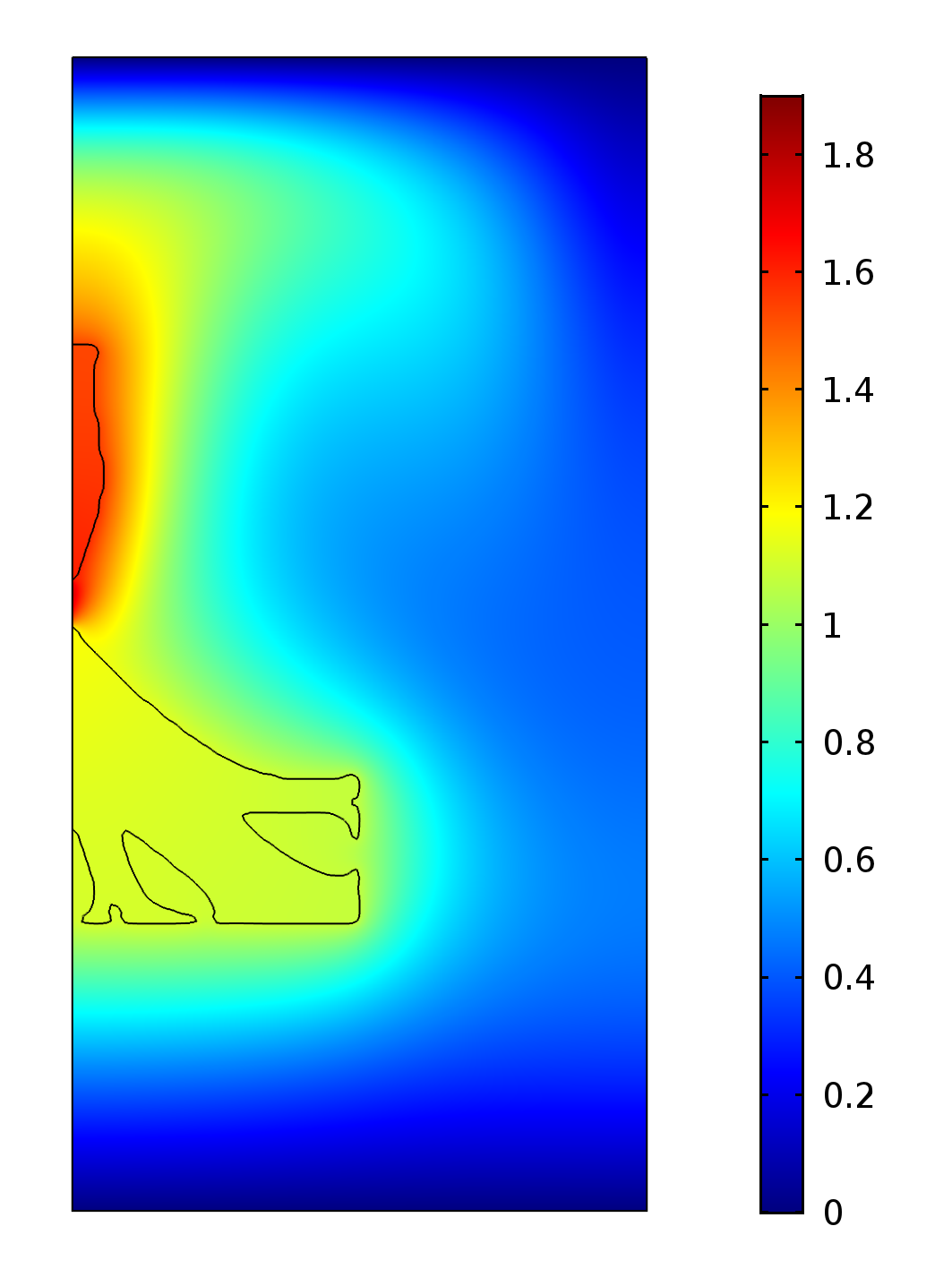}\hfill
\includegraphics[width=0.3\textwidth]{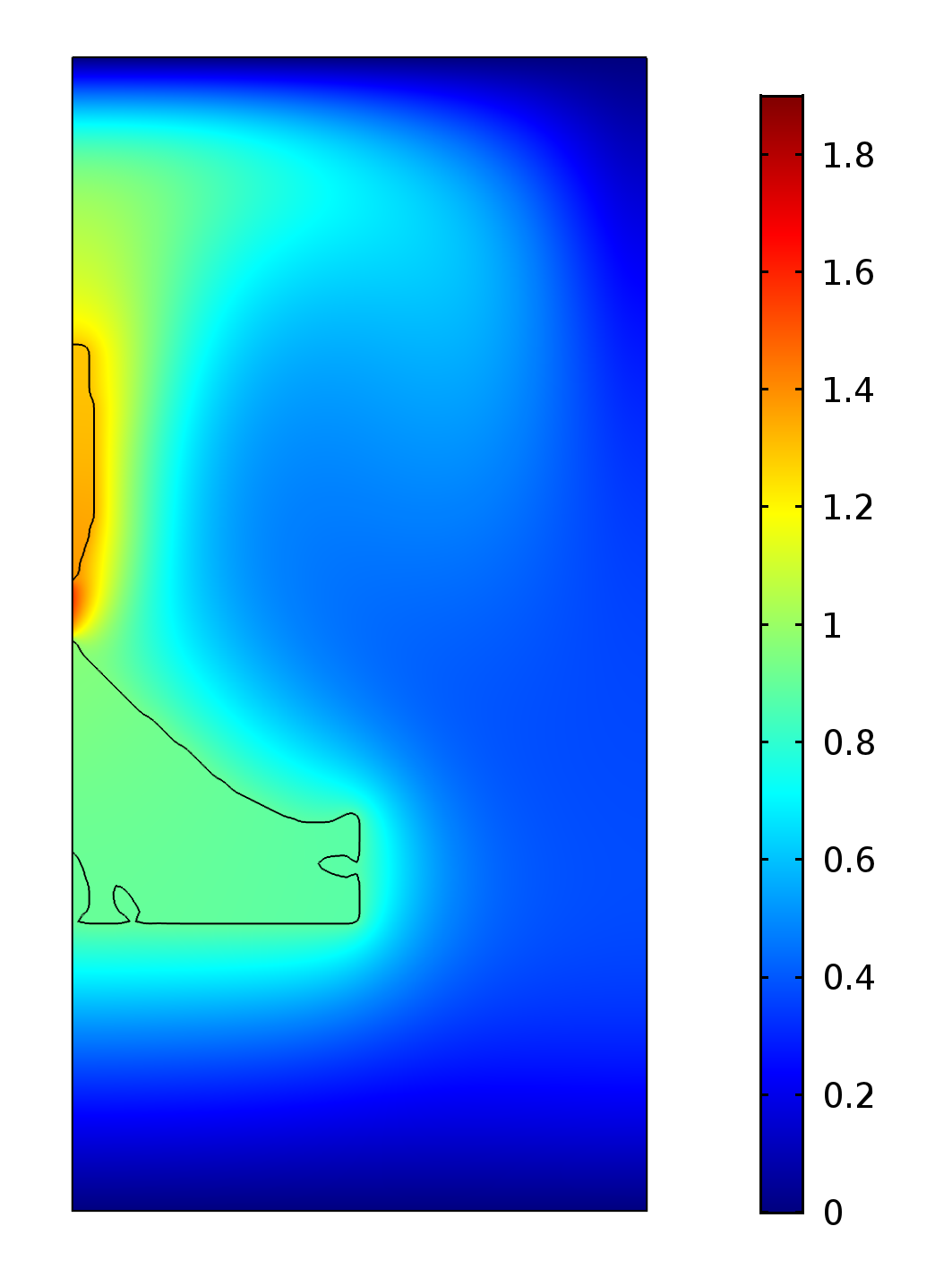}\\
\includegraphics[width=0.3\textwidth]{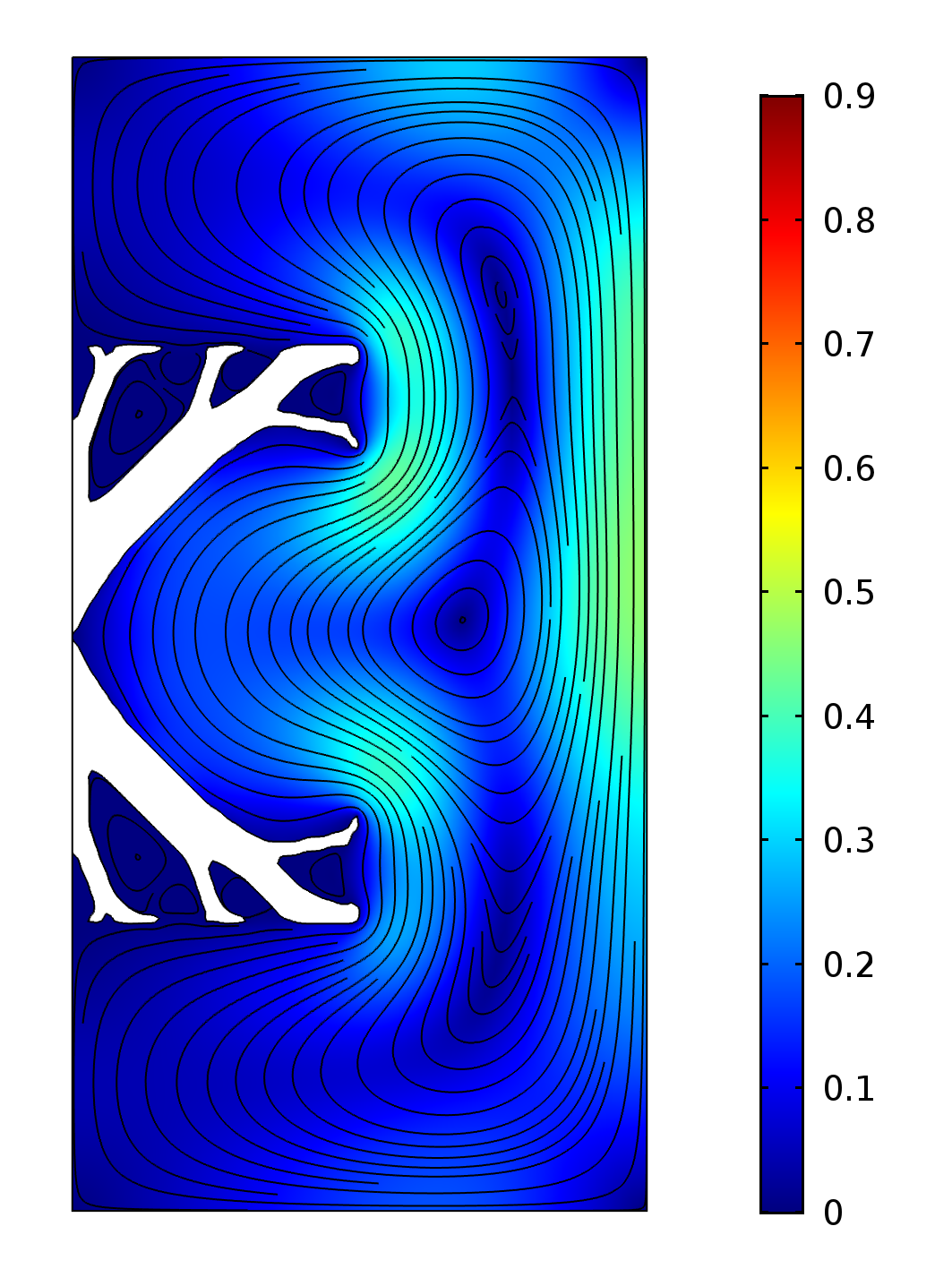}\hfill
\includegraphics[width=0.3\textwidth]{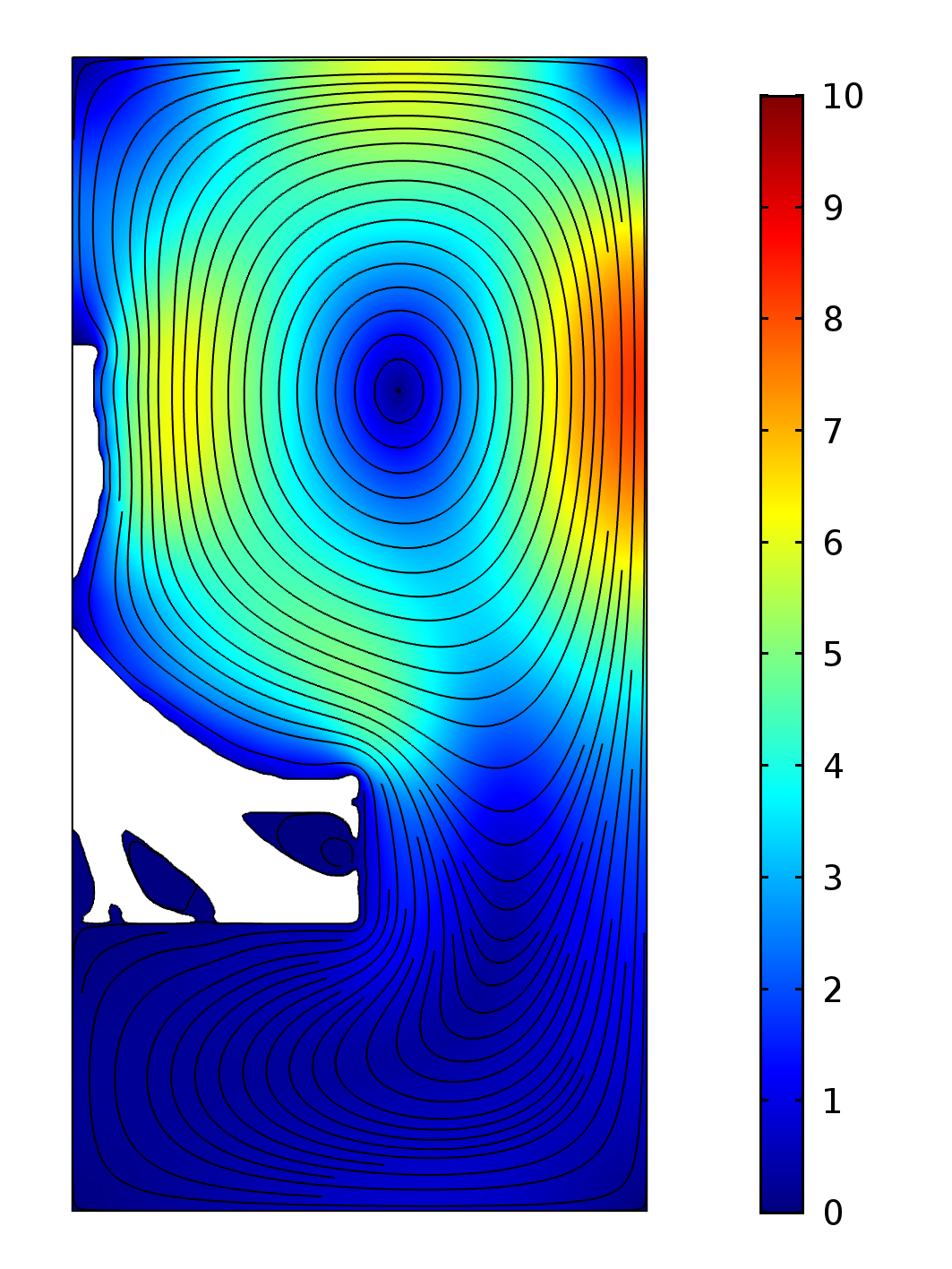}\hfill
\includegraphics[width=0.3\textwidth]{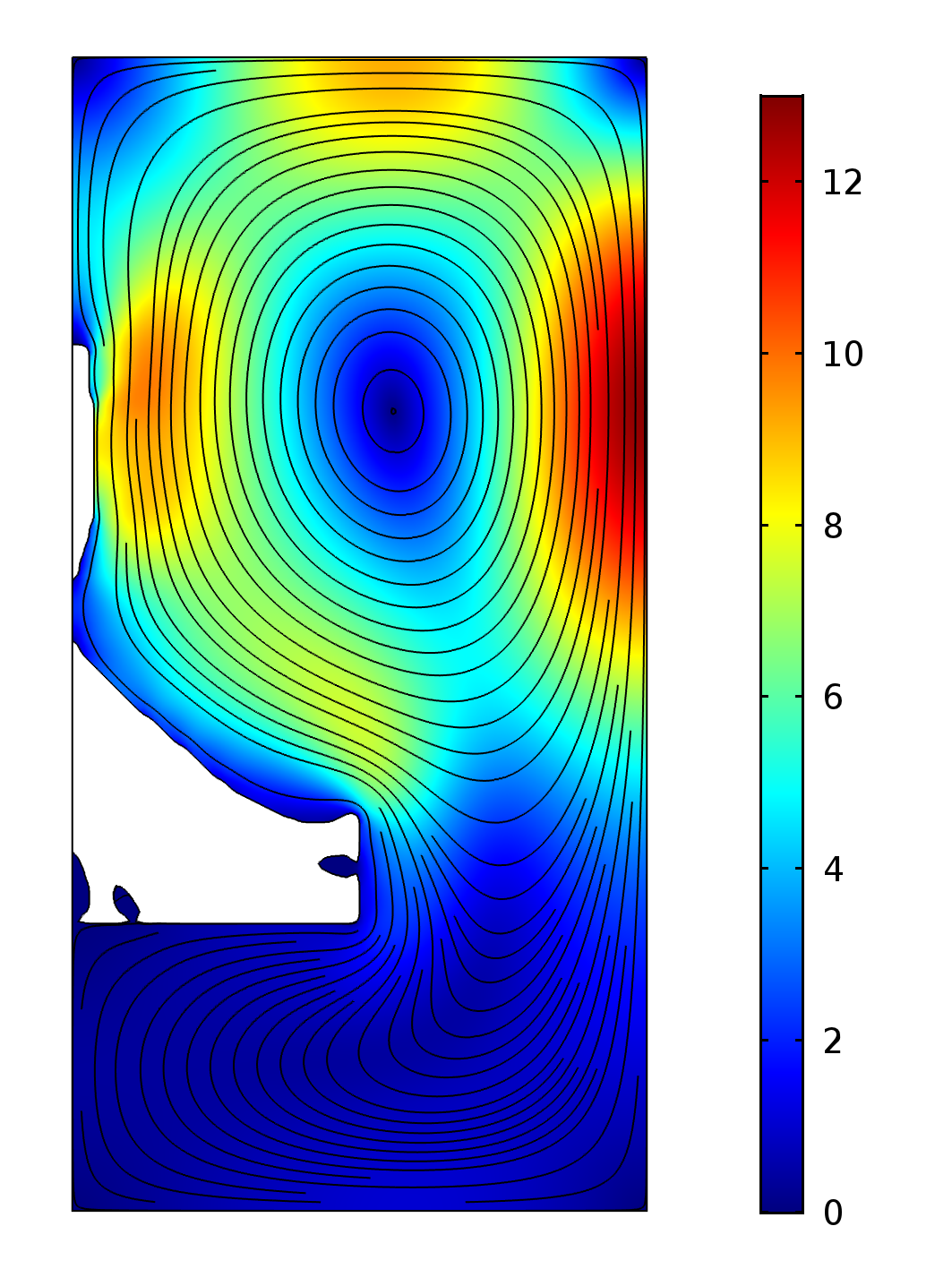}
\caption{Reevaluation of designs using COMSOL Multiphysics 5.3. Designs thresholded at $\gamma=0.5$ and extracted using and isocontour.  All boundaries are subject to the slip condition. Top row: Temperature plots. Bottom row: Velocity magnitude and streamlines. Temperature and velocity ranges correspond to Fig. \ref{fig:ex2_designs}}
\label{fig:ex2_NS_reeval}
\end{figure*}

The obtained designs from the reduced-order model have been thresholded at $\gamma=0.5$, smoothed using an isocontour and exported to COMSOL. Using $\gamma=0.5$ as the threshold value yields exported designs with volume fractions close to the optimized. The visualisations of the results are shown in Fig. \ref{fig:ex2_NS_reeval} and the obtained performance, solid volume fraction and maximum temperatures are listed in Tab. \ref{tab:ex2_NS_performance} for all conditions. Here the boundary conditions have been maintained as slip boundaries as for the reduced order model. This also applies to the solid-fluid interface.

For the lowest Grashof numbers the temperature and velocity distributions are very similar which is attributed to the diffusion dominance. The velocities show the expected behaviour and the streamlines of the two models are similar. The colours show that the local maximum velocity at the fluid-solid interface differs from the NS model where the maximum is attained at a certain distance from the solid due to viscous boundary effects.

When the Grashof number is increased it is seen that the difference in the temperature distributions increases. The localisation of the velocity seems to limit the ability to convect the fluid. Comparing the flow speed for $\textrm{Gr}=10240$, it is evident that the maximum velocity near the top part of the heat sink for both models is around 6-7. As the high velocity area in the full model is much more distributed, this also means that more momentum is available which in turn convects more heat resulting in a more mushroom-shaped thermal plume. This also applies to the $\textrm{Gr}=51200$ case where the plume is increasingly mushroom-shaped, but still lacks momentum in comparison to the full model. For all cases, the diffusion dominated behaviour at the bottom of the domain seems similar for both models.

\begin{table*}[h]
 \begin{tabular}{l c c c c c c c}
 &\multicolumn{3}{c}{Evaluated at Gr} &  & \multicolumn{3}{c}{$\Delta T_{max}$ at Gr}\\
 \noalign{\smallskip}\hline\noalign{\smallskip}
  Designed for Gr & 5120 & 10240 & 51200 & Solid vol.frac. & 5120 & 10240 & 51200\\
  \noalign{\smallskip}\hline\noalign{\smallskip}
  5120 & 11.36          & 10.11 & 8.52 & 0.287 & 1.93 & 1.77 & 1.49\\
  10240 & 11.31          & 8.24 & 7.27 & 0.296 & 2.33 & 1.76 & 1.58\\
  51200& \textbf{11.17} & \textbf{8.07} & \textbf{7.09} & 0.294 & 2.36 & 1.77 & 1.58\\
  \noalign{\smallskip}\hline
 \end{tabular}
\caption{Performance for reevaluated designs using design extraction and full NS model cf. Fig. \ref{fig:ex2_NS_reeval}. The volume fraction of the extracted solid wrt. design domain is given. All designs are originally attaining the 0.3 limit.}
\label{tab:ex2_NS_performance}
\end{table*}

The post processed performance listed in Tab. \ref{tab:ex2_NS_performance} shows that the thresholded and smoothed $\textrm{Gr}=51200$ design outperform all the others using full NS modeling. However, it must be noted that the difference in thermal compliance is very small, in the order of $2\%$ if one compares the performance of a design optimized at a certain setting with that obtained using the  ``$\textrm{Gr}=51200$ design`` evaluated at that setting. This small difference may of course be due to the lower fidelity of the reduced-order model and the tuning hereof, but may also be a consequence of the better geometry representation obtained using a smoothed geometry represented in a body-fitted mesh. 

\section{Discussion and conclusions}
The paper has demonstrated a novel method for order reduction of models for optimization of natural convection problems. The reduction in dimensionality from the Navier-Stokes equations to a potential flow model has been shown to be efficient and work relatively well as a vehicle for the modeling of convection. The major obstacle for general applicability may be to find a suitable test case for tuning the material parameter $\bar{\mu}$ to yield representative state fields under different conditions and during design evolution. The lower computational complexity of the reduced-order model has shown to speed up the topology optimization design procedure for fully coupled natural buoyancy problems. 

The obtained designs have been compared to those obtained using the full model and the general tendencies and performance is maintained in high-fidelity simulations. Another challenge of employing this model is clearly the lack of a boundary layer, which tends to overpredict local velocities near the boundary, while underpredicitng those further away. In the second example, this resulted in the reduced-order model to underpredict the convection and thus predict a slightly lower performance. For other cases it may be opposite, e.g. in the case of narrow channels in the design, which may cause overprediction of the convective performance in the reduced-order model due to the lack of viscous friction.

 In comparison to the model utilizing Newton's law of cooling, which is one step further down the order-reduction-path, the potential flow model is superior in predicting convection surfaces, as it does include the ability to model the natural convection in narrow and closed regions of the heat sink, which is clearly a problem for the simplified convection model.
 
 The presented reduced-order model enables designers to shorten the computational time for design synthesis. The obtained designs come close to those of a full Navier-Stokes-Brinkman model and may be postprocessed using CAD directly or used as a close-to-optimal initial design for a method that models the fully coupled Navier-Stokes equations.\\

\textbf{Acknowledgements}
\textit{The work has been partly funded by the TopTEN project granted by Independent Research Fund Denmark. The authors would like to thank the TopOpt group for fruitfull discussions. }

\bibliographystyle{spbasic_csan}      
\bibliography{csan}   

\begin{thebibliography}{64}
\providecommand{\natexlab}[1]{#1}
\providecommand{\url}[1]{{#1}}
\providecommand{\urlprefix}{URL }
\expandafter\ifx\csname urlstyle\endcsname\relax
  \providecommand{\doi}[1]{DOI~\discretionary{}{}{}#1}\else
  \providecommand{\doi}{DOI~\discretionary{}{}{}\begingroup
  \urlstyle{rm}\Url}\fi
\providecommand{\eprint}[2][]{\url{#2}}

\bibitem[{Alexandersen(2011)}]{Alexandersen2011}
Alexandersen J (2011) Topology optimisation of convection problems. {B.Eng.}
  thesis, Technical University of Denmark, \doi{10.13140/RG.2.2.24635.72485}

\bibitem[{Alexandersen(2013)}]{Alexandersen2013}
Alexandersen J (2013) Topology optimization of coupled conveciton problems.
  Master's thesis, Technical University of Denmark

\bibitem[{Alexandersen(2016)}]{AlexandersenThesis}
Alexandersen J (2016) Efficient topology optimisation of multiscale and
  multiphysics problems. PhD thesis, Technical University of Denmark

\bibitem[{Alexandersen et~al(2014)Alexandersen, Aage, Andreasen, and
  Sigmund}]{Alexandersen2014}
Alexandersen J, Aage N, Andreasen CS, Sigmund O (2014) {Topology optimisation
  for natural convection problems}. International Journal for Numerical Methods
  in Fluids 76(10):699--721, \doi{10.1002/fld.3954}

\bibitem[{Alexandersen et~al(2015)Alexandersen, Sigmund, and
  Aage}]{Alexandersen2015}
Alexandersen J, Sigmund O, Aage N (2015) Topology optimisation of passive
  coolers for light-emitting diode lamps. In: 11th World Congress on Structural
  and Multidisciplinary Optimization, \doi{10.13140/RG.2.1.3906.5446}

\bibitem[{Alexandersen et~al(2016)Alexandersen, Sigmund, and
  Aage}]{Alexandersen2016a}
Alexandersen J, Sigmund O, Aage N (2016) {Large scale three-dimensional
  topology optimisation of heat sinks cooled by natural convection}.
  International Journal of Heat and Mass Transfer 100:876--891,
  \doi{10.1016/j.ijheatmasstransfer.2016.05.013}

\bibitem[{Alexandersen et~al(2018)Alexandersen, Sigmund, Meyer, and
  Lazarov}]{Alexandersen2018}
Alexandersen J, Sigmund O, Meyer K, Lazarov BS (2018) Design of passive coolers
  for light-emitting diode lamps using topology optimisation. International
  Journal of Heat and Mass Transfer 122:138--149,
  \doi{10.1016/j.ijheatmasstransfer.2018.01.103}

\bibitem[{Andreasen et~al(2009)Andreasen, Gersborg, and
  Sigmund}]{Andreasen2009}
Andreasen CS, Gersborg AR, Sigmund O (2009) {Topology optimization of
  microfluidic mixers}. International Journal for Numerical Methods in Fluids
  61(5):498--513, \doi{10.1002/fld.1964}

\bibitem[{Angot et~al(1999)Angot, Bruneau, and Fabrie}]{Angot1999}
Angot P, Bruneau CH, Fabrie P (1999) A penalization method to take into account
  obstacles in incompressible viscous flows. Numerische Mathematik
  81(4):497--520, \doi{10.1007/s002110050401}

\bibitem[{Bends{\o}e and Kikuchi(1988)}]{Bendsoe1988}
Bends{\o}e MP, Kikuchi N (1988) {Generating optimal topologies in structural
  design using a homogenization method}. Computer Methods in Applied Mechanics
  and Engineering 71(2):197--224, \doi{10.1016/0045-7825(88)90086-2}

\bibitem[{Bends{\o}e and Sigmund(2003)}]{Bendsoe2003}
Bends{\o}e MP, Sigmund O (2003) {Topology Optimization - Theory, Methods, and
  Applications}. Springer Verlag, Berlin Heidelberg

\bibitem[{Borrvall and Petersson(2003)}]{Borrvall2003}
Borrvall T, Petersson J (2003) {Topology optimization of fluids in Stokes
  flow}. International Journal for Numerical Methods in Fluids 41(1):77--107,
  \doi{10.1002/fld.426}

\bibitem[{Bourdin(2001)}]{Bourdin2001}
Bourdin B (2001) {Filters in topology optimization}. Int J Numer Meth Engng
  50(9):2143--2158, \doi{10.1002/nme.116}

\bibitem[{Brinkman(1947)}]{Brinkman1947}
Brinkman HC (1947) {A Calculation of the Viscous Force Exerted By A Flowing
  Fluid On A Dense Swarm of Particles}. Applied Scientific Research Section
  A-mechanics Heat Chemical Engineering Mathematical Methods 1(1):27--34

\bibitem[{Brooks and Hughes(1982)}]{Brooks1982}
Brooks AN, Hughes TJ (1982) Streamline upwind/petrov-galerkin formulations for
  convection dominated flows with particular emphasis on the incompressible
  navier-stokes equations. Computer Methods in Applied Mechanics and
  Engineering 32(1):199 -- 259, \doi{10.1016/0045-7825(82)90071-8}

\bibitem[{Bruns(2007)}]{Bruns2007}
Bruns TE (2007) {Topology optimization of convection-dominated, steady-state
  heat transfer problems}. International Journal of Heat and Mass Transfer
  50(15-16):2859--2873, \doi{10.1016/j.ijheatmasstransfer.2007.01.039}

\bibitem[{Bruns and Tortorelli(2001)}]{BruTor01}
Bruns TE, Tortorelli DA (2001) {Topology optimization of non-linear elastic
  structures and compliant mechanisms}. Computer Methods in Applied Mechanics
  and Engineering 190(26-27):3443--3459, \doi{10.1016/S0045-7825(00)00278-4}

\bibitem[{Coffin and Maute(2016{\natexlab{a}})}]{Coffin2016a}
Coffin P, Maute K (2016{\natexlab{a}}) A level-set method for steady-state and
  transient natural convection problems. Structural and Multidisciplinary
  Optimization 53(5):1047--1067, \doi{10.1007/s00158-015-1377-y}

\bibitem[{Coffin and Maute(2016{\natexlab{b}})}]{Coffin2016}
Coffin P, Maute K (2016{\natexlab{b}}) Level set topology optimization of
  cooling and heating devices using a simplified convection model. Structural
  and Multidisciplinary Optimization 53(5):985--1003,
  \doi{10.1007/s00158-015-1343-8}

\bibitem[{Deaton and Grandhi(2014)}]{Deaton2014}
Deaton JD, Grandhi RV (2014) A survey of structural and multidisciplinary
  continuum topology optimization: post 2000. Structural and Multidisciplinary
  Optimization 49(1):1--38, \doi{10.1007/s00158-013-0956-z}

\bibitem[{Dede(2009)}]{Dede2009}
Dede E (2009) Multiphysics topology optimization of heat transfer and fluid
  flow systems. In: Proceedings of the COMSOL Conference 2009 Boston

\bibitem[{Dilgen et~al(2018)Dilgen, Dilgen, Fuhrman, Sigmund, and
  Lazarov}]{Dilgen2018}
Dilgen SB, Dilgen CB, Fuhrman DR, Sigmund O, Lazarov BS (2018) {Density based
  topology optimization of turbulent flow heat transfer systems}. Structural
  and Multidisciplinary Optimization 57(5):1905--1918,
  \doi{10.1007/s00158-018-1967-6}

\bibitem[{Donea and Huerta(2003)}]{Donea2003}
Donea J, Huerta A (2003) {Finite Element Methods for Flow Problems}. John Wiley
  {\&} Sons, Ltd, Chichester, UK, \doi{10.1002/0470013826}

\bibitem[{Donoso and Sigmund(2004)}]{Donoso2004c}
Donoso A, Sigmund O (2004) {Topology optimization of multiple physics problems
  modelled by Poisson ' s equation}. Latin American Journal of Solids and
  Structures 1(2):169--189

\bibitem[{Dugast et~al(2018)Dugast, Favennec, Josset, Fan, and
  Luo}]{Dugast2018}
Dugast F, Favennec Y, Josset C, Fan Y, Luo L (2018) {Topology optimization of
  thermal fluid flows with an adjoint Lattice Boltzmann Method}. Journal of
  Computational Physics 365:376--404, \doi{10.1016/J.JCP.2018.03.040}

\bibitem[{Evgrafov(2006)}]{Evgrafov2006}
Evgrafov A (2006) {Topology optimization of slightly compressible fluids}. ZAMM
  86(1):46--62, \doi{10.1002/zamm.200410223}

\bibitem[{Fries and Matthies(2004)}]{Fries2004}
Fries TP, Matthies HG (2004) {A Review of Petrov-Galerkin Stabilization
  Approaches and an Extension to Meshfree Methods A Review of Petrov-Galerkin
  Stabilization Approaches and an Extension to Meshfree Methods}. Tech. rep.,
  Institute of Scientific Computing, Technical University Braunschweig,
  Braunschweig

\bibitem[{Gersborg-Hansen et~al(2005)Gersborg-Hansen, Sigmund, and
  Haber}]{Gersborg-Hansen2005}
Gersborg-Hansen A, Sigmund O, Haber R (2005) {Topology optimization of channel
  flow problems}. Structural and Multidisciplinary Optimization 30(3):181--192,
  \doi{10.1007/s00158-004-0508-7}

\bibitem[{Gersborg-Hansen et~al(2006)Gersborg-Hansen, Bends{\o}e, and
  Sigmund}]{Gersborg-Hansen2006a}
Gersborg-Hansen A, Bends{\o}e MP, Sigmund O (2006) {Topology optimization of
  heat conduction problems using the finite volume method}. Structural and
  Multidisciplinary Optimization 31(4):251--259,
  \doi{10.1007/s00158-005-0584-3}, \eprint{arXiv:1011.1669v3}

\bibitem[{Guest and Pr{\'{e}}vost(2006)}]{Guest2006a}
Guest JK, Pr{\'{e}}vost JH (2006) {Topology optimization of creeping fluid
  flows using a Darcy–Stokes finite element}. International Journal for
  Numerical Methods in Engineering 66(3):461--484, \doi{10.1002/nme.1560}

\bibitem[{Haertel and Nellis(2017)}]{Haertel2017}
Haertel JH, Nellis GF (2017) A fully developed flow thermofluid model for
  topology optimization of 3d-printed air-cooled heat exchangers. Applied
  Thermal Engineering 119:10 -- 24,
  \doi{https://doi.org/10.1016/j.applthermaleng.2017.03.030}

\bibitem[{Haertel et~al(2018)Haertel, Engelbrecht, Lazarov, and
  Sigmund}]{Haertel2018}
Haertel JH, Engelbrecht K, Lazarov BS, Sigmund O (2018) Topology optimization
  of a pseudo 3d thermofluid heat sink model. International Journal of Heat and
  Mass Transfer 121:1073 -- 1088,
  \doi{https://doi.org/10.1016/j.ijheatmasstransfer.2018.01.078}

\bibitem[{Haertel et~al(2015)Haertel, Engelbrecht, Lazarov, and
  Sigmund}]{Haertel2015}
Haertel JHK, Engelbrecht K, Lazarov BS, Sigmund O (2015) Topology optimization
  of thermal heat sinks. In: Proceedings of COMSOL conference 2015

\bibitem[{Iga et~al(2009)Iga, Nishiwaki, Izui, and Yoshimura}]{Iga2009}
Iga A, Nishiwaki S, Izui K, Yoshimura M (2009) {Topology optimization for
  thermal conductors considering design-dependent effects, including heat
  conduction and convection}. International Journal of Heat and Mass Transfer
  52(11-12):2721--2732, \doi{10.1016/J.IJHEATMASSTRANSFER.2008.12.013}

\bibitem[{Joo et~al(2017)Joo, Lee, and Kim}]{Joo2017}
Joo Y, Lee I, Kim SJ (2017) {Topology optimization of heat sinks in natural
  convection considering the effect of shape-dependent heat transfer
  coefficient}. International Journal of Heat and Mass Transfer 109:123--133,
  \doi{10.1016/j.ijheatmasstransfer.2017.01.099}

\bibitem[{Joo et~al(2018)Joo, Lee, and Kim}]{Joo2018}
Joo Y, Lee I, Kim SJ (2018) {Efficient three-dimensional topology optimization
  of heat sinks in natural convection using the shape-dependent convection
  model}. International Journal of Heat and Mass Transfer 127:32--40,
  \doi{10.1016/J.IJHEATMASSTRANSFER.2018.08.009}

\bibitem[{Koga et~al(2013)Koga, Lopes, Nova, de~Lima, and Silva}]{Koga2013}
Koga AA, Lopes ECC, Nova HFV, de~Lima CR, Silva ECN (2013) Development of heat
  sink device by using topology optimization. International Journal of Heat and
  Mass Transfer 64:759--772, \doi{10.1016/j.ijheatmasstransfer.2013.05.007}

\bibitem[{Laniewski-Wollk and Rokicki(2016)}]{Laniewski-Wollk2016}
Laniewski-Wollk L, Rokicki J (2016) Adjoint lattice boltzmann for topology
  optimization on multi-gpu architecture. Computers \& Mathematics with
  Applications 71(3):833 -- 848,
  \doi{http://dx.doi.org/10.1016/j.camwa.2015.12.043}

\bibitem[{Lazarov et~al(2014)Lazarov, Alexandersen, and Sigmund}]{Lazarov2014}
Lazarov BS, Alexandersen J, Sigmund O (2014) Topology optimized designs of
  steady state conduction heat transfer problems with convection boundary
  conditions. In: EngOpt 2014, \doi{10.13140/RG.2.2.29361.68966}

\bibitem[{Lazarov et~al(2018)Lazarov, Sigmund, Meyer, and
  Alexandersen}]{Lazarov2018}
Lazarov BS, Sigmund O, Meyer K, Alexandersen J (2018) Experimental validation
  of additively manufactured optimized shapes for passive cooling. Applied
  Energy 226:330--339, \doi{10.1016/j.apenergy.2018.05.106}

\bibitem[{Lei et~al(2018)Lei, Alexandersen, Lazarov, Wang, Haertel, Angelis,
  Sanna, Sigmund, and Engelbrecht}]{Lei2018}
Lei T, Alexandersen J, Lazarov BS, Wang F, Haertel JH, Angelis SD, Sanna S,
  Sigmund O, Engelbrecht K (2018) Investment casting and experimental testing
  of heat sinks designed by topology optimization. International Journal of
  Heat and Mass Transfer 127:396 -- 412,
  \doi{10.1016/j.ijheatmasstransfer.2018.07.060}

\bibitem[{Marck et~al(2013)Marck, Nemer, and Harion}]{Marck2013}
Marck G, Nemer M, Harion JL (2013) Topology optimization of heat and mass
  transfer problems: Laminar flow. Numerical Heat Transfer, Part B:
  Fundamentals 63(6):508--539, \doi{10.1080/10407790.2013.772001}

\bibitem[{Moon et~al(2004)Moon, Kim, and Wang}]{Moon2004}
Moon H, Kim C, Wang S (2004) Reliability-based topology optimization of thermal
  systems considering convection heat transfer. In: Proceedings of the 10th
  AIAA/ISSMO Multidisciplinary Analysis and Optimization Conference

\bibitem[{Okkels and Bruus(2007)}]{Okkels2007}
Okkels F, Bruus H (2007) {Scaling behavior of optimally structured catalytic
  microfluidic reactors}. Physical Review E 75(1):016,301,
  \doi{10.1103/PhysRevE.75.016301}

\bibitem[{Olesen et~al(2006)Olesen, Okkels, and Bruus}]{Olesen2006}
Olesen LH, Okkels F, Bruus H (2006) {A high-level programming-language
  implementation of topology optimization applied to steady-state
  {\{}N{\}}avier-{\{}S{\}}tokes flow}. International Journal for Numerical
  Methods in Engineering 65(7):975--1001

\bibitem[{Rodrigues and Fernandes(1995)}]{Rodrigues1995}
Rodrigues H, Fernandes P (1995) A material based model for topology
  optimization of thermoelastic structures. International Journal for Numerical
  Methods in Engineering 38(12):1951--1965, \doi{10.1002/nme.1620381202}

\bibitem[{Saglietti(2018)}]{Saglietti2018a}
Saglietti C (2018) On optimization of natural convection flows. PhD thesis, KTH
  Royal Institute of Technology, iSBN: 978-91-7729-820-5

\bibitem[{Saglietti et~al(2018)Saglietti, Wadbro, Berggren, and
  Henningson}]{Saglietti2018}
Saglietti C, Wadbro E, Berggren M, Henningson DS (2018) Heat transfer
  maximization in a three-dimensional conductive differentially heated cavity
  by means of topology optimization. In: Proceedings of the Seventh European
  Conference on Computational Fluid Dynamics (ECCM-ECFD) 2018

\bibitem[{Shakib et~al(1991)Shakib, Hughes, and Johan}]{Shakib1991}
Shakib F, Hughes TJ, Johan Z (1991) {A new finite element formulation for
  computational fluid dynamics: X. The compressible Euler and Navier-Stokes
  equations}. Computer Methods in Applied Mechanics and Engineering
  89(1-3):141--219, \doi{10.1016/0045-7825(91)90041-4}

\bibitem[{Sigmund(2001)}]{Sigmund2001}
Sigmund O (2001) {Design of multiphysics actuators using topology optimization
  – Part I: One-material structures}. Computer Methods in Applied Mechanics
  and Engineering 190(49-50):6577--6604, \doi{10.1016/S0045-7825(01)00251-1}

\bibitem[{Subramaniam et~al(2018)Subramaniam, Dbouk, and
  Harion}]{Subramaniam2018}
Subramaniam V, Dbouk T, Harion JL (2018) {Topology optimization of conductive
  heat transfer devices: An experimental investigation}. Applied Thermal
  Engineering 131:390--411, \doi{10.1016/J.APPLTHERMALENG.2017.12.026}

\bibitem[{Svanberg(1987)}]{Svanberg1987}
Svanberg K (1987) {The Method Of Moving Asymptotes - A New Method For
  Structural Optimization}. International Journal for Numerical Methods in
  Engineering 24(2):359--373, \doi{10.1002/nme.1620240207}

\bibitem[{Thellner(2005)}]{Thellner2005}
Thellner M (2005) {Multi-parameter topology optimization in continuum
  mechanics}. PhD thesis, Link{\"{o}}ping University, The Institute of
  Technology

\bibitem[{Wiker et~al(2007)Wiker, Klarbring, and Borrvall}]{Wiker2007}
Wiker N, Klarbring A, Borrvall T (2007) {Topology optimization of regions of
  Darcy and Stokes flow}. International Journal for Numerical Methods in
  Engineering 69(7):1374--1404, \doi{10.1002/nme.1811}

\bibitem[{Yaji et~al(2016)Yaji, Yamada, Yoshino, Matsumoto, Izui, and
  Nishiwaki}]{Yaji2016}
Yaji K, Yamada T, Yoshino M, Matsumoto T, Izui K, Nishiwaki S (2016) Topology
  optimization in thermal-fluid flow using the lattice boltzmann method.
  Journal of Computational Physics 307:355 -- 377,
  \doi{10.1016/j.jcp.2015.12.008}

\bibitem[{Yaji et~al(2018)Yaji, Ogino, Chen, and Fujita}]{Yaji2018}
Yaji K, Ogino M, Chen C, Fujita K (2018) {Large-scale topology optimization
  incorporating local-in-time adjoint-based method for unsteady thermal-fluid
  problem}. Structural and Multidisciplinary Optimization 58(2):817--822,
  \doi{10.1007/s00158-018-1922-6}

\bibitem[{Yamada et~al(2011)Yamada, Izui, and Nishiwaki}]{Yamada2011}
Yamada T, Izui K, Nishiwaki S (2011) A level set-based topology optimization
  method for maximizing thermal diffusivity in problems including
  design-dependent effects. ASME Journal of Mechanical Design 133(3):1--9,
  \doi{10.1115/1.4003684}

\bibitem[{Yan et~al(2018)Yan, Wang, and Sigmund}]{Yan2018}
Yan S, Wang F, Sigmund O (2018) On the non-optimality of tree structures for
  heat conduction. International Journal of Heat and Mass Transfer 122:660 --
  680, \doi{10.1016/j.ijheatmasstransfer.2018.01.114}

\bibitem[{Yin and Ananthasuresh(2002)}]{Yin2002}
Yin L, Ananthasuresh G (2002) {A novel topology design scheme for the
  multi-physics problems of electro-thermally actuated compliant
  micromechanisms}. Sensors and Actuators A: Physical 97-98:599--609,
  \doi{10.1016/S0924-4247(01)00853-6}

\bibitem[{Yoon(2010{\natexlab{a}})}]{Yoon2010}
Yoon GH (2010{\natexlab{a}}) Topological design of heat dissipating structure
  with forced convective heat transfer. Journal of Mechanical Science and
  Technology 24(6):1225--1233, \doi{10.1007/s12206-010-0328-1}

\bibitem[{Yoon(2010{\natexlab{b}})}]{Yoon2010a}
Yoon GH (2010{\natexlab{b}}) {Topological design of heat dissipating structure
  with forced convective heat transfer}. Journal of Mechanical Science and
  Technology 24(6):1225--1233, \doi{10.1007/s12206-010-0328-1}

\bibitem[{Zeng et~al(2018)Zeng, Kanargi, and Lee}]{Zeng2018}
Zeng S, Kanargi B, Lee PS (2018) Experimental and numerical investigation of a
  mini channel forced air heat sink designed by topology optimization.
  International Journal of Heat and Mass Transfer 121:663 -- 679,
  \doi{10.1016/j.ijheatmasstransfer.2018.01.039}

\bibitem[{Zhao et~al(2018)Zhao, Zhou, Sigmund, and Andreasen}]{Zhao2018}
Zhao X, Zhou M, Sigmund O, Andreasen CS (2018) {A “poor man's approach” to
  topology optimization of cooling channels based on a Darcy flow model}.
  International Journal of Heat and Mass Transfer 116:1108--1123,
  \doi{10.1016/j.ijheatmasstransfer.2017.09.090}

\bibitem[{Zhou et~al(2016)Zhou, Alexandersen, Sigmund, and {W.
  Pedersen}}]{Zhou2016}
Zhou M, Alexandersen J, Sigmund O, {W Pedersen} CB (2016) {Industrial
  application of topology optimization for combined conductive and convective
  heat transfer problems}. Structural and Multidisciplinary Optimization
  54(4):1045--1060, \doi{10.1007/s00158-016-1433-2}

\end{thebibliography}

\begin{appendices}

\section{Sensitivity analysis} \label{app:sensitivity}

The sensitivity of the objective function with respect to the design variables $\bm{\gamma}$ is determined using the adjoint sensitivity method. The objective function is augmented by the product of the residual and an adjoint vector:
\begin{align}
\hat{\psi} = \psi + \bm{\lambda}^{\textrm{T}} \mathbf{R}(\mathbf{s}),
\end{align}
where $\bm{\lambda}$ is the adjoint vector.

Differentiating with respect to the design variables $\bm{\gamma}$ yields:
\begin{equation}
\frac{\difd \hat{\psi}}{\difd \bm{\gamma}} = \frac{\partial \psi}{\partial \bm{\gamma}} + \frac{\partial \psi}{\partial \mathbf{s}} \frac{\difd \mathbf{s}}{\difd \bm{\gamma}} + \bm{\lambda}^{\textrm{T}} \left( \frac{\partial \mathbf{R}}{\mathbf{\partial} \bm{\gamma}} + \frac{\partial \mathbf{R}}{\partial \mathbf{s}} \frac{\difd \mathbf{s}}{\difd \bm{\gamma}}  \right)
\end{equation}
~\\
The adjoint vector is chosen such that terms involving $\frac{\difd \mathbf{s}}{\difd \bm{\gamma}}$ are eliminated. The following adjoint problem is thus solved:
\begin{equation}
\bm{\lambda}^{\textrm{T}} \frac{\partial \mathbf{R}}{\partial \mathbf{s}}= \frac{\partial \phi}{\partial \mathbf{s}}
\label{eq:adjointSystem}
\end{equation}
The matrix $\frac{\partial \mathbf{R}}{\partial \mathbf{s}}$ is recognized as the tangent matrix utilized in the Newton-Raphson algorithm which more conveniently may be determined as:
\begingroup
\renewcommand*{\arraystretch}{1.6}
\begin{equation}
\frac{\partial \mathbf{R}}{\partial \mathbf{s}} = \begin{bmatrix} \frac{\partial \mathbf{R}_p}{\partial \mathbf{P}} & \frac{\partial \mathbf{R}_p}{\partial \mathbf{T}} \\ \frac{\partial \mathbf{R}_t}{\partial \mathbf{P}} & \frac{\partial \mathbf{R}_t}{\partial \mathbf{T}} \end{bmatrix}
\end{equation}	
\endgroup
\noindent
When using the thermal compliance as objective function we have:
\begin{equation}
\frac{\partial \psi}{\partial \mathbf{s}} = \begin{pmatrix}
\mathbf{0} \\ \mathbf{f}_t
\end{pmatrix}^{\textrm{T}}
\end{equation}
~\\
After solving \eqref{eq:adjointSystem} for $\bm{\lambda}$ the sensitivities can be determined as:
\begin{equation}
\frac{\difd \psi}{\difd \bm{\gamma}} = \frac{\partial \psi}{\partial \bm{\gamma}} - \bm{\lambda}^{\textrm{T}} \frac{\partial \mathbf{R}}{\partial \bm{\gamma}} = - \bm{\lambda}^{\textrm{T}} \frac{\partial \mathbf{R}}{\partial \bm{\gamma}} 
\end{equation}

\section{Simplified convection model} \label{app:simplified}

The simplified convection model used for comparison herein, is based on Newton's law of cooling applied on the solid-fluid interface using a design field gradient approach. This was first presented for a density-based topology optimisation approach by \citep{Lazarov2014}. Instead of applying a design-dependent convection boundary condition on interfaces between all elements based on the jump in the physical element design field \citep{Bruns2007,Alexandersen2011}, a mathematically consistent and convergent volumetric formulation is posed based on the gradient of the physical design field. The approach can be seen as loosely equivalent to that presented for a level set approach by \citep{Yamada2011}, in that the gradient of the design field is equivalent to the gradient of the level set field, which defines the surface. For an overview of this approach, please see the work by \citep{Alexandersen2011,Coffin2016}.

Shortly, the governing partial differential equation becomes:
\begin{equation}
-\frac{\partial}{\partial x_{i}}\left( k(\gamma) \frac{\partial T}{\partial x_{i}} \right) + \norm{ \frac{\partial \gamma}{\partial x_{i}} }_{2} h \left( T - T_0 \right) - Q = 0
\end{equation}
where $h$ is the convection coefficient and in the discrete case $\norm{ \frac{\partial \gamma}{\partial x_{i}} }_{2} = n_{i} d\Gamma_{fs}$, with $n_{i}$ being the surface normal and $\Gamma_{fs}$ being the fluid-solid interface.
For the full derivation, please see \cite{Lazarov2014}.

The optimization process is carried out with a continuation approach for the filter radius. The filter radius starts large to smooth out boundary effects initially and then gradually decreased using the sequence $r_{\text{min}} \in \{0.48,0.36,0.24,0.12\}$ switching after 50 iterations or when the objective functional change has been under $10^{-3}$ for 10 consecutive iterations. The final length scale is double the size of that used for the fluid model results because the simplified model favours very small features of high complexity with many internal voids. Therefore, the larger length scale results in a fairer comparison as it reduces the non-physical artifacts. The conductivity is interpolated using the modified SIMP approach, $k(\gamma) = k_{\text{min}} + (k_s - k_{\text{min}})\gamma^{p}$, with $k_s = 100$, $k_{\text{min}} = 10^{-6}$ and a constant penalization factor of $p=6$. The relatively high penalization factor produces better performing results, but is not necessary to yield well-defined 
topologies, as the surface convection term introduces automatic penalisation of intermediate densities
\end{appendices}

\end{document}